\documentclass[aip,graphicx,amsmath,amssymb,preprint]{revtex4-1}

\usepackage{graphicx}
\usepackage{dcolumn}
\usepackage{bm}
\usepackage{xcolor}
\usepackage{soul}  
\usepackage{cancel}

\usepackage{subcaption}
\usepackage{float}
\usepackage{lineno,hyperref}

\newcommand{\der}[2]{\dfrac{\partial #1}{\partial #2}}
\newcommand{\Ipdemi}[1]{{#1}_{i+\frac{1}{2},j}}

\usepackage{enumitem}
\usepackage{ stmaryrd }
\definecolor{rltgreen}{rgb}{0,0.5,0}
\newcommand{\CLG}        [1] {}
\newcommand{\CR}        [1] {}
\newcommand{\CG}        [1] {}
\newcommand{\CO}        [1] {}
\newcommand{\CB}        [1] {}

\providecommand\bnabla{\boldsymbol{\nabla}}

\usepackage[utf8]{inputenc}
\usepackage[T1]{fontenc}
\usepackage{mathptmx}

\begin{document}


\title{Numerical investigation of three-dimensional partial cavitation in a Venturi geometry}



\author{Camille GOUIN}
 \email{camille.psb@hotmail.fr}
 \affiliation{Dynfluid laboratory, ENSAM, 151 Boulevard de l'Hôpital, 75013 Paris, France.}
\author{Carlos JUNQUEIRA-JUNIOR}%
\affiliation{Dynfluid laboratory, ENSAM, 151 Boulevard de l'Hôpital, 75013 Paris, France.}
\author{Eric GONCALVES DA SILVA}%
\affiliation{Institut Pprime, UPR 3346 CNRS, ISAE-ENSMA, 1 avenue Clément Ader, 86961 Futuroscope Chasseneuil cedex, France.}
\author{Jean-Christophe ROBINET}%
\affiliation{Dynfluid laboratory, ENSAM, 151 Boulevard de l'Hôpital, 75013 Paris, France.}

\date{\today}


\begin{abstract}
Sheet cavitation appears in many hydraulic applications and can lead to technical issues. Some fundamental outcomes such as the complex topology of 3-Dimensional cavitation pockets and their associated dynamics need to be carefully visited. In the paper, the dynamics of partial cavitation developing in a 3-D Venturi geometry and the interaction with sidewalls are numerically investigated. The simulations are performed using a one-fluid compressible Reynolds-Averaged Navier–Stokes (RANS) solver associated to a non-linear turbulence model and a void ratio transport-equation model. A detailed analysis of this cavitating flow is carried out using innovative tools such as Spectral Proper Orthogonal Decompositions. Particular attention is paid in the study of 3-D effects by comparing numerical results obtained with sidewalls and periodic conditions. A three-dimensional dynamics of the sheet cavitation, unrelated to the presence of sidewalls, is identified and discussed.
\end{abstract}


\maketitle 



\section{Introduction}



Cavitation is the formation of vapor cavities in a liquid due to a pressure drop. The phenomenon occurs in hydraulic systems or turbomachinery and can, eventually, cause structural damage, noise and degrade the performance of the apparatus. Such effects drive the study of the different types of cavitation, and in particular considerable efforts have been made to explore the dynamics of partial cavities appearing along solid bodies. Such cavitation pockets are characterised by a fluctuating closure region leading to cavity length oscillations and the shedding of vapor structures. Partial cavity can be classified in two main forms of appearance: closed or quasi-stable cavity and open cavity or cloud cavitation, depending on the flow in the cavity closure region \citep{Laberteaux2001}. A quasi-stable cavity presents only small shedding at its closure region with a relatively stable cavity length. In opposite, cloud cavitation is a highly unsteady phenomenon presenting a periodically varying length that is associated to the large shedding of vapor clouds. Both types of cavities have been studied, experimentally and numerically, to describe the physical mechanism, the internal structure of cavities, the turbulence-cavitation interaction and to investigate the transition from quasi-stable to cloud cavitation \citep{Lange1994,Kawanami1997,Reisman98,Gopalan00,Callenaere2001,Coutier-Delgosha2006,Hayashi2014,Kravtsova2014,Jahangir2018}. 
Two main mechanisms have been identified for the break-off cycles: the development of a liquid re-entrant jet and the propagation of pressure waves created by the cloud collapses \citep{Arndt2000,Stanley2011,Stanley2014,Ganesh2016,Charriere2017}.

The structures of partial cavities have a fully three-dimensional topology as observed on hydrofoils with high-speed imaging method. The re-entrant jet does not progress only on the streamwise direction and a spanwise component was depicted \citep{Lange1997}. To distinguish between various directions of the re-entrant flow, the term side-entrant jet was introduced. This term refers to the part of the jet that has a strong spanwise velocity component directed into the cavity originating from the sidewalls. The term re-entrant jet or middle jet is reserved for the flow originating from the part of the cavity where the closure is more or less perpendicular to the incoming flow and is thus mainly directed upstream. Foeth et al.\@ (2006)\cite{Foeth2006} investigated the cavitating flow structures on the Delft twisted hydrofoil and reported the joint action of the re-entrant and side-entrant jets in the shedding. Dular et al.\@ (2007)\cite{Dular2007} studied hydrofoils with swept leading edges and their experiments showed that the re-entrant jet velocity has a spanwise component if the closure line of the cavity is inclined. The numerical simulation of Schnerr et al.\@ (2008)\cite{Schnerr2008} on a twisted hydrofoil showed that the interactions between the re-entrant jet and the spanwise velocity component also cause separation of sheet cavities. The spanwise pressure gradients greatly affect the three-dimensional structure of the cavity, which cause the U-shaped feature observed and discussed by many authors \citep{Kubota1989,Peng2016}. Recently, experimental observations on hydrofoils highlighted the important role played by the two side-entrant jets which propagated diagonally upstream to the leading edge affecting the attached part of the cavity \citep{Kadivar2019,Che2019}. Such comments was also reported in the numerical study of Long et al.\@ (2018)\cite{Long2018} on a twisted hydrofoil using a Lagrangian method.

Another discussion on three-dimensional cavity structures concerns the existence of an oblique mode of the pocket oscillations. This mode was firstly discussed by Decaix and Goncalves (2013)\cite{Decaix2013} on a quasi-stable cavity appearing on a Venturi geometry using Scale-Adaptive Simulation (SAS). Later, Timoshevskiy et al.\@ (2016)\cite{Timoshevskiy2016} maintained that the oblique mode associated with the development of the spanwise instability exists for all test objects independent of their shape. Yet, the existing visualisations did not clearly report such alternating sheet movement. On the other hand, various experimental studies showed different cavity shedding appearances and behaviors due to the influence of the scale of the geometry and the surface effects \citep{Kawakami2008, Dular2012}. Authors observed that at a certain ratio between the length of the sheet cavity and the channel width, an irregular break-off pattern occured.

The experimental study of complex three-dimensional sheet and cloud cavitation still suffers from the limitation in experimental technique, thus the numerical simulation appears to be an attractive tool for a better understanding of the two-phase structures and their dynamics. The framework for such turbulent cavitating flows is usually the one-fluid mixture approach based on an average statistical treatment with local thermodynamics assumption. Two main families are often used: the Homogeneous Equilibrium Model (HEM) with a suitable equation of state for the liquid-vapor mixture \citep{Delannoy90,Clerc00,Sinibaldi06} or the Transport Equation Model (TEM) or Homogeneous Relaxation Model (HRM) involving a transport equation for the void ratio \citep{Downar-Zapolski1996,Kunz2000,Zwart2004,Helluy06}. This equation includes a source term modeling the mass transfer between phases. Another crucial point for cavitating flow simulation is the turbulence modeling. Different approaches have been investigated in order to capture the finer-scale dynamics. Firstly, computations were performed solving the unsteady Reynolds-Averaged Navier–Stokes (URANS) equations naturally adapted to the two-phase averaged models \citep{Saito2007,Park2013} and more recently using advanced models such as SAS \citep{Sedlar2016,Hidalgo2019}, Filter-Based turbulence Model (FBM) \cite{Sun2019} or Partially-Averaged Navier–Stokes (PANS) \citep{Ji2013}. 
As the URANS approach did not fully account for the turbulent-cavitation interactions, Large Eddy Simulations were tested on both hydrofoil and Venturi geometries \citep{Huang2014, Ji2013a, Gnanaskandan2016a, Chen2019, Sun2020}. Yet, due to the large Reynolds number of usual applications (greater than 1 million), the necessity to consider the channel with sidewalls, the use of very small time step and the problem of the statistics convergence for low-frequency periodic phenomenon, the cost a well-resolved simulation remains prohibitive even using supercomputers. To circumvent such difficulties, micro-channel cavitating flows has been considered by Egerer et al.\@ (2014)\cite{Egerer2014}.\\

The present study focuses on a 3-D quasi-stable cavity flow on a 4$^o$ divergent angle Venturi geometry, used in the experiment of Barre et al.\@(2009)\cite{Barre2009}. The selection of the configuration is motivated by the modest number of experiments with quantitative data and by a configuration compatible with URANS approach. Particular attention is paid to sidewalls effects and the 3-D topology of the pocket. In the first part, the system of equations and the numerical formulation is developed. Numerical simulations are performed, in a second part, to compare results with experimental data. Then, the results of 3-D computations with sidewalls and with periodic boundary conditions are investigated. Numerical tools such as Power Spectral Densities or Spectral Proper Orthogonal Decompositions are used to compare both cases and analyse the flow dynamics. In the last part, the authors' interpretation of the encountered phenomenon is discussed regardings the dominant mechanisms of sheet cavitation flows.



\section{Governing equations}

\subsection{The 1-fluid homogeneous approach}

There are several ways to simulate two-phase flows, the most straightforward one is to use a two-fluid model.
Nevertheless, in the case of sheet cavitation simulation, this choice would lead to unaffordable computational costs and difficulties related to the interface tracking with the creation and the destruction of vapor pockets or the transfer terms computation due to the phase change. 
%
Therefore, a one-fluid homogeneous approach is selected in the present study and hypothesis over thermodynamical and mechanical equilibrium between the liquid and vapor phases are applied \citep{Merkle98}. The flow is considered as a mixture and the phases are assumed to share the same pressure, velocity and temperature. The averaged fraction of presence $\alpha_k$, for a given $k$ phase, is introduced to define the conservative form of mixture properties as:
\begin{eqnarray}
 \rho_m &=& \sum_k \alpha_k \rho_k\,\mbox{,}\\
 \rho_m u_{i,m} &=& \sum_k \alpha_k \rho_k u_{i,k}\,\mbox{,}\\
 \rho_m e_m &=& \sum_k \alpha_k \rho_k e_k.
\end{eqnarray}

\subsection{The compressible RANS equations}

The compressible Reynolds-Averaged Navier-Stokes (RANS) system of equations is used
to calculate the two-phase flow in the present work. The $k-\ell$ two-equations model of Smith \citep{Smith90,Smith94} is selected to calculate turbulence quantities. The choice of the turbulence model is motivated by previous works over a panel of models \cite{Goncalves2012,Charriere2015,Charriere2017}.\\
A limiter term is applied to the calculation of the turbulent viscosity property of mixture
fluid, $\mu_{tm}$. The correction is motivated by previous results \citep{Reboud98,Decaix12}, which indicate an overestimation of such a quantity for the two-phase flow configurations of interest. The limitation, for the $k_m-\ell_m$ model, is computed using a function over $\rho_m$ and is here written as :
\begin{eqnarray}
 \mu_{tm}&=&f(\rho_m)\frac{\Phi \sqrt{2k_m} \ell_m}{B_1^{1/3}}, \label{eqn:limiteur_viscosite_turbulente}\\
f(\rho_m)&=&\rho_v+\left(\dfrac{\rho_v-\rho_m}{\rho_v-\rho_l}\right)^n(\rho_l-\rho_v)\,\mbox{,}\label{eqn:reboud}
\end{eqnarray}
where $\rho_v$ and $\rho_l$ stand for the saturation vapor and liquid density, respectively. $\Phi$ and $B_1$ come from the $k-\ell$ model described by Smith (1994)\cite{Smith94}. The limitation is controlled by the parameter $n >>1$ which is precised in section \ref{sec:numerical_val}. Furthermore, the work of Dandois (2014)\cite{Dandois2014} indicates the possibility of non-physical flow results related to an overestimation of the turbulent viscosity in corners. Hence, the Quadratic Constitutive Relation (QCR) correction \citep{Spalart2000} is applied into the Reynolds stress tensor:
\begin{equation}
 \tau_{m,ij}^{QCR} = \tau_{m,ij} - c_{nl1}(O_{ik}\tau_{m,jk} + O_{jk}\tau_{m,ik})\,\mbox{,}
\end{equation}
where $c_{nl1}=0.3$ is an empirical constant and $O_{ik}$ is the normalised rotation tensor:
\begin{equation}
 O_{ik}=\frac{\der{u_i}{x_k}-\der{u_k}{x_i}}{\sqrt{\der{u_n}{x_p}\der{u_p}{x_n} } }\,\mbox{,}
\end{equation}

\noindent The mixture viscous stress tensor $\sigma_m$ and the heat flux vector $q_m$ are defined as
\begin{equation}
\sigma_{m,ij}=\mu_m \left[ \frac{\partial u_{m,i}}{\partial x_j} + \frac{\partial u_{m,j}}{\partial x_i} -\frac{2}{3}\frac{\partial u_{m,n}}{\partial x_n} \delta_{ij} \right]\,\mbox{;} \ \quad
q_{m,i}=\lambda_m \frac{\partial T_m}{\partial x_i}\,\mbox{;}
\end{equation}
where $\lambda_m = \sum_k \alpha_k\lambda_k$, is the thermal conductivity of the mixture and 
$\lambda_k$ is the thermal conductivity of the $k$-th phase.
Moreover, the mixture turbulent stress tensor $\tau_m$ and the turbulent heat flux vector $q^t_m$ are formulated using the Boussinesq relation and the Fourier law, respectively defined as:
\begin{equation}
 \tau_{m,ij} = \mu_{tm} \left[ \der{u_{m,i}}{x_j} + \der{u_{m,j}}{x_i} - \frac{2}{3} \der{u_{m,n}}{x_n} \delta_{ij} \right] - \dfrac{2}{3}\rho_mk_m\delta_{ij}\,\mbox{,}
\end{equation}
\begin{equation}
	q_{m,j}^{t} = \lambda_{tm} \der{T_m}{x_j} \approx \frac{\mu_{tm} C_{pm}}{Pr_t} \der{T_m}{x_j}\,\mbox{,}
\end{equation}
with turbulent Prandtl number $Pr_t=1$. Due to the lack of data over turbulent two-phase flow, the value of the Prandtl number is transposed from aerodynamic studies for monophasic flow. The thermal capacity of the mixture $C_{p_m}$ is defined based on $C_{p_v}$ and $C_{p_l}$,
which stand for the thermal capacity of the vapor and the liquid, respectively,
\begin{equation}
  \rho_m C_{p_m}(\alpha) = \alpha \rho_v C_{p_v} + (1-\alpha)\rho_l C_{p_l}\,\mbox{.}
\label{eqn:rho_mCpm}
\end{equation}

\subsection{The four-equation cavitation model}

The cavitation modeling approach used in the current work combines the mass conservation, momentum, and energy equations of the Navier-Stokes formulation to another transport equation over the fraction of presence of
phases. Moreover, an appropriate set of equations of state is used to model the cavitation.

\subsubsection{Void ratio transport equation}

The void ratio $\alpha$ is defined as the averaged fraction of presence for the vapor phase. A transport equation for the void ratio is added to complete the cavitation model:
\begin{equation}
\der{\alpha}{t}+u_{m,j}\der{\alpha}{x_j} = K\der{u_{m,j}}{x_j} + \dfrac{\dot{m}}{\rho_I}\,\mbox{.} \label{eqn:ttv_1}
\end{equation}
The formulation of the transport equation is based on the work of Saurel et al.\@ (2008)\cite{Saurel2008} and written by Goncalves (2013)\cite{Goncalves2013} in a four-equation model. The mass flow rate $\dot m$ from liquid to vapor\cite{Goncalves2013} can be written as:
\begin{equation}
    \dot m = \dfrac{\rho_l\rho_v}{\rho_l - \rho_v} \left( 1- \dfrac{c_m^2}{c^2_{Wallis}} \right) \der{u_{m,j}}{x_j}\, \mbox{.}
\end{equation}
The calculation is based on the Wallis speed of sound  $c_{Wallis}$, which is expressed as a weighted harmonic mean of each one of the two flow phases speed of the sound \citep{Wallis67}:
\begin{equation}
    \dfrac{1}{\rho_m c_{Wallis}^2}=\dfrac{\alpha}{\rho_v c_v^2}+\dfrac{(1-\alpha)}{\rho_l c_l^2}\,\mbox{,}
    \label{Wallis}
\end{equation}
where $c_k$ stands for the pure phase speed of sound. The interface density $\rho_I$ and the constant $K$ are respectively defined as 
\begin{equation}
\rho_I = \dfrac{\dfrac{\rho_lc_l^2}{1-\alpha}+\dfrac{\rho_vc_v^2}{\alpha}}{\dfrac{c_l^2}{1-\alpha}+\dfrac{c_v^2}{\alpha}}\quad  \text{and}\quad K = \dfrac{\rho_lc_l^2 - \rho_vc_v^2}{\dfrac{\rho_vc_v^2}{\alpha} + \dfrac{\rho_lc_l^2}{1-\alpha}}\,\mbox{.}
\end{equation}

\subsubsection{Equations of state}

Two different equations of state (EOS) are used for the mixture temperature $T_m$ and the mixture pressure $p_m$ depending on a pressure threshold. Pressure and temperature are defined by the stiffened gas EOS for the pure phase while sinusoidal EOS are applied for computing the mixture part of the flow \citep{Charriere2015}. The threshold is calculated from the vaporisation pressure $P^{vap}$ and a delta pressure based on a chosen parameter $c_{min}$, the minimal speed of sound in the mixture:
\begin{equation}
\Delta p_m = \left(\dfrac{\rho_l-\rho_v}{2}\right)c_{min}^2\dfrac{\pi}{2} \mbox{.}
\label{eq:delta_p}
\end{equation}
The selection of $c_{min}$ is based on the study of Charriere (2015)\cite{these_Charriere2015} and set to $0.472 \ m.s^{-1}$. This parameter allows the activation in advance of the phase change in order to smooth the density gradient around the interface. 
%
The density jump between the liquid and the vapor is stiff for a mixture problem. Thus, the mixture pressure is computed according to the relation :
\begin{align}
    \left\{
    \begin{array}{ll}
            p_{SG} & \qquad \mathrm{if}\quad p_m \geq P^{vap}+\Delta p \\
            p_{sinus} & \qquad \mathrm{otherwise}\,\mbox{,} \\
    \end{array}
    \right.
\end{align}
with $p_{SG}$ and $p_{sinus}$ the pressure respectively defined by the stiffened gas EOS and the sinusoidal EOS:
\begin{equation}
 p_{SG}(\rho_m,e_m) = (\gamma_m-1)\rho_k(e_m-\hat q_m) - \gamma_m p_{m,\infty}\,\mbox{,}
\label{eq_phase_gaz_raide_pression}
\end{equation}
 \begin{equation}
   p_{sinus}(\alpha) = P_{vap} + \left(\dfrac{\rho_l - \rho_v}{2}\right)c^2_{min} \arcsin(1-2\alpha)\,\mbox{,}
\label{eqn:sinus} 
\end{equation}
where the mixture energy of formation $\rho_m\hat q_m=\alpha \rho_v\hat q_v +(1-\alpha) \rho_l \hat q_l$ is calculated from $\hat q_v$ and $\hat q_l$, which stand respectively to the vapor and the liquid energies of formation. The mixture temperature is set equally above and below the pressure threshold :
\begin{equation}
T_m(\rho_m,e_m) = \dfrac{h_m(\alpha)-\hat q_m(\alpha)}{C_{p_m}(\alpha)}=\dfrac{e_m(\alpha)+p_m(\alpha) / \rho_m(\alpha) - \hat q_m(\alpha)}{C_{p_m}(\alpha)}\,\mbox{,}
\label{eq_melange_gaz_raide_temp}
\end{equation}
where $h_m$ and $e_m$ are the specific mixture enthalpy and internal energy respectively. The mixture speed of sound $c_m$ is processed following the same approach. Above the pressure threshold, the Wallis speed of sound, Eq.\@ \eqref{Wallis}, is considered while below the threshold, the speed of sound is computed using the sinusoidal EOS, Eq.\@ \ref{speed_of_sound}. 
The reader can find more details on the study of the speed of sound development performed by Charriere (2015)\cite{these_Charriere2015}. For the current case, the phase change does not affect the temperature of the mixture. Therefore, the phase enthalpy $h_k$ and phase density $\rho_k$ are defined as constants for a reference temperature $T_{ref}$.
%
\begin{equation}
 c_m^2=(\gamma_m-1)\dfrac{\rho_v\rho_l(h_v^{ref}-h_l^{ref})}{\rho_m(\rho_l-\rho_v)}+\dfrac{A\> c_{min}^2}{\sqrt{1-(A(1-2\alpha))^2}}\,\mbox{.}
\label{speed_of_sound}
\end{equation}
The phase enthalpy is $h_k^{ref}= C_{p_k}T_{ref} + q_k$. 
The $A$ coefficient is added in order to guarantee the velocity fitting with Wallis speed of sound above the pressure threshold. 
In the current case, $A$ is fixed to $0.9999$\cite{Goncalves2009}.




\section{Numerical formulation}

The global system is the four-equation model coupled with the turbulence model:
\begin{equation}
 \label{eqn:systeme_1}
 \der{\bm{w}}{t} + \bnabla\cdot\left[\bm{F}_c(\bm{w}) - \bm{F}_v(\bm{w})\right] = \bm{S}(\bm{w}),
\end{equation}
with
\[
 \bm{w}=\left(\begin{array}{c}
            \rho_m\\
            \rho_m u_{m,i}\\
            \rho_m E_m \\
            \alpha\\
	    \rho_m \psi_k
          \end{array}
          \right),
 \
  \bm{F}_v=\left(\begin{array}{c}
            0\\
	    \sigma_{m,ij}+\tau_{m,ij}\\
            (\sigma_{m,ij} + \tau_{m,ij})u_{m,j} - q_{m,j} - q_{m,j}^t\\
            0\\
	    \left(\mu_m+\dfrac{\mu_m^t}{\sigma_{\psi_k}}\right)\der{\psi_k}{x_j}
          \end{array}
          \right),
\]
\[
  \bm{F}_c=\left(\begin{array}{c}
            \rho_mu_{m,j}\\
            \rho_m u_{m,i}u_{m,j} + p_m\\
            \rho_m E_m +p_m\\
            \alpha u_{m,j} \\
            \rho_m \psi_k u_{m,j}
          \end{array}
          \right),
\
\bm{S} = \left(\begin{array}{c}
            0\\
            0\\
            0\\
            (K+\alpha)\der{u_{m,j}}{x_j} + \dfrac{\dot{m}}{\rho_I}\\
	    C_{\psi_k}
          \end{array}
          \right).
\]
The variable $\psi_k$ depends on the choice of the turbulence model of $k$ equations. The turbulent source terms $C_{\psi_k}$ and the constant $\sigma_{\psi_k}$ also relies on this model.

\subsection{Low Mach number preconditionning}

Some parts of the flow into the venturi configuration are supposed to be incompressible (Mach number around 0.1) despite using a compressible hypothesis. Therefore, it is necessary to use a low Mach number preconditioning method to deal with numerical errors and stiffness of the equation system. A preconditioning matrix is computed based on the work of Turkel (1987)\cite{Turkel87} using a $\beta$ all-speed flow parameter proportional to the Mach number
\citep{Choi1993},
\begin{equation}
\beta^2 = \min[\max(M^2, \theta M^2_{\infty}) , 1].
\end{equation}
with the constant $\theta$ set to 3. The preconditioning is applied only on the dissipation terms to preserve the time discretisation consistency.

\subsection{Time integration}

An explicit third-order Strong Stability Preserving Runge-Kutta method (SSPRK3), described by Spiteri and Ruuth (2002)\cite{Spiteri2002} and Gottlieb (2005)\cite{Gottlieb2005}, is used as a time-marching scheme.
The explicit equation is given by 
\begin{equation}
    \bm{w}^{n+1} = \bm{w}^n + h\left(\frac{1}{6}\bm{k}_1 + \frac{1}{6}\bm{k}_2 + \frac{2}{3}\bm{k}_3\right),
    \label{eq:rk}
\end{equation}
with
\begin{equation*}
\left\lbrace
 \begin{array}{lll}
  \bm{k}_1 &=& \bm{F}(t_n,\bm{w}_n),\\
  \bm{k}_2 &=& \bm{F}(t_n+h,\bm{w}_n+h\bm{k}_1),\\
  \bm{k}_3 &=& \bm{F}(t_n+\frac{h}{2},\bm{w}_n+\frac{h}{4}(\bm{k}_1+\bm{k}_2)) \, \mbox{,}
 \end{array}
  \right.
\end{equation*}
in which $h$ is the time-step and $\bm{F}$ represents the numerical fluxes and the source terms of Eq.\@\eqref{eqn:systeme_1}.

\subsection{Spatial discretisation}

A cell-centered finite-volume technique is used for the spatial discretisation of the RANS equations. The numerical fluxes are calculated using a centered scheme coupled with an artificial dissipative term. 
%
The chosen scheme is based on the 2$^{\text{nd}}$ order Jameson-Schmidt-Turkel \citep{Jameson1981}
and, it is extended to the precision of the 3$^{\text{rd}}$ order. 
%
Furthermore, an additional term is added to the dissipation, with a density sensor $\eta_i^{(I)}$, to allow dissipation around the phase interface:
\begin{equation}
 \eta_i^{(I)} \ = \ \frac{|\rho_{i+1}-2\rho_i+\rho_{i-1}|}{\rho_{i+1}+2\rho_i+\rho_{i-1}}.
\end{equation}
The global scheme formulation is developed in Appendix \ref{ap:JST3}.

\subsection{Boundary conditions}

The venturi-type geometry configuration used in the current article requires the use of wall and inlet/outlet boundary conditions. The former is implemented here using wall-functions regarding a less expensive representation of the boundary layer while the latter is calculated using Euler characteristic equations.

\paragraph{Wall function:}

The boundary condition for a wall is defined by the following wall function:
\begin{equation}
 \begin{array}{cclcc}
  u^{+}&=&y^{+} &\text{if}&\quad y^{+}<11.13,\\
 u^{+}&=&\dfrac{1}{\kappa}\ln y^{+} + 5.25 &\text{if}&\quad y^{+}>11.13,
 \end{array}
\end{equation}
with the Von Karman constant $\kappa = 0.41$. This no slip boundary condition combined  with the adiabatic hypothesis for walls results in normal derivatives of the void ratio, the density and the pressure are set to zero at the wall boundary.
\paragraph{Inlet and outlet boundaries:} The void ratio $\alpha$, the phase density $\rho_k$ and the velocity components are imposed at the inlet boundary. Then, the pressure is computed using the Euler's characteristic equations:
\begin{equation}
    \begin{array}{rcl}
         -c^2 (\rho^b - \rho^s) + (P^b-P^s) & = & 0,  \\
          v^b - v^s & = & 0, \\
          \rho(\alpha^b - \alpha^s) - K (\rho^b-\rho^s) & = & 0, \\
          (\lambda_+-u)(P^b-P^s)+\rho \beta^2 c^2(u^b-u^s) & = & 0, \\
          (\lambda_--u)(P^b-P^s)+\rho \beta^2 c^2(u^b-u^s) & = & 0,
    \end{array}
    \label{caracteristiques}
\end{equation}
where $b$ index stands for boundary variables and $s$ index stands for variables computed with the numerical scheme. $\lambda_\pm$ are the highest and the lowest eigenvalues of the preconditionning system. The static pressure is selected while other variables are calculated using the characteristic equations at the outlet boundary condition.




\section{Ventury configuration and comparison}\label{sec:num_val} \label{sec:numerical_val}

The present section is devoted to introduce the studied case and to compare numerical results with experimental data. Previous computations and comparisons with literature using the current cavitation model, for different configurations (expansion tube, underwater explosion with cavitation, compression of a vapour bubble, venturis, shock tubes, ...), has already been published \citep{Goncalves2013,Goncalves2014,Charriere2015,Goncalves2018}.

\subsection{Case set up}

\begin{table}[H]
\begin{ruledtabular}
\begin{tabular}{cccc}
        $L_x$ & $h_{in}$ & $h_{throat}$ & $L_y$ \\
        \hline
        $1.512 \ m$ & $0.05 \ m$ & $0.0437 \ m$ & $0.044 \ m$ \\
\end{tabular}
\end{ruledtabular}
\caption{Geometric dimensions.}
\label{tab:expe_parameters}
\end{table}

\begin{figure*}
    \centering
	\includegraphics[width=0.95\textwidth]{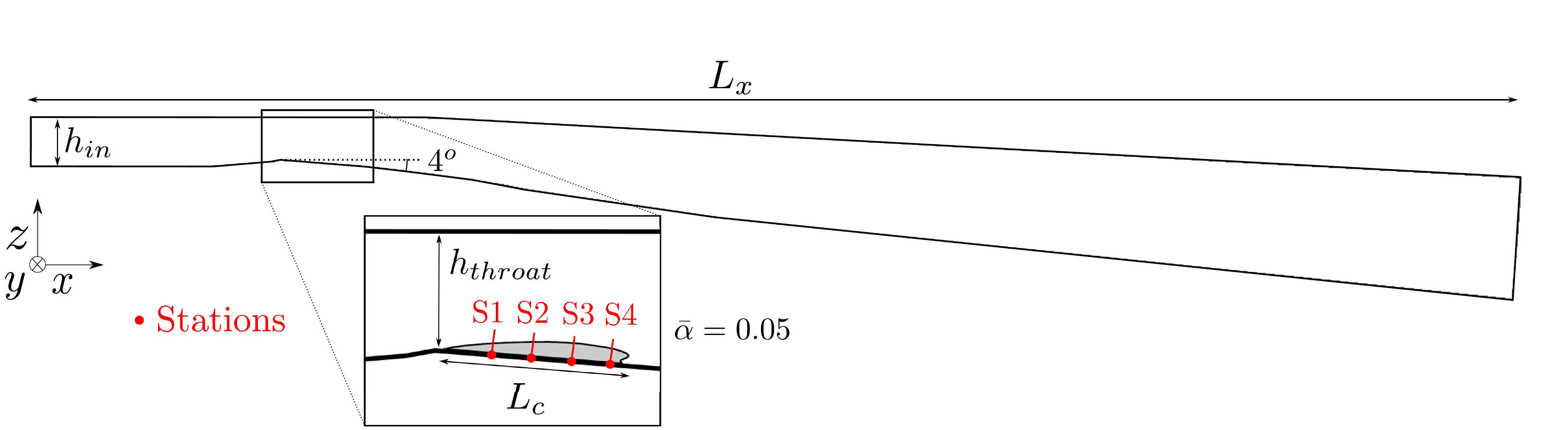}
    \caption{Schematic view of the venturi used in the computation.}
    \label{fig:geometry}
\end{figure*}

A 4$^\text{o}$ divergence angle venturi configuration, as the one used in the experiment of Barre et al.\@ (2009)\cite{Barre2009}, is selected for the study. Figure \ref{fig:geometry} illustrates the venturi geometry, and Tab.\@ \ref{tab:expe_parameters} indicates flow and sections parameters used in the current case. Probes positioning is calibrated to capture data adjacent to the cavitation pocket at four stations (S1 to S4 in Fig.\@\ref{fig:geometry}) located at $20.9 \ mm$, $38.4 \ mm$, $55.8 \ mm$ and $73.9 \ mm$ from the venturi throat. The inflow parameters are set as followed: the streamwise velocity $u_{in}=10.8 m.s^{-1}$, the temperature $T_{in}=293 \ K$, the void ratio $\alpha_{in}=10^{-10}$, the density $\rho_{in}=1000.831 \ kg.m^{-3}$ and the vaporisation pressure $P_{vap}=2339 \ Pa$. The inflow cavitation number $\sigma_{in}$ is $0.55$ and the inflow Reynolds number is $Re_{in}=\rho_{in}u_{in}h_{in}/\mu_{in}=5.4\times 10^5$. The outflow pressure is calibrated to correspond with this cavitation number. The study is focused on one operating point corresponding to the selected experiment set up.

\begin{figure*}
    \centering
	\includegraphics[width=0.95\textwidth]{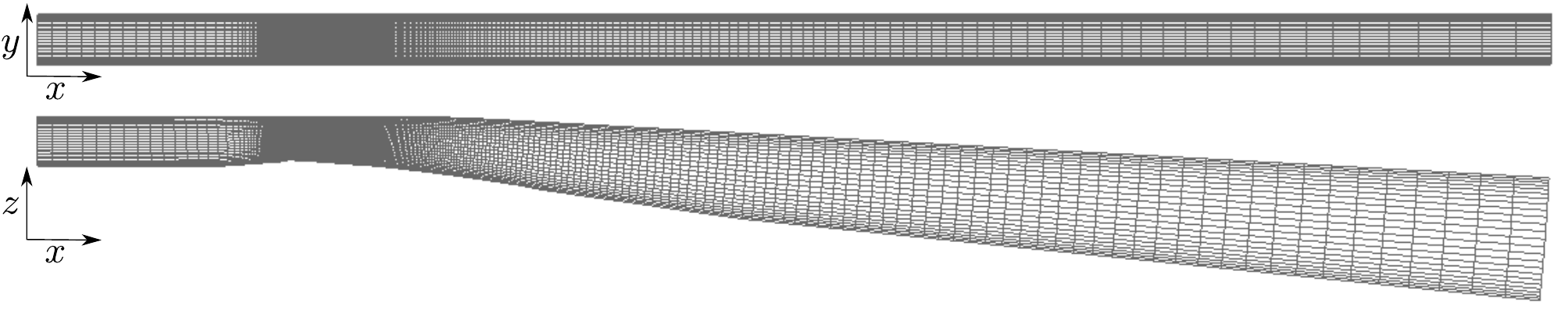}
    \caption{Mesh generation of the 3-D case represented with one visible mesh point out of three in the $y$ and $z$ directions.}
    \label{fig:mesh}
\end{figure*}

Calculations over the 4$^\text{o}$ divergent angle venturi apply 2-D and 3-D computational domains with $340\times72$ and $340\times72\times72$ mesh cells represented in Fig.\@\ref{fig:mesh}, respectively. The grids are designed in a structured fashion concerning maximum normal distance to the walls of wall-bounded cells, $z^+$ according to Fig.\@ \ref{fig:geometry} coordinates, between 10 and 15 in the area of interest. The $n$ parameter of the Reboud limiter, from Eq.\@ \ref{eqn:reboud}, is set to $10$ in the 2-D calculation \citep{Goncalves2012} and $19$ in the 3-D calculation. As observed in other works\cite{Zhou2008,Ducoin2012,Ji2014}, the choice of $n$ in the 3-D configuration is motivated by an under-prediction of the re-entrant jet development for the k-l model using $n=10$. For that reason, the $n$ parameter is calibrated to correctly capture the re-entrant jet by comparing with experimental data. The dissipative terms parameters of the extended Jameson-Schmidt-Turkel scheme $k_2$, $k_2^I$ and $k_4$ are respectively set to $1.0$, $1.5$ and $0.045$. Furthermore, the time step is fixed to $4.58 \times 10^{-6} \ s$  and $2.29 \times 10^{-7} \ s$ for the 2-D and 3-D simulations, respectively, and a total of $2.06 \ s$ physical time is run for the two numerical studies. Table \ref{tab:simu_parameters} presents the required parameters for the computation of the cavitation model for both phases. Another 3-D computation is carried out on the same geometry with the same parameters except for a twice larger width and periodic side boundary conditions.
\begin{table}[H]
\begin{ruledtabular}
    \begin{tabular}{cccccc}
         & $ \rho^{sat} \ (kg/m^3)$ & $\gamma$ & $p_{\infty} \ (Pa)$ & $q \ (J/kg)$ & $Cp \ (J/K\ kg)$  \\
        \hline
        liquid & $998.16$ & $1.01$ & $1.211 \times 10^7$ & $-1.142 \times 10^6$ & $4183$ \\
        vapor & $0.0173$ & $1.32$ & $0$ & $1.985 \times 10^6$ & $1883$ \\
    \end{tabular}
\end{ruledtabular}
\caption{Parameters of the cavitation model.}
\label{tab:simu_parameters}
\end{table}
\subsection{Comparison with experimental data}

The numerical results of the in-flow simulation are then compared to the experimental data from Barre et al.\@ (2009)\cite{Barre2009}. This experiment provides measures of time-averaged velocity, void ratio, and wall pressure profiles at stations located in the midspan of the venturi.

\subsubsection{Velocity and void ratio profiles}

\begin{figure}[H]
    \centering
    \includegraphics[width=0.9\textwidth]{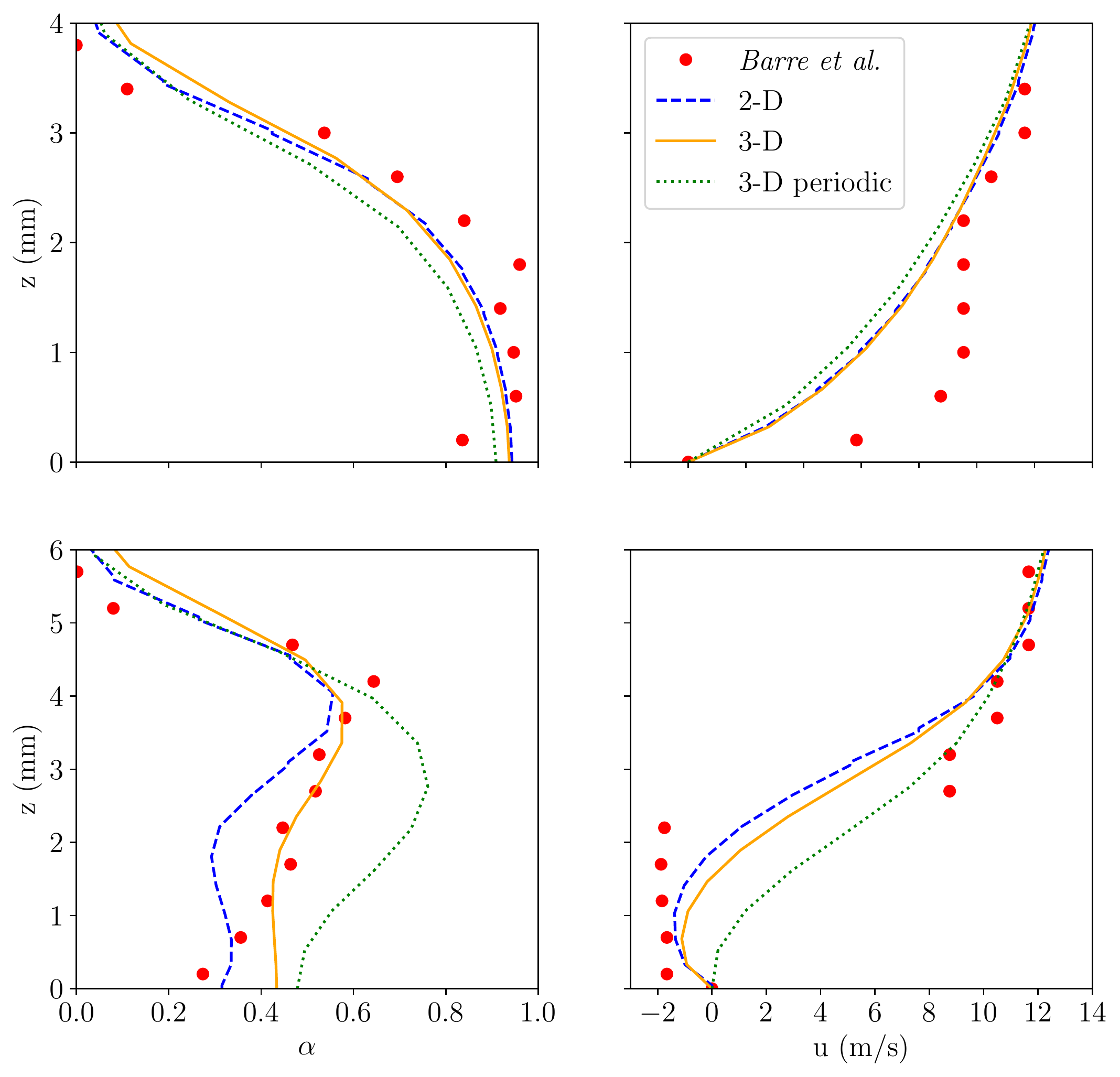}
    \caption{Time-averaged comparison at midspan between experiment, 2-D, 3-D and 3-D periodic for void ratio (left) and velocity (right) at stations S1 (top) and S2 (bottom).}
    \label{fig:comp_velo_alpha_1}
\end{figure}

\begin{figure}[H]
    \centering
    \includegraphics[width=0.9\textwidth]{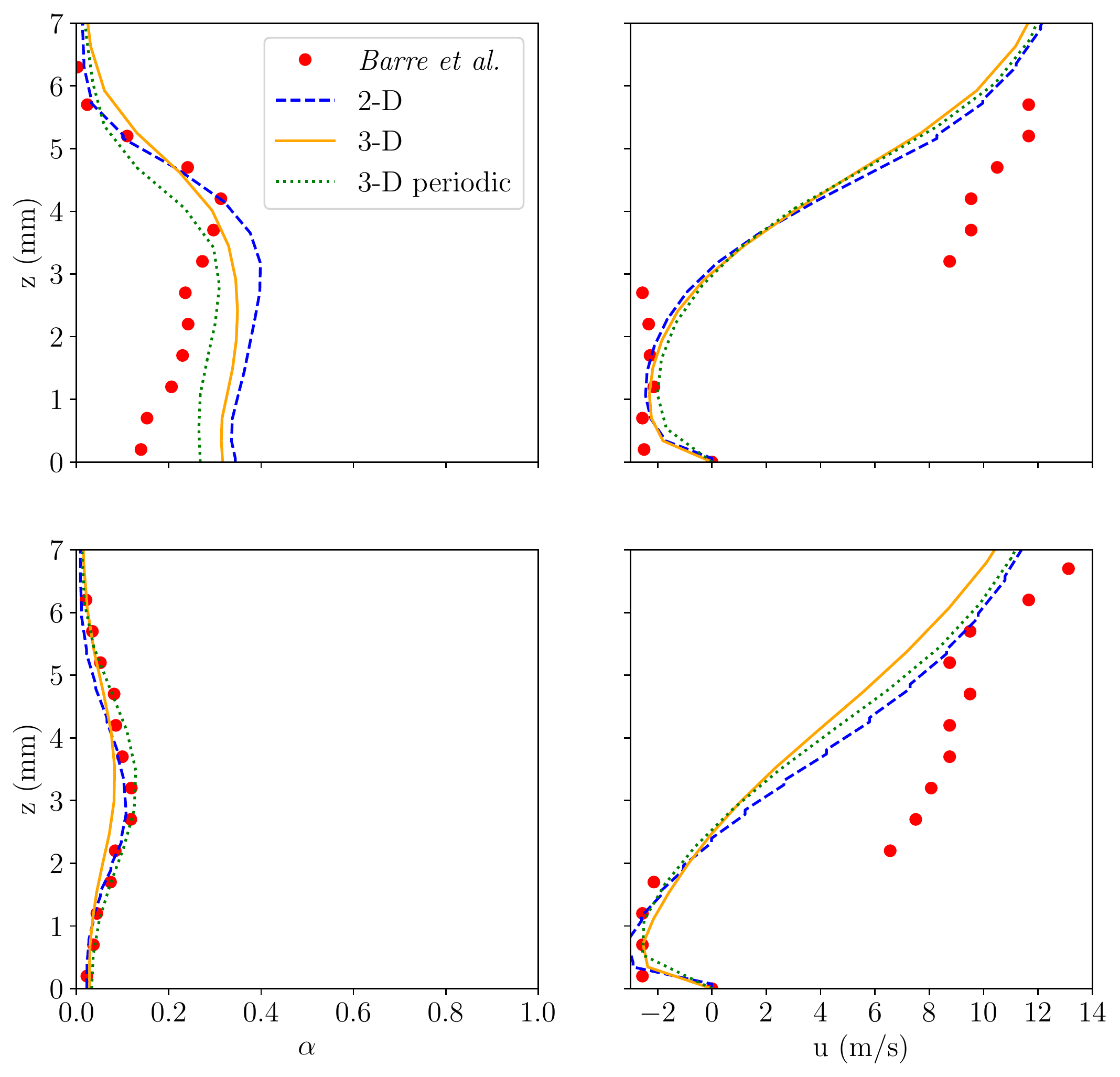}
    \caption{Time-averaged comparison at midspan between experiment, 2-D, 3-D and 3-D periodic for void ratio (left) and velocity (right) at stations S3 (top) and S4 (bottom).}
    \label{fig:comp_velo_alpha_2}
\end{figure}

Profiles of time-averaged velocity and time-averaged void ratio from 2-D and both 3-D simulation results are compared with experimental data at different positions in Figs.\@ \ref{fig:comp_velo_alpha_1} and \ref{fig:comp_velo_alpha_2}. Numerical results have similar behavior for the first station $S1$. The capture of the re-entrant jet is in a good match with the experiment for the numerical results, apart from the 3-D periodic case at $S2$, since the negative values of the velocity are correctly determined, as observed in the velocity profiles at the three other stations ($S2$, $S3$ and $S4$). The void ratio profiles are in good accordance for all computations but are in better agreement for the 3-D case with sidewalls, whose results indicate a better representation of the pocket size and shape when compared to the 2-D and 3-D periodic calculations. The 3-D computation with sidewalls correctly captures the physical behavior of the cavitating flow. Moreover, the time-averaged results of this case are sensibly conformed with the experimental results. Differences between 3-D computations with sidewalls and periodic boundary conditions are discussed later in the paper.

\subsubsection{Wall pressure profiles}

\begin{figure}[H]
    \centering
    \includegraphics[width=0.9\textwidth]{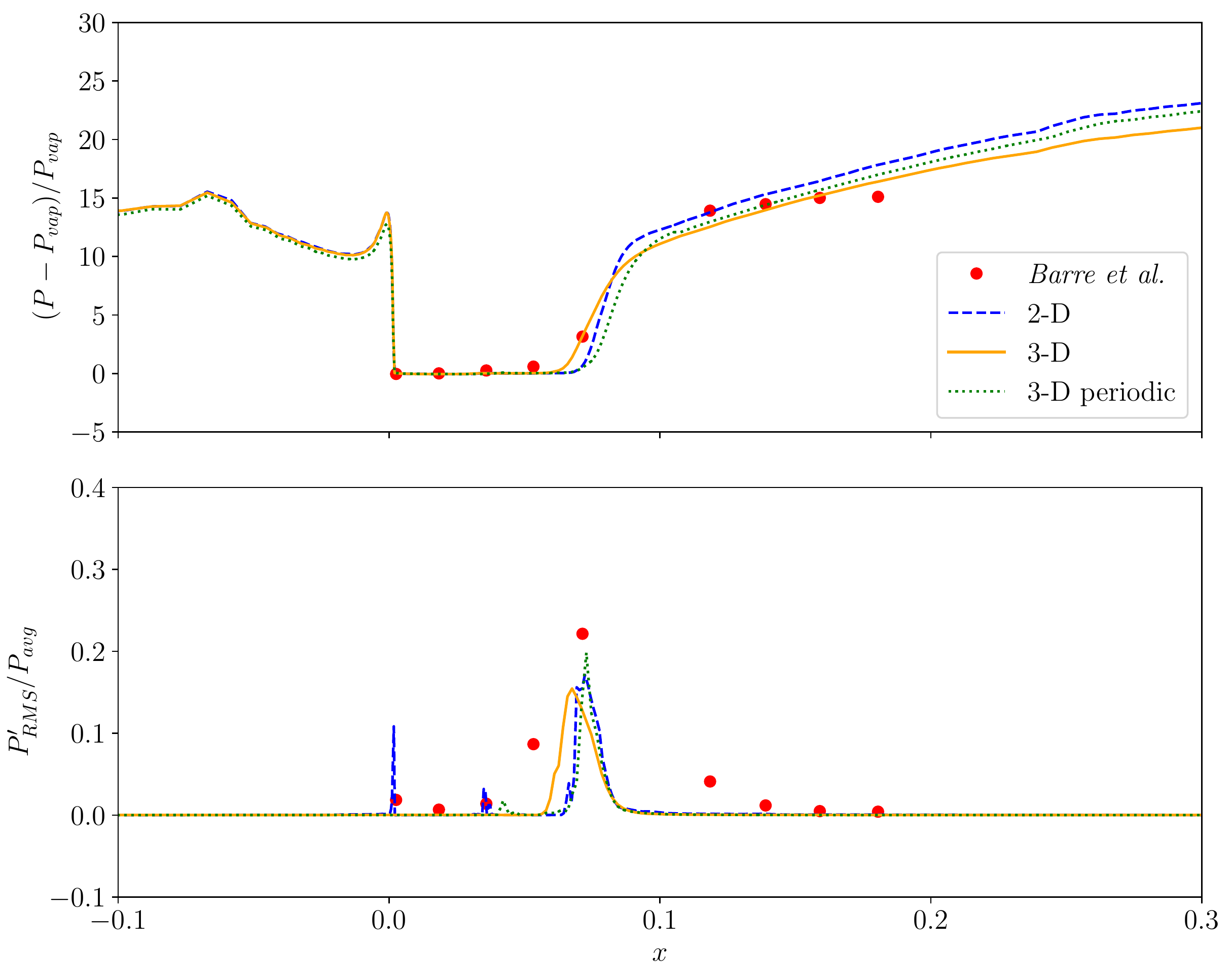}
    \caption{Time-averaged comparison between experiment, 2-D, 3-D and 3-D periodic for wall pressure and wall pressure RMS over wall pressure.}
    \label{fig:comp_pressure}
\end{figure}

The mean wall pressure and pressure fluctuations profiles are extracted from computations and are presented in Fig.\@ \ref{fig:comp_pressure} in comparison with experimental data. The wall pressure profile from the 2-D and 3-D calculations are in good agreement with experimental data along the cavity and reasonably fits the experimental data downstream the cavity. Nevertheless, the root mean square (RMS) fluctuations are slightly underestimated for all cases. Moreover, oscillations detected in the 2-D computation are not observed in 3-D computations.

\section{Global behavior}

The current section is dedicated to the first interpretation of numerical results from computations over 3-D configurations with and without sidewalls. A statistical analysis of the calculation data is performed to investigate the time-averaged and dynamical global behavior of such flow configuration. Data are extracted at a $2.3\times 10^{-3} \ s$ timestep.

\subsection{Time-averaged data analysis}

\begin{figure}[H]
  \begin{subfigure}{0.49\textwidth}
    \includegraphics[width=\linewidth]{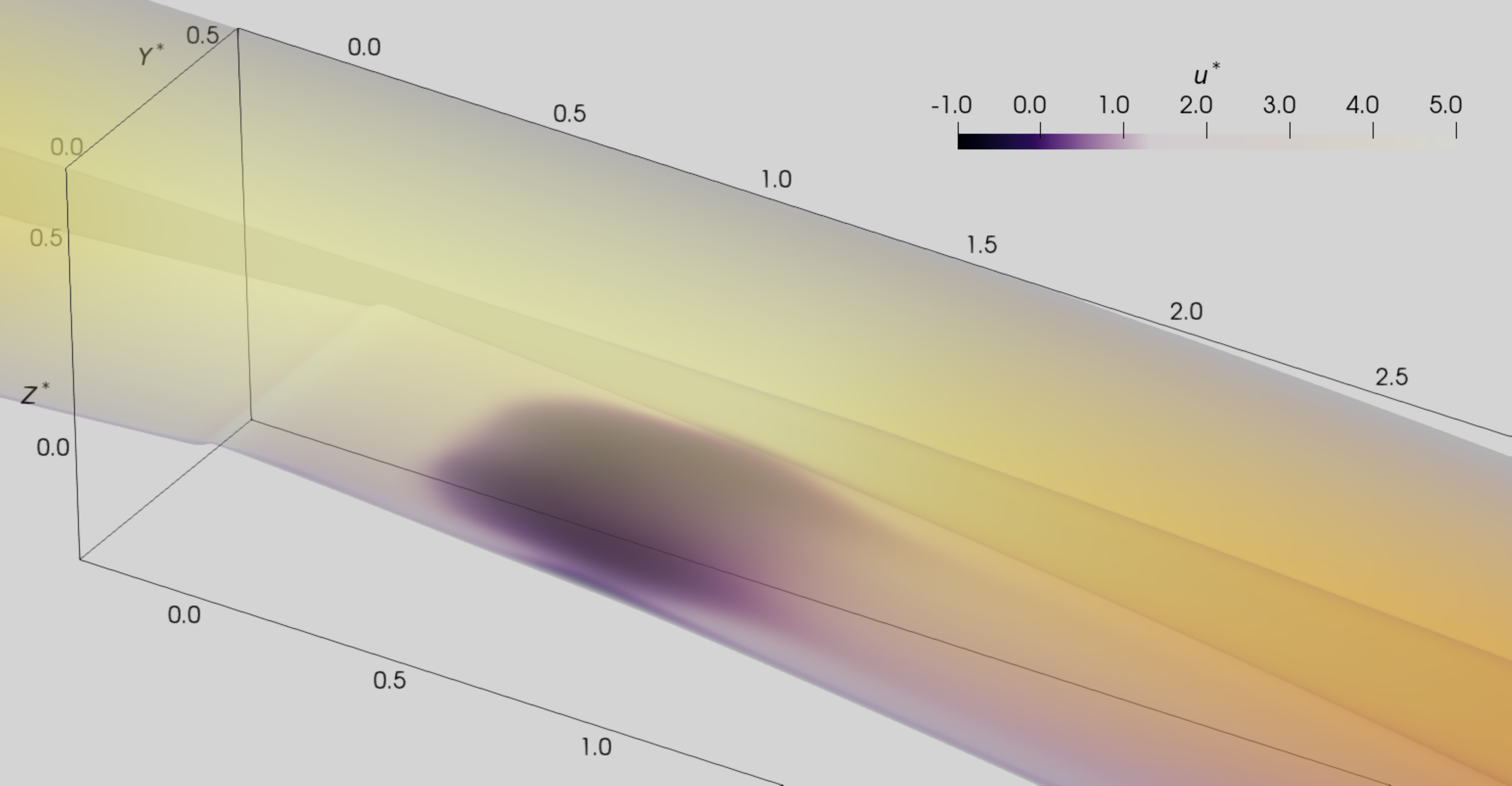}
    \caption{} \label{fig:mean_u}
  \end{subfigure}%
  \hspace*{\fill}   
  \begin{subfigure}{0.49\textwidth}
    \includegraphics[width=\linewidth]{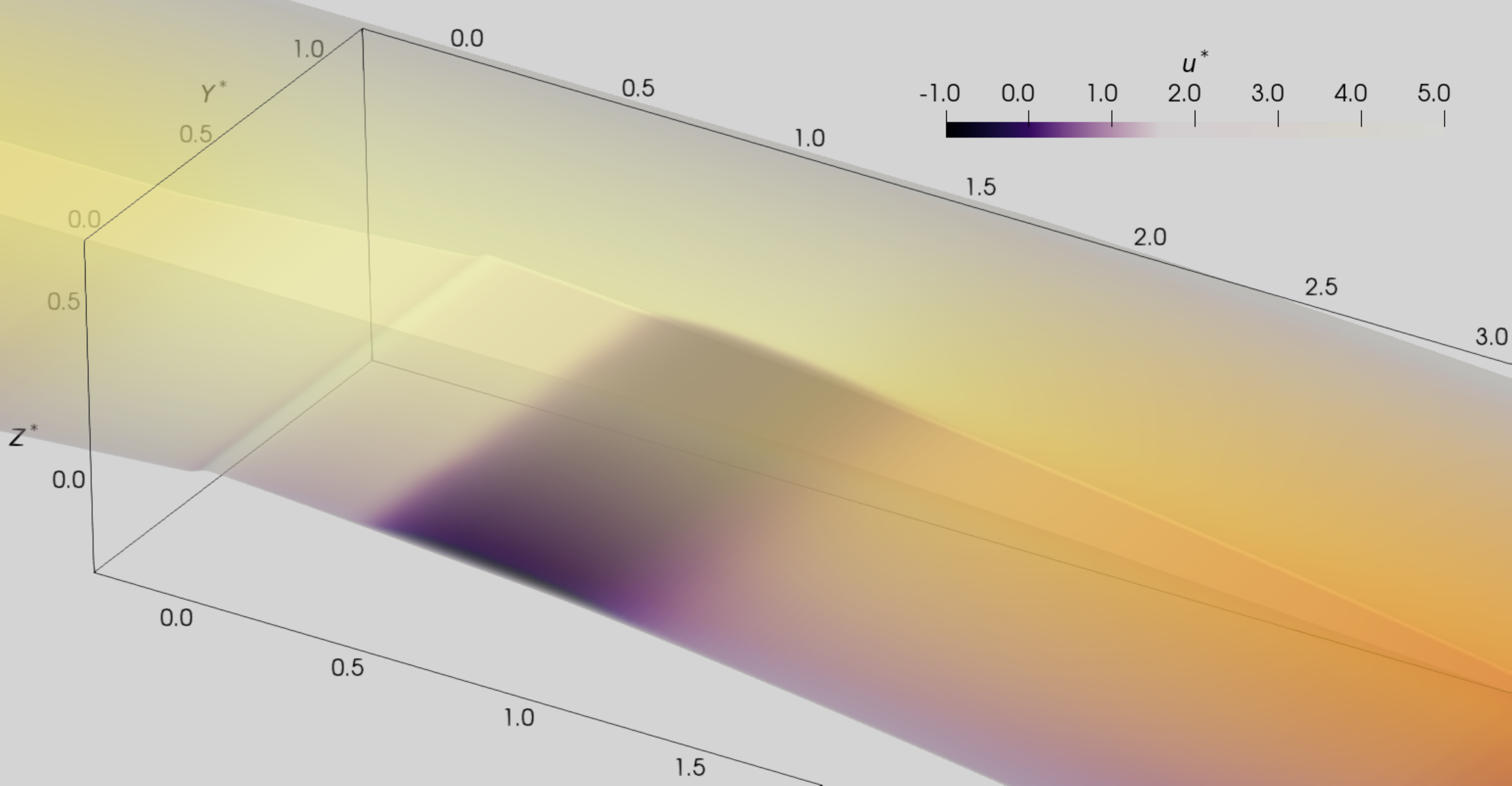}
    \caption{} \label{fig:mean_u_perio}
  \end{subfigure}%
  \caption{Volume rendering of the time-averaged streamwise velocity : (a) for the venturi with sidewalls; (b) for the venturi with side periodic boundary conditions}
\end{figure}

According to experimental observations of the Venturi \citep{Barre2009}, a weakly fluctuating cavity emerges without any large vapor shedding process. A time-averaged cavity length $L_c$ between $70$ and $85 \ mm$, estimated with an $\alpha$ contour of $0.05$, is observed in the experiment. Numerical results for the case with sidewalls present a pocket length of $L_c=78.8 \ mm$, which is consistent with the experiment. This length is selected to be the characteristic length for the current study. The maximum value of the time-averaged re-entrant jet velocity is also used as the characteristic velocity, $u_{max}^{jet}=2.38 \ m.s^{-1}$. The selection of these two characteristic variables is motivated by the observation of the cavitating flow behavior and the identification of the leading mechanism. 
Moreover, the study of Dular and Bachert (2009)\cite{Dular2009} defines the re-entrant jet velocity at the cavity closure and the length of the attached vapor pocket as the most correct values to investigate the cavitating flow over a hydrofoil. The maximum reverse flow is also considered as the characteristic velocity in non-cavitating flows with separation bubble, Hammond and Redekopp (1998)\cite{Hammond1998} or Rist and Maucher (2002)\cite{Rist2002} show the important role of the reverse flow in the triggering of instabilities. The Strouhal number is then defined as :
\begin{equation}
    St=\frac{L_c f}{u_{max}^{jet}}.
\end{equation}
In the literature, for sheet to cloud cavitation cases, the Strouhal number is mostly defined with inlet velocity\citep{Gnanaskandan2016,Ganesh2016,Budich2018}. The choice of the characteristic velocity will be justified later in the paper by investigating the resulting Strouhal number. Variables with a superscript $^*$ in the manuscript are dimensionless and are calculated using the characteristic length $L_c$ and characteristic velocity $u_{max}^{jet}$.\\
\begin{figure}[H]
  \begin{subfigure}{0.49\textwidth}
    \includegraphics[width=\linewidth]{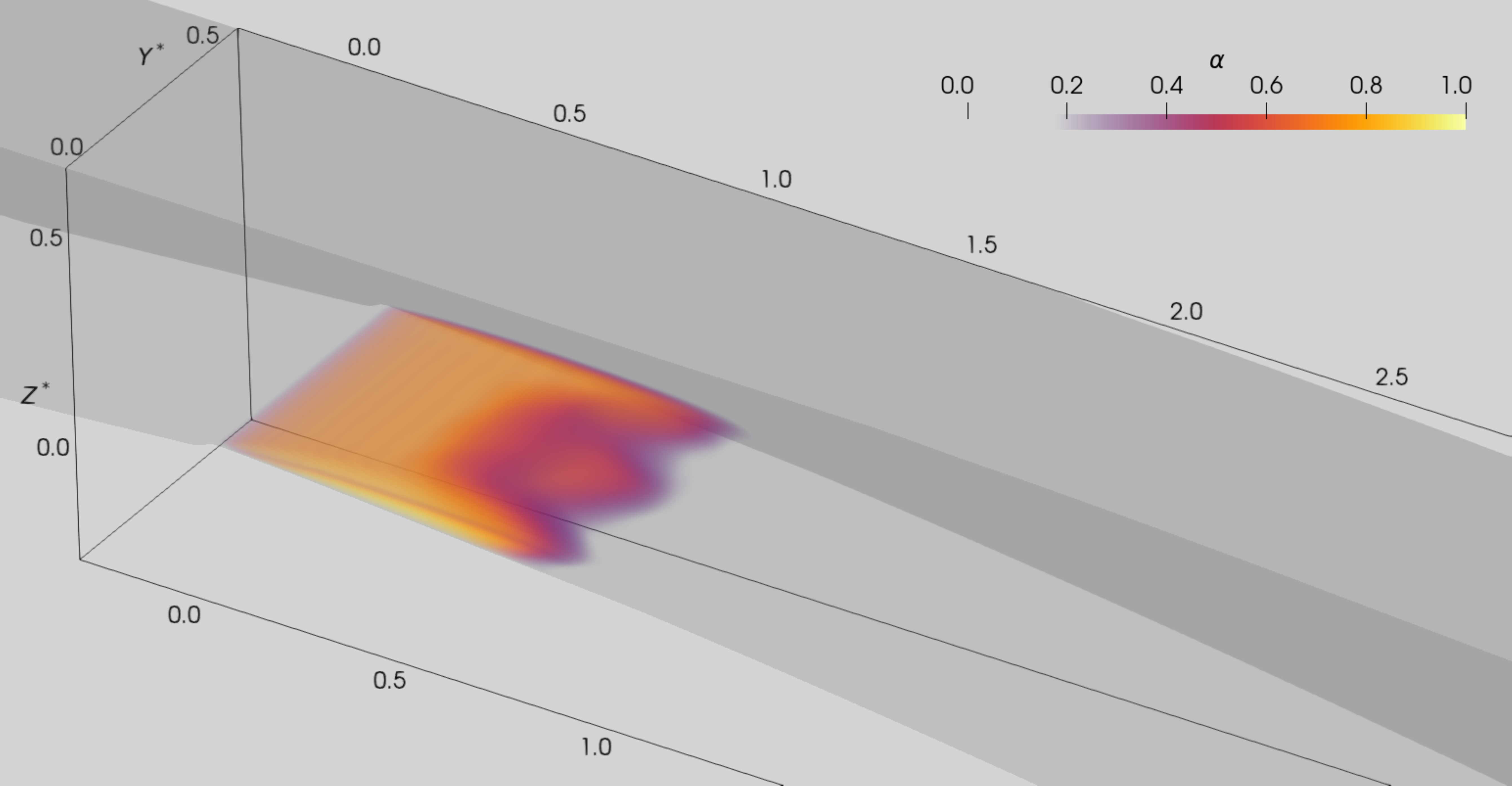}
    \caption{} \label{fig:mean_alpha}
  \end{subfigure}%
  \hspace*{\fill}  
  \begin{subfigure}{0.49\textwidth}
    \includegraphics[width=\linewidth]{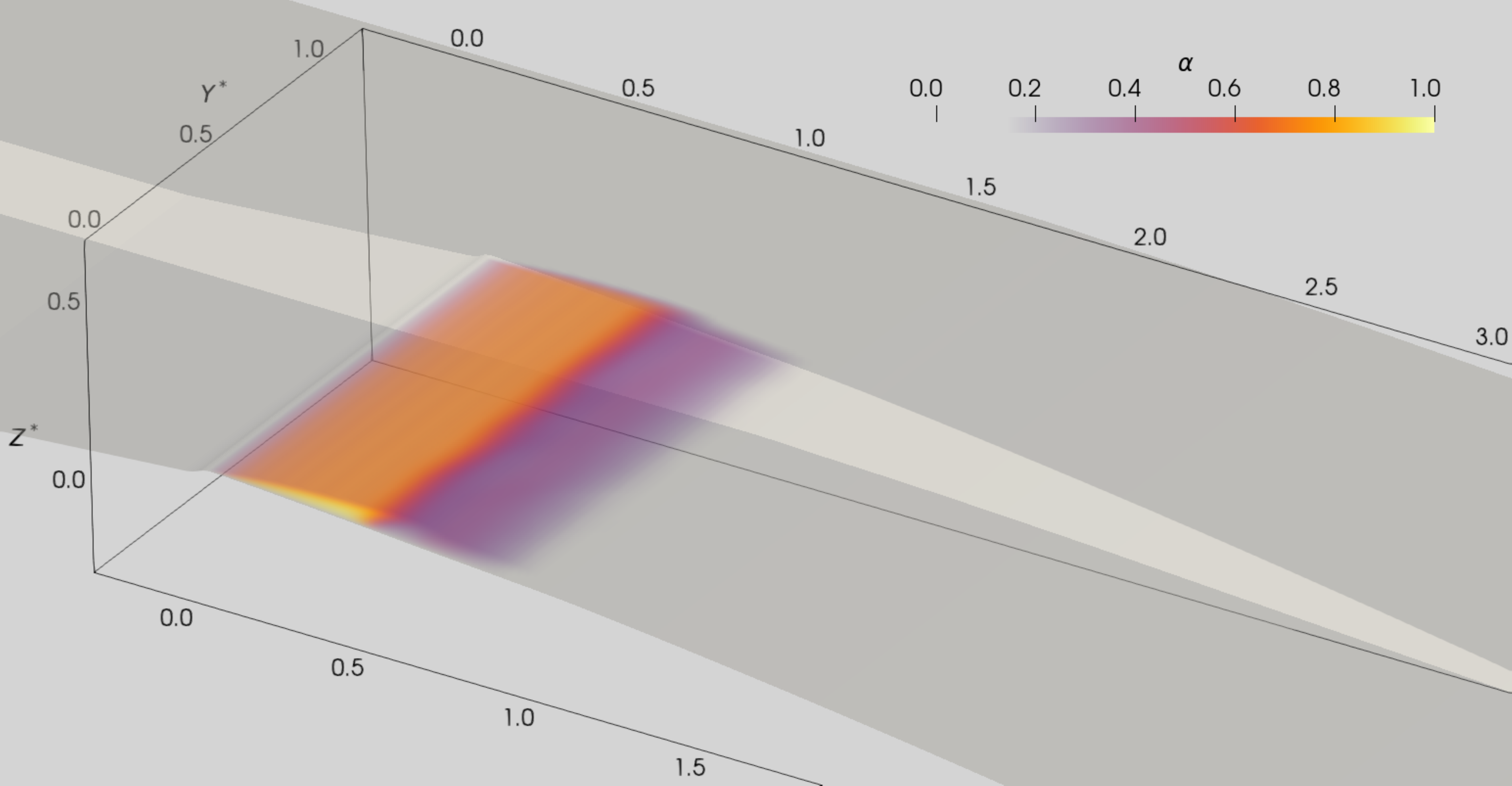}
    \caption{} \label{fig:mean_alpha_perio}
  \end{subfigure}%
  \caption{Volume rendering of the time-averaged void ratio: (a) for the venturi with sidewalls; (b) for the venturi with side periodic boundary conditions}
\end{figure}
A symmetrical attached cavity is detected in Fig.\@ \ref{fig:mean_alpha} with a longer cavitation pocket length near sidewalls than near of the midplane of the Venturi. The U-shape of the cavitation pocket is detected as described in many references \citep{Kubota1989,Peng2016}. The lower amount of void ratio suggests vapor release and/or pocket oscillations around the mid-width. Figure \ref{fig:mean_alpha_perio} presents the cavitation pocket shape for the periodic case. Unlike the case with sidewalls, the cavity length is constant in all the venturi width. Then, the observation of the flow direction velocity in Fig.\@ \ref{fig:mean_u} and \ref{fig:mean_u_perio} underlines the presence of the re-entrant jet along the wall. The jet geometry is symmetric and is not present close to sidewalls whereas it is localised in all the venturi width for the periodic case. Moreover, for the venturi with sidewalls, the vertical velocity is also symmetric, while the spanwise velocity is anti-symmetric. 

\subsection{Dynamic analysis}

\begin{figure}[H]
    \begin{center}
    \begin{subfigure}[b]{\textwidth}
     \includegraphics[width=\textwidth]{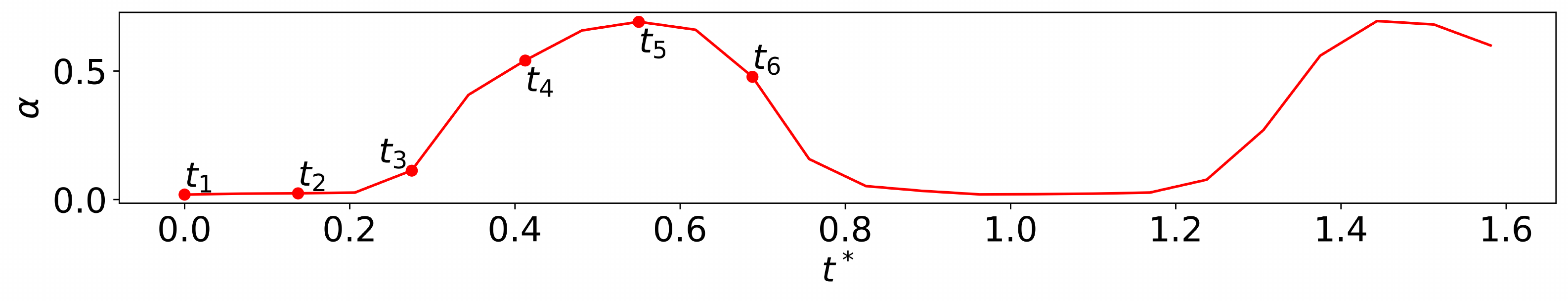}
     \caption{Void ratio signal.}\label{subfig:signal}
    \end{subfigure}
    \begin{subfigure}{0.31\textwidth}
    \vspace{0.3cm}
    \includegraphics[width=\textwidth]{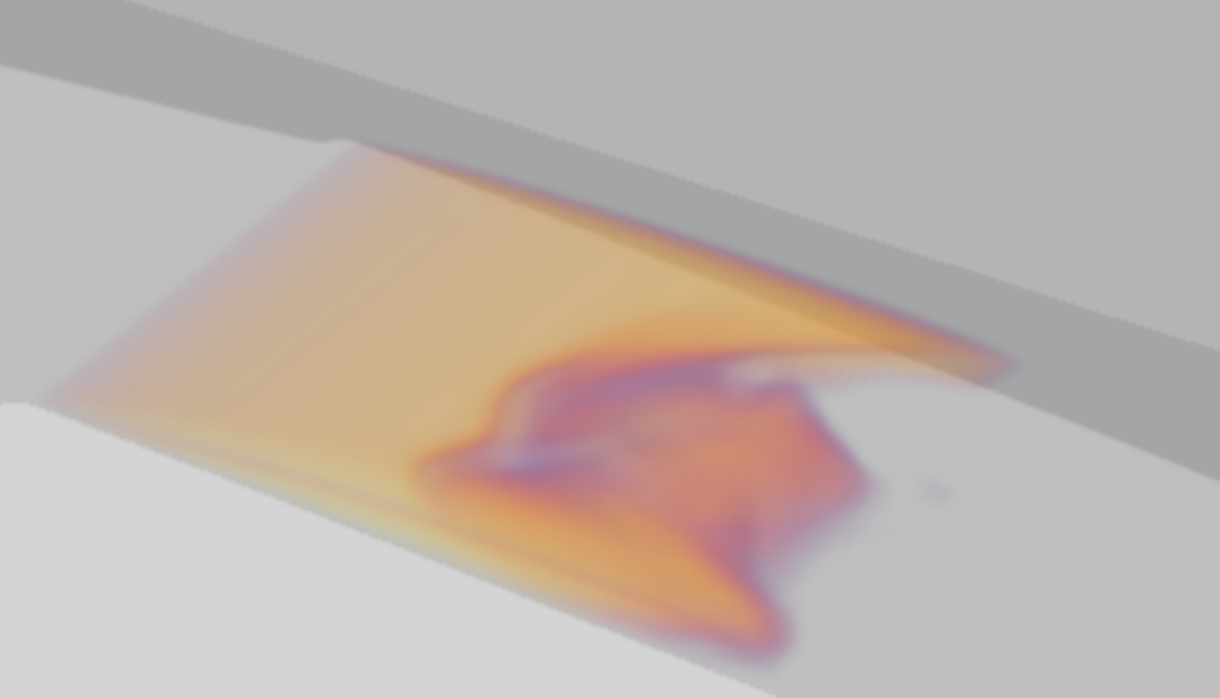}
    \caption{$t_1$}\label{subfig:t1}
    \end{subfigure}
    \hfill
    \begin{subfigure}{0.31\textwidth}
    \vspace{0.3cm}
    \includegraphics[width=\textwidth]{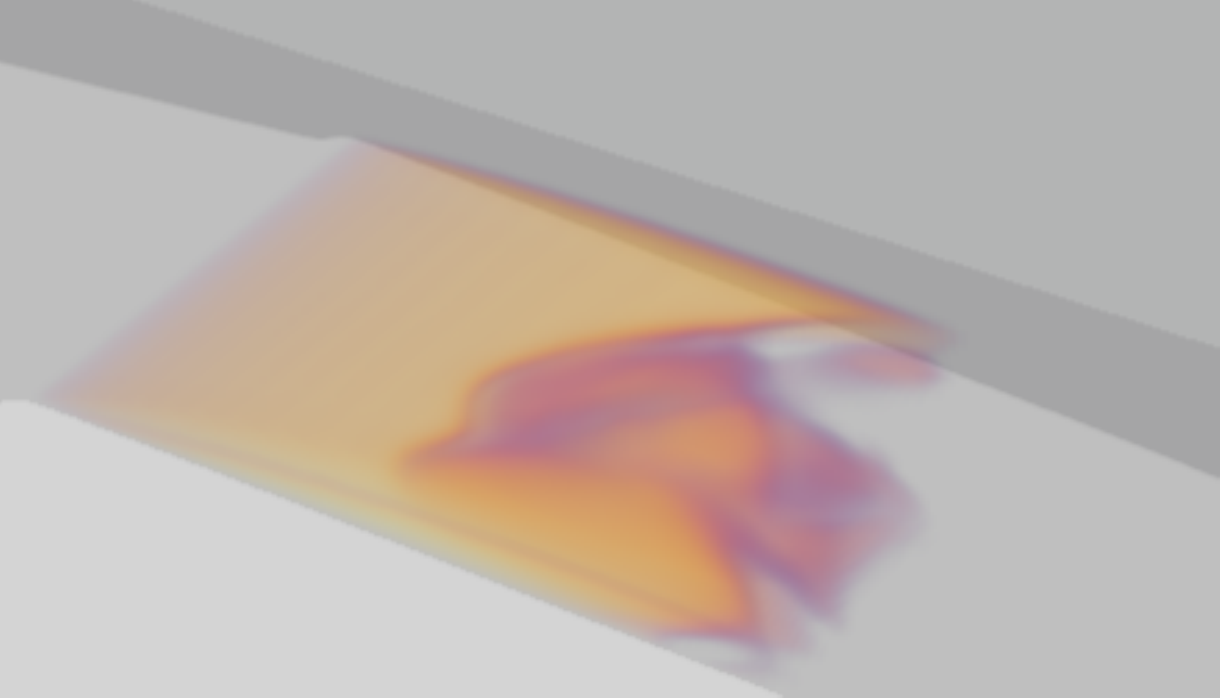}
    \caption{$t_2$}\label{subfig:t2}
    \end{subfigure}
    \hfill
    \begin{subfigure}{0.31\textwidth}
    \vspace{0.3cm}
    \includegraphics[width=\textwidth]{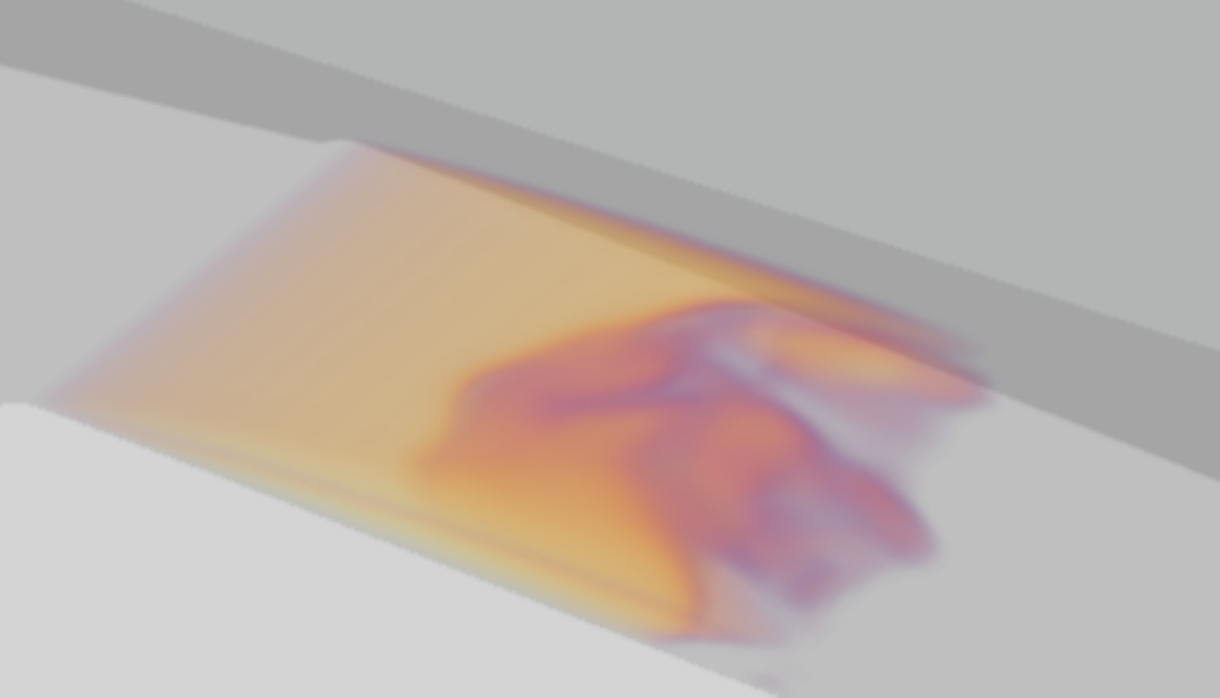}
    \caption{$t_3$}\label{subfig:t3}
    \end{subfigure}
    \hfill
    \begin{subfigure}{0.31\textwidth}
    \vspace{0.3cm}
    \includegraphics[width=\textwidth]{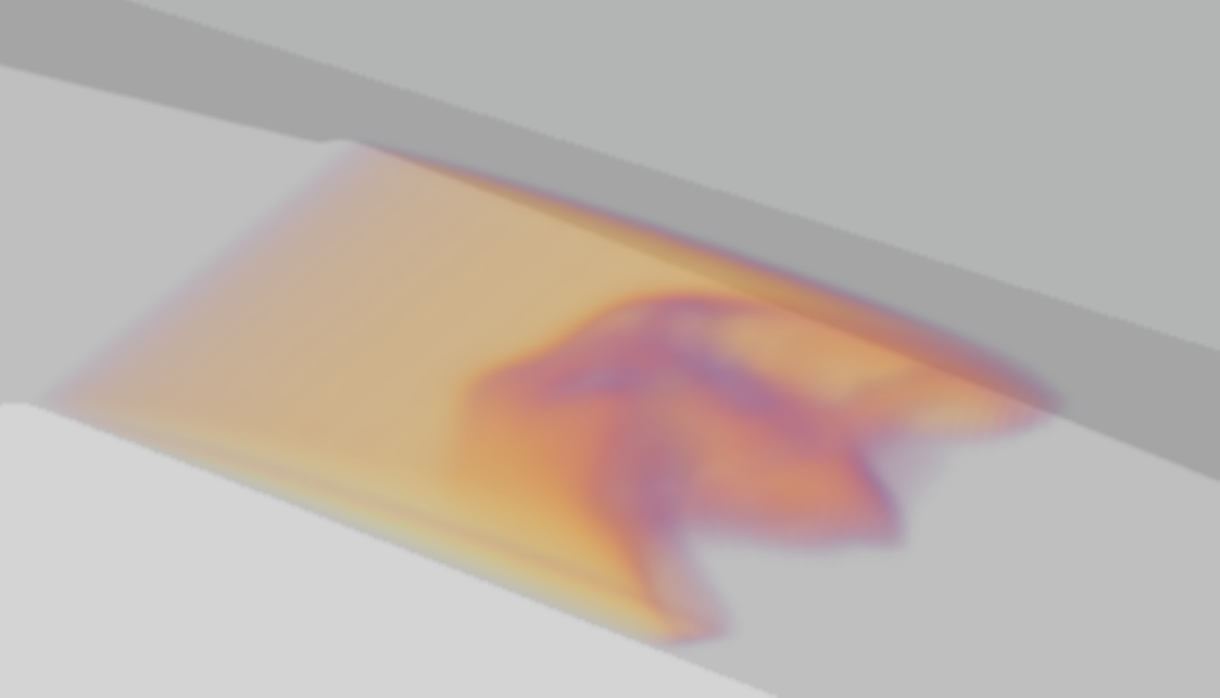}
    \caption{$t_4$}\label{subfig:t4}
    \end{subfigure}
    \hfill
    \begin{subfigure}{0.31\textwidth}
    \vspace{0.3cm}
    \includegraphics[width=\textwidth]{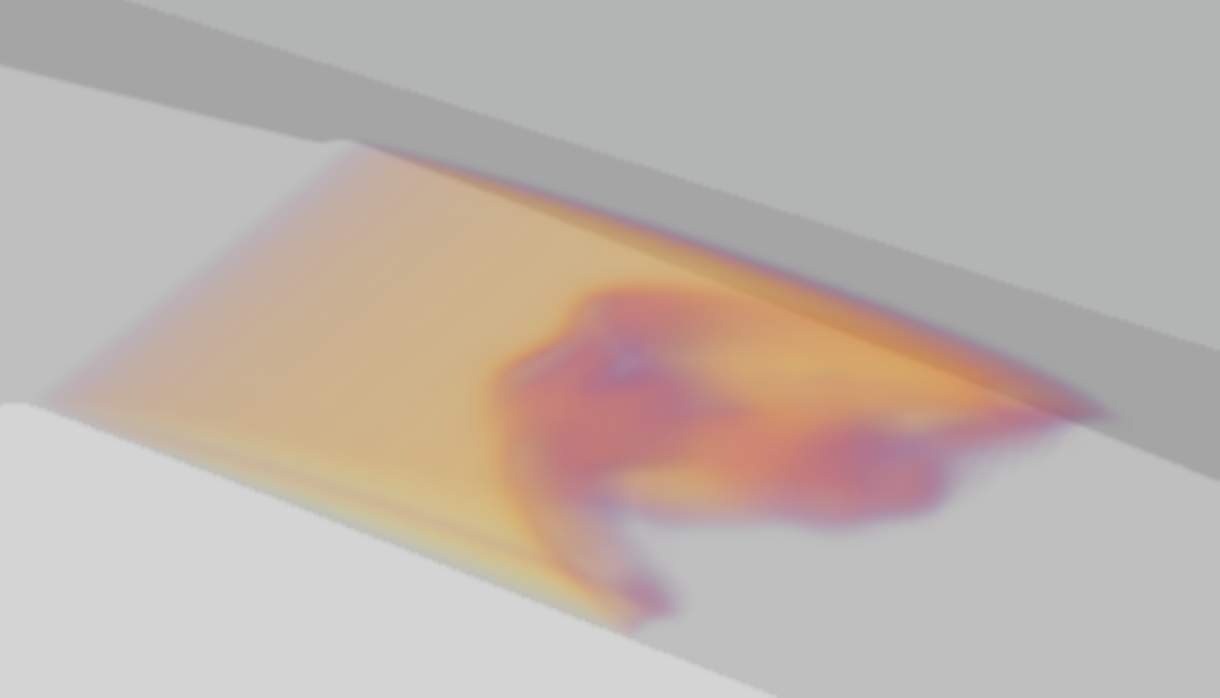}
    \caption{$t_5$}\label{subfig:t5}
    \end{subfigure}
    \hfill
    \begin{subfigure}{0.31\textwidth}
    \vspace{0.3cm}
    \includegraphics[width=\textwidth]{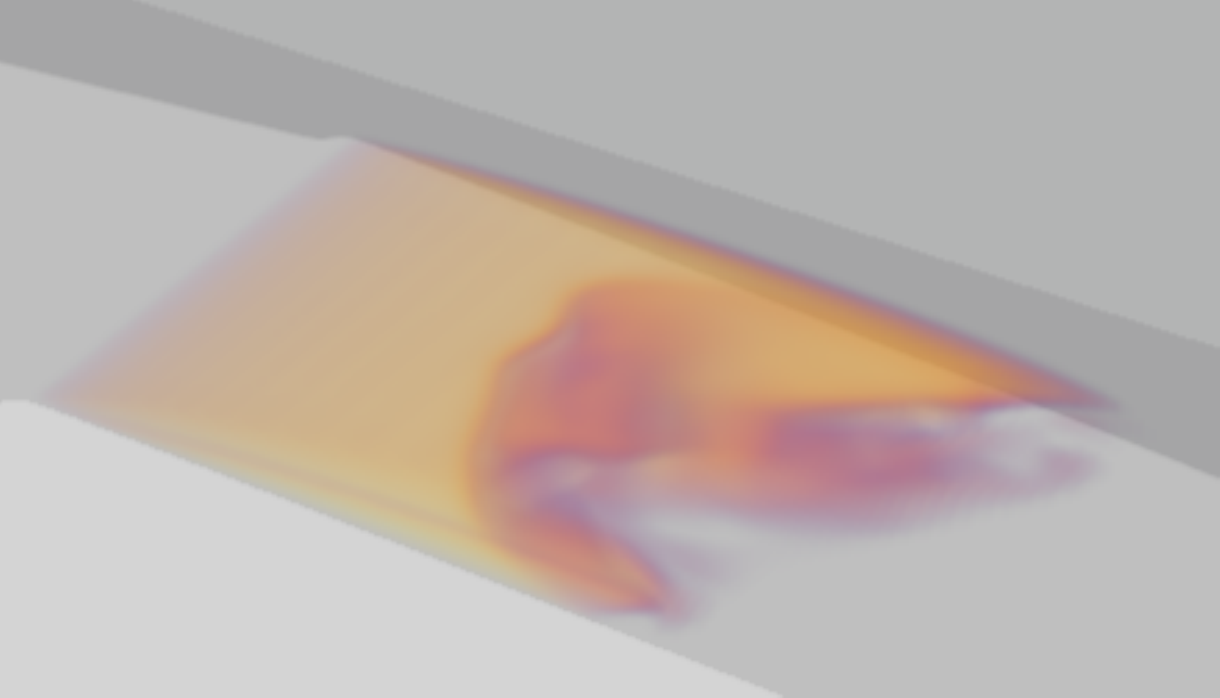}
    \caption{$t_6$}\label{subfig:t6}
    \end{subfigure}
    \end{center}
    \caption{Time evolution of the attached cavity with snapshots separated by $\Delta t = 4.58 \ 10^{-3} \ s$ with a volume rendering; (b)-(g) Snapshots extracted at different time represented in the void ratio signal (a).}
    \label{fig:screenshot_oblic}
\end{figure}

\begin{figure}[h]
    \centering
    \includegraphics[width=0.9\textwidth]{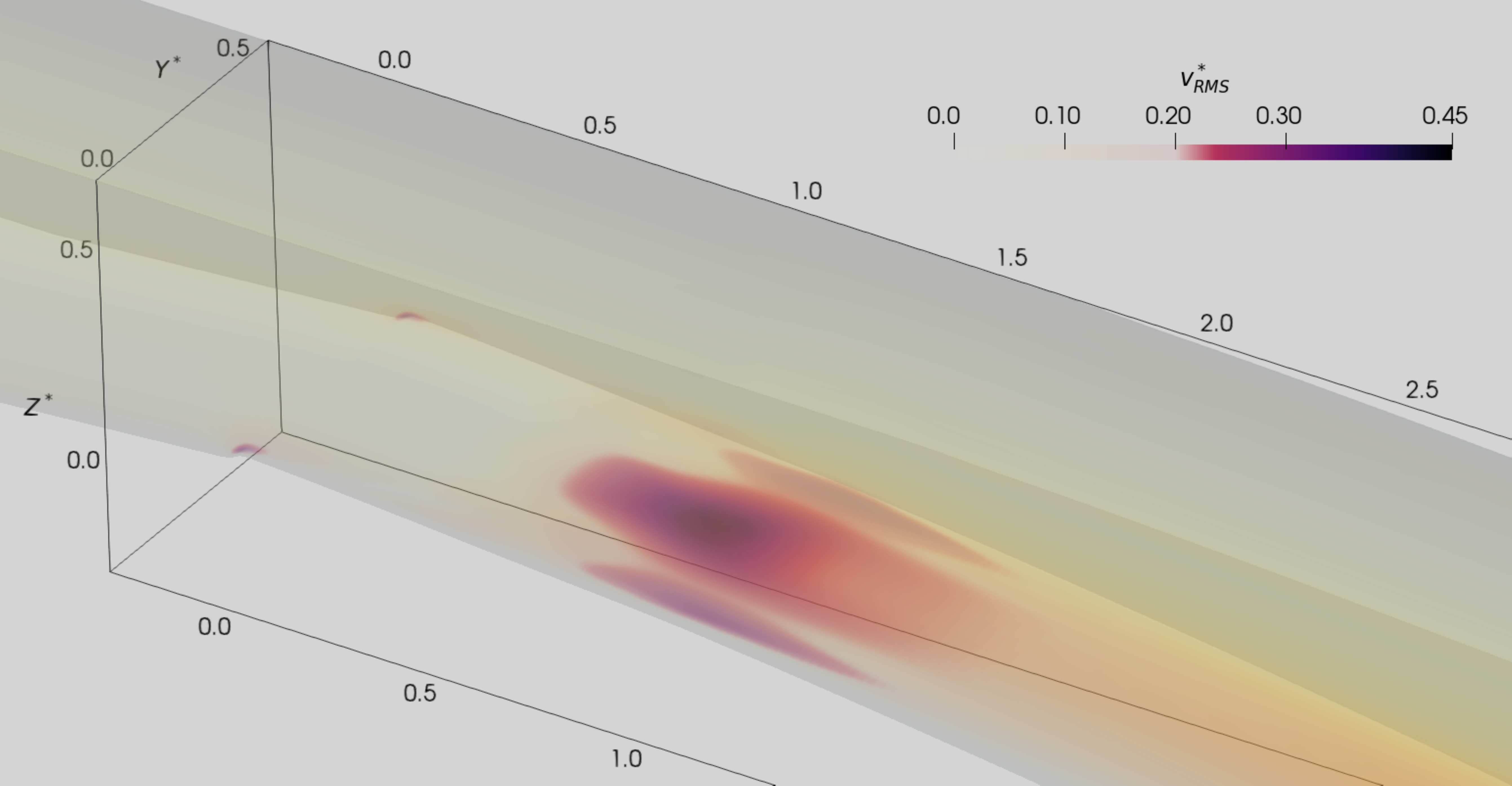}
    \caption{Volume rendering of the RMS fluctuation of the spanwise velocity $v$. }
    \label{fig:rms_v}
\end{figure}

\begin{figure}[H]
    \begin{center}
    \begin{subfigure}{0.49\textwidth}
    \includegraphics[width=\textwidth]{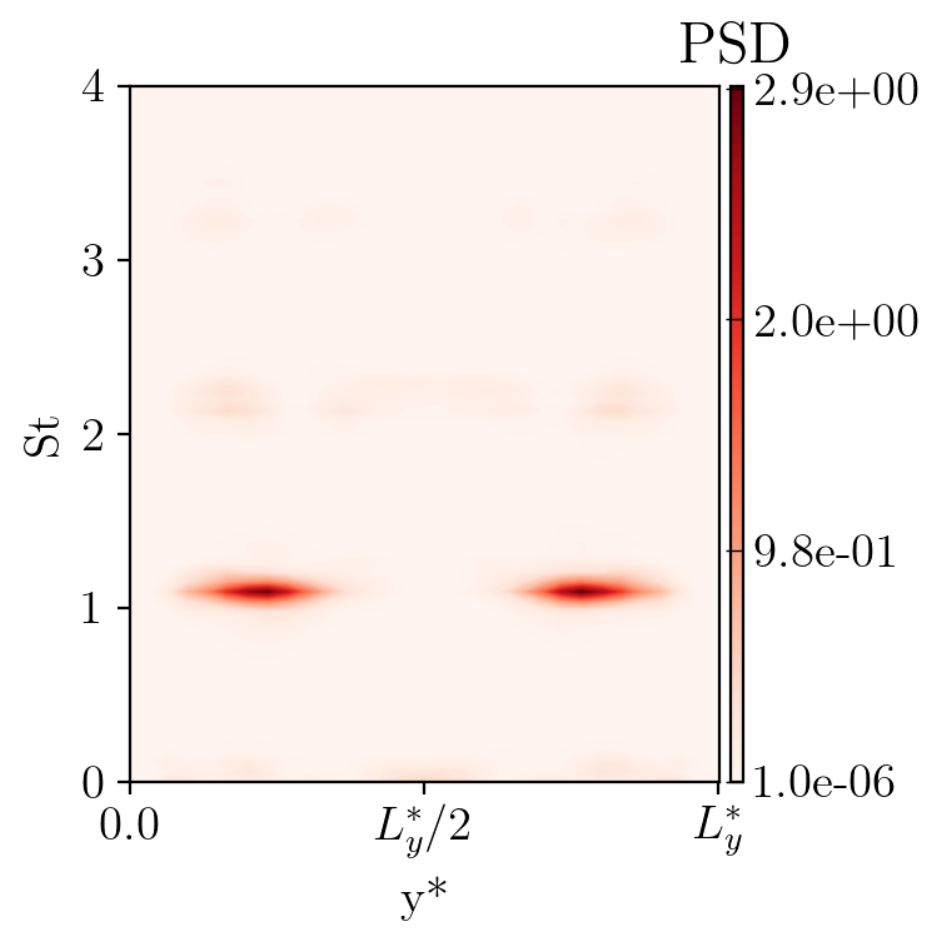}
    \caption{}\label{subfig:PSD_x_alpha}
    \end{subfigure}
    \begin{subfigure}{0.49\textwidth}
    \includegraphics[width=\textwidth]{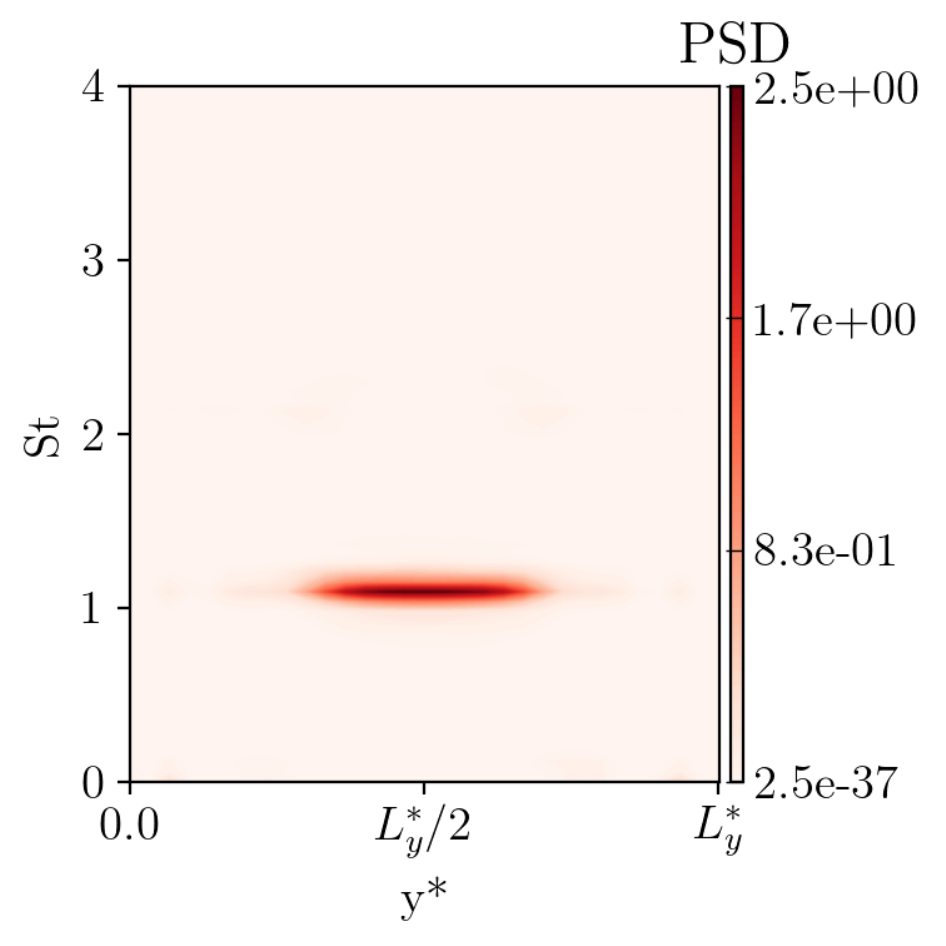}
    \caption{}\label{subfig:PSD_x_v}
    \end{subfigure}
    \end{center}
    \caption{PSD maps along spanwise axis at $x^*=0.64$ and at a vertical distance $z_p^*=0.032$ from the bottom wall : (a) for the void ratio $\alpha$; (b) for the spanwise velocity $v$.}
    \label{fig:Mapping_alpha_v_x}
\end{figure}

\begin{figure}[H]
    \centering
    \includegraphics[width=0.8\textwidth]{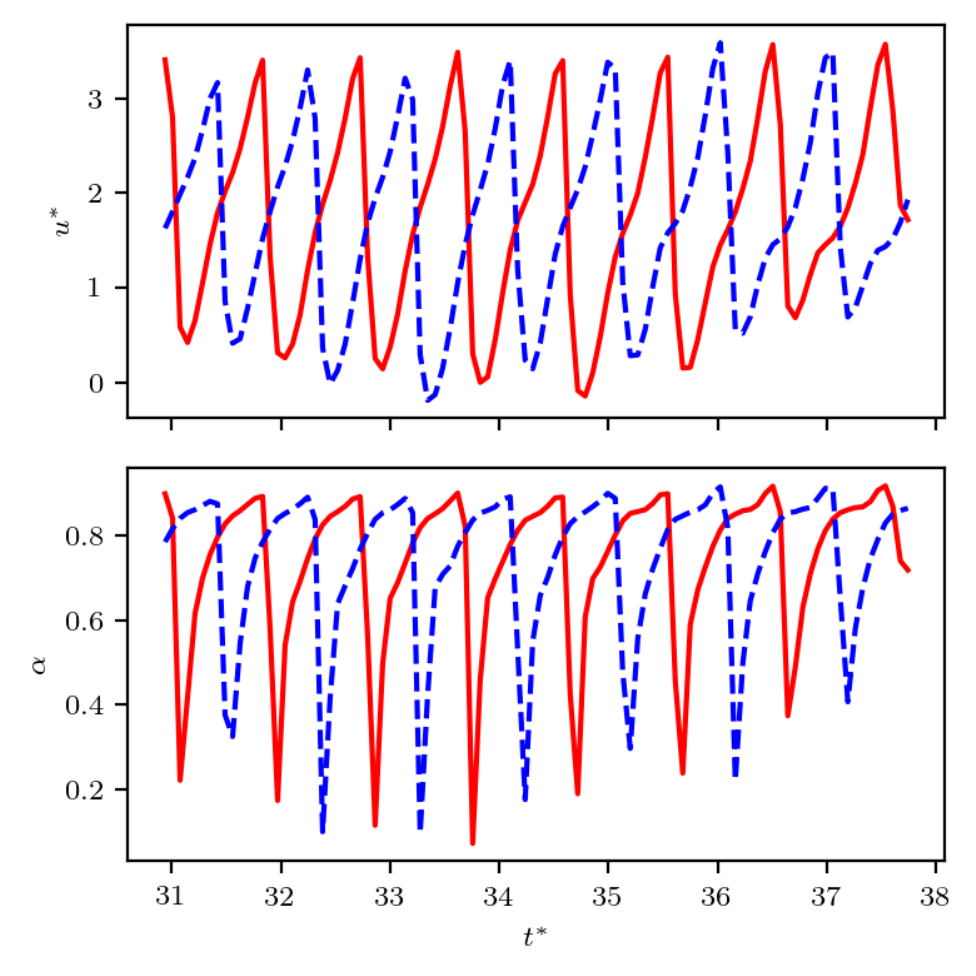}
    \caption{Temporal evolution of the flow direction velocity $u$ and the void ratio $\alpha$ at $x^*=0.64$ along both sidewalls (one in red, the other in blue) for the case with sidewalls.}
    \label{fig:correlation}
\end{figure}

A study of the flow dynamics is carried out to statistically interpret the behavior of the cavitation pocket and velocity components over time for configurations with sidewalls and with side periodic boundaries. First, the analysis is focused on the case with sidewalls. Figure \ref{fig:screenshot_oblic} shows the void ratio $\alpha$ into the flow at six different times. It is worth to notice that the pocket shape is not symmetric and evolves with time. A small high-frequency vapor shedding appears around the cavity closure while the pocket seems to oscillate in the spanwise direction. A statistical analysis is carried out to check any data fluctuations within the venturi flow. The RMS results over the spanwise velocity $v$ are presented in Fig.\@ \ref{fig:rms_v}. The highest values of fluctuation are located at the pocket closure, mostly around the mid-width. Therefore, it corroborates the occurrence of a spanwise oscillation of the flow nearby this area.\\
%
Subsequently, the time evolution of flow variables is extracted for points in the flow direction and in the spanwise direction to perform Power Spectral Densities (PSD). The result is presented as a map of PSD along the longitudinal and spanwise axis. PSD maps provide information to identify any high energy frequencies and locate the associated phenomenon into the venturi. Figure \ref{fig:Mapping_alpha_v_x} presents the PSD map over the spanwise direction, respectively for the void ratio and the spanwise velocity, positioning at almost the two thirds of the mean cavity length. A sample of signal used for the PSD computations can be observed in Fig.\@\ref{fig:correlation}. It is worth to remark that no particular dynamics are detected inside the attached cavitation pocket. A dominant Strouhal number of $1.09$ is highlighted around the cavity closure by detecting the highest PSD energy values. These are underlined in the mid-width of the venturi for velocity and near sidewalls for the void ratio. Similar behavior is observed for PSD maps downstream the cavity but with the appearance of a low frequency for the void ratio. Regarding previous remarks over snapshots of Fig.\@ \ref{fig:screenshot_oblic} and the RMS of the spanwise velocity, the Strouhal number $St_0=1.09$ seems to be linked to a spanwise oscillation of the cavitation pocket. Firsts harmonics of $St_0$ also emerge from the PSD map for the void ratio.\\

\begin{figure}[H]
    \centering
    \includegraphics[width=0.8\textwidth]{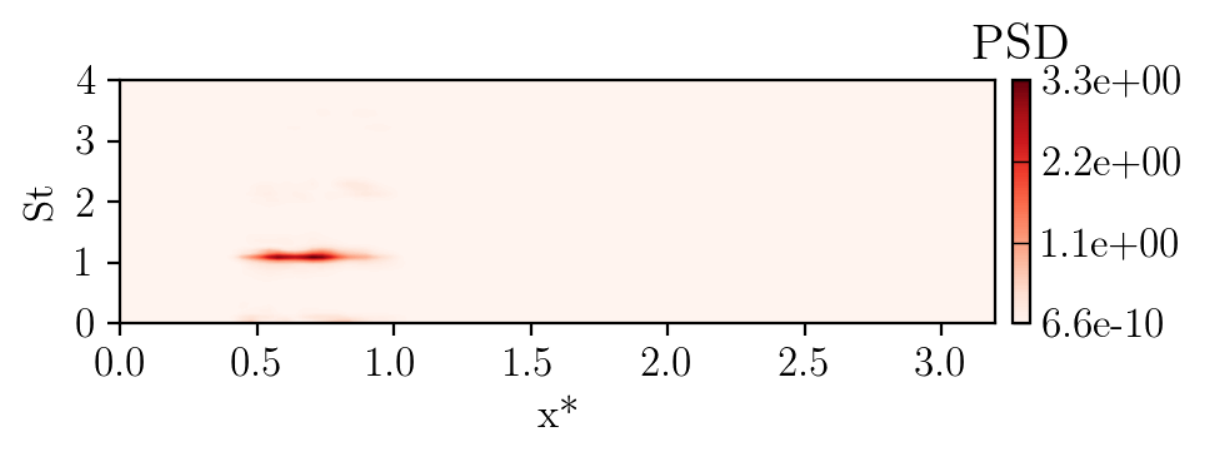}
    \caption{PSD map for $\alpha$ along longitudinal axis at $y^*=L_y^*/4$ and at a vertical distance $z_p^*=0.032$ from the bottom wall.}
    \label{fig:Mapping_alpha_y}
\end{figure}

\begin{figure}[H]
    \begin{center}
    \begin{subfigure}{0.8\textwidth}
    \includegraphics[width=\textwidth]{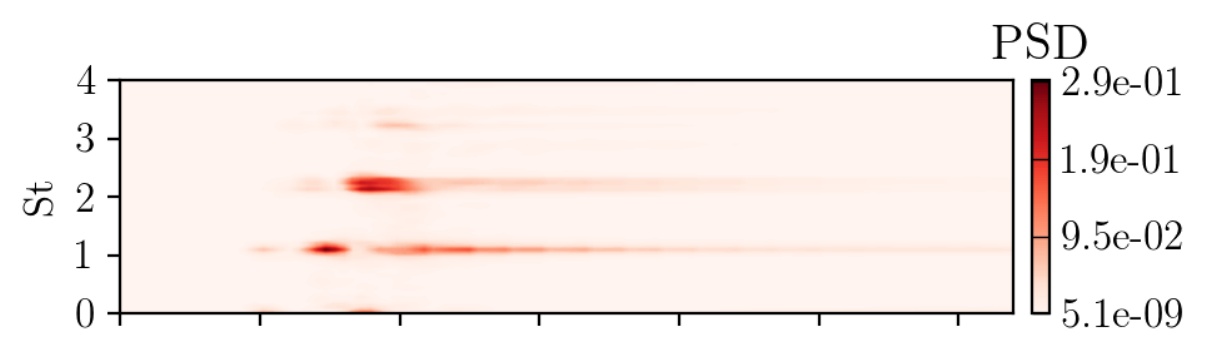}
    \caption{}\label{subfig:PSD_y_8_v}
    \end{subfigure}
    \begin{subfigure}{0.8\textwidth}
    \includegraphics[width=\textwidth]{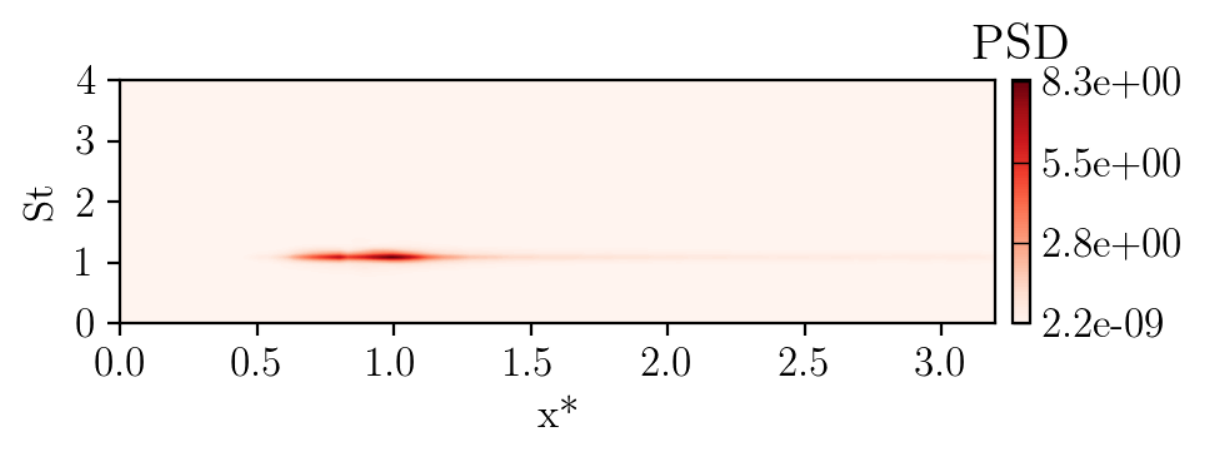}
    \caption{}\label{subfig:PSD_y_2_v}
    \end{subfigure}
    \end{center}
    \caption{PSD maps for spanwise velocity $v$ along longitudinal axis at a vertical distance $z_p^*=0.032$ from the bottom wall : (a) at $y^*=L_y^*/8$; (b) at $y^*=L_y^*/2$.}
    \label{fig:Mapping_v_y}
\end{figure}

Figure \ref{fig:Mapping_alpha_y} shows the PSD map over the streamwise direction for the void ratio at the quarter width. As previously noticed, neither particular dynamic is detected in the mid-width for the void ratio. Nevertheless, at the quarter width, the highest values of the PSD are observed around the cavity closure at the same Strouhal number $St_0$.
Figure \ref{fig:Mapping_v_y} presents the PSD map over the streamwise direction for the spanwise velocity, at two positions on the spanwise axis: one located in the mid-width ($y^*=L_y^*/2$) and another one in the eighth width ($y^*=L_y^*/8$). The same behavior as for the void ratio is underlined at both positions but with also a propagation of the dynamics downstream. Furthermore, at the eighth width positioning, the two first harmonics are also detected around the cavity closure and downstream.\\

The PSD analysis identified a dominant dynamics at the Strouhal number $St_0$. This phenomenon appears nearby the cavity closure and is propagated downstream. Furthermore, a motion of the cavitation pocket has been highlighted close to sidewalls. A spanwise velocity variation also emerges at the mid-width of the venturi. The dominant flow fluctuations and its location has been determined. However, a correlation study is carried out by extracting the flow variables over time close to both sidewalls to specify the cavity behavior. One can remark that, in Fig.\@ \ref{fig:correlation}, data are in opposition of phase, which can lead to a conclusion that the cavitation pocket motion is assimilated to a periodic oscillation from one sidewall to another.\\

\begin{figure}[H]
    \begin{center}
    \begin{subfigure}{0.8\textwidth}
    \includegraphics[width=\textwidth]{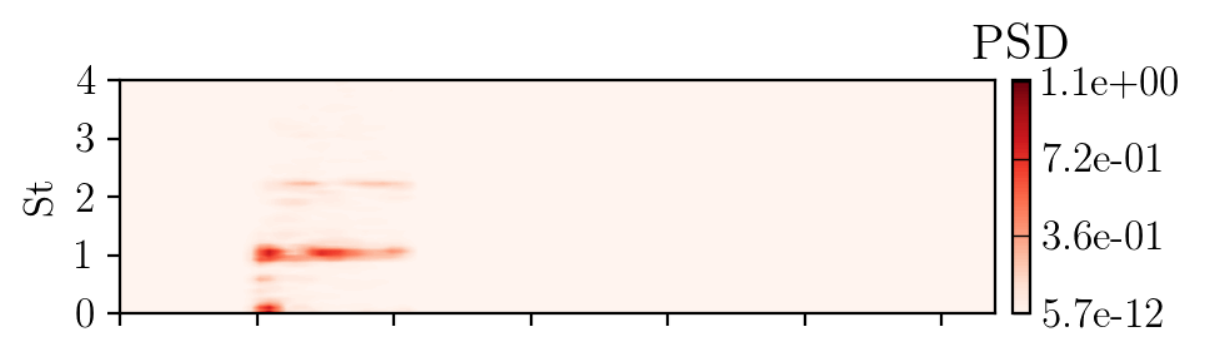}
    \caption{}\label{subfig:PSD_y_alpha_perio}
    \end{subfigure}
    \begin{subfigure}{0.8\textwidth}
    \includegraphics[width=\textwidth]{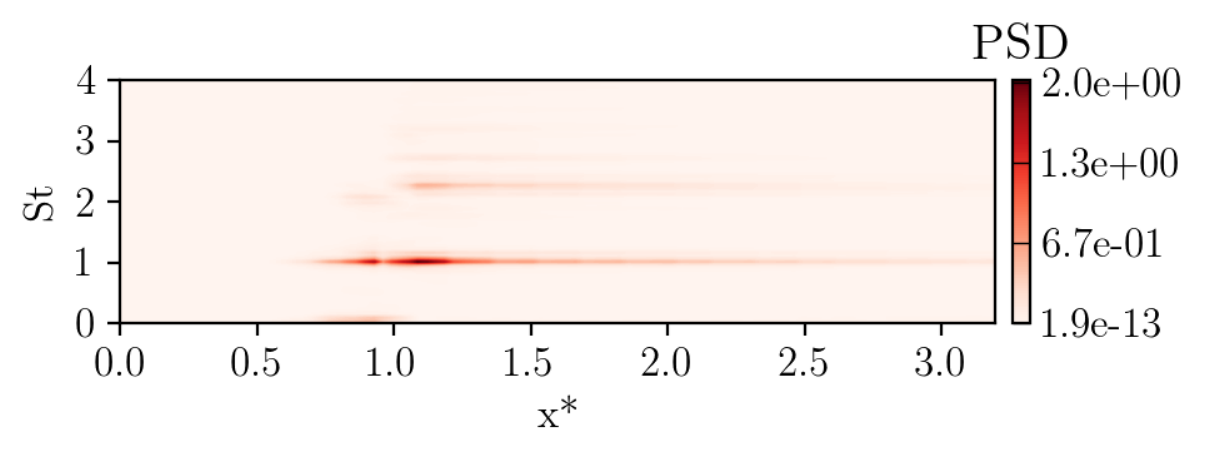}
    \caption{}\label{subfig:PSD_y_v_perio}
    \end{subfigure}
    \end{center}
    \caption{PSD maps of the periodic case along longitudinal axis at the mid-width and at a vertical distance $z_p^*=0.032$ from the bottom wall : (a) for the void ratio $\alpha$; (b) for the spanwise velocity $v$.}
    \label{fig:Mapping_perio}
\end{figure}

Dynamics analysis of the case with periodic boundaries is then carried out to invastigate the sidewall effects. First observations of snapshots do not allow to identify any periodic oscillations of the cavitation pocket. However, a three-dimensional dynamic behavior of the cavity is observed around the closure. Figure \ref{fig:Mapping_perio} shows PSD maps along the longitudinal axis for the void ratio and the spanwise velocity component. The same Strouhal number $St_0$ and its first harmonic are extracted around the cavity closure for both variables and propagated downstream for the spanwise velocity component. The same Strouhal number is obtained from PSD maps over the results of the periodic case and the case with sidewalls. Hence, it suggests that the cavitation pocket fluctuations are not dependant of the presence of sidewalls. A deepened analyse is performed in Sec. \ref{sec:modal} to confirm this observation. 

\section{Re-entrant jet}

\begin{figure}[h]
    \begin{center}
    \begin{subfigure}[b]{\textwidth}
     \includegraphics[width=\textwidth]{./plot_overtime.pdf}
     \caption{}
    \end{subfigure}
    \begin{subfigure}{0.31\textwidth}
    \vspace{0.3cm}
    \includegraphics[width=\textwidth]{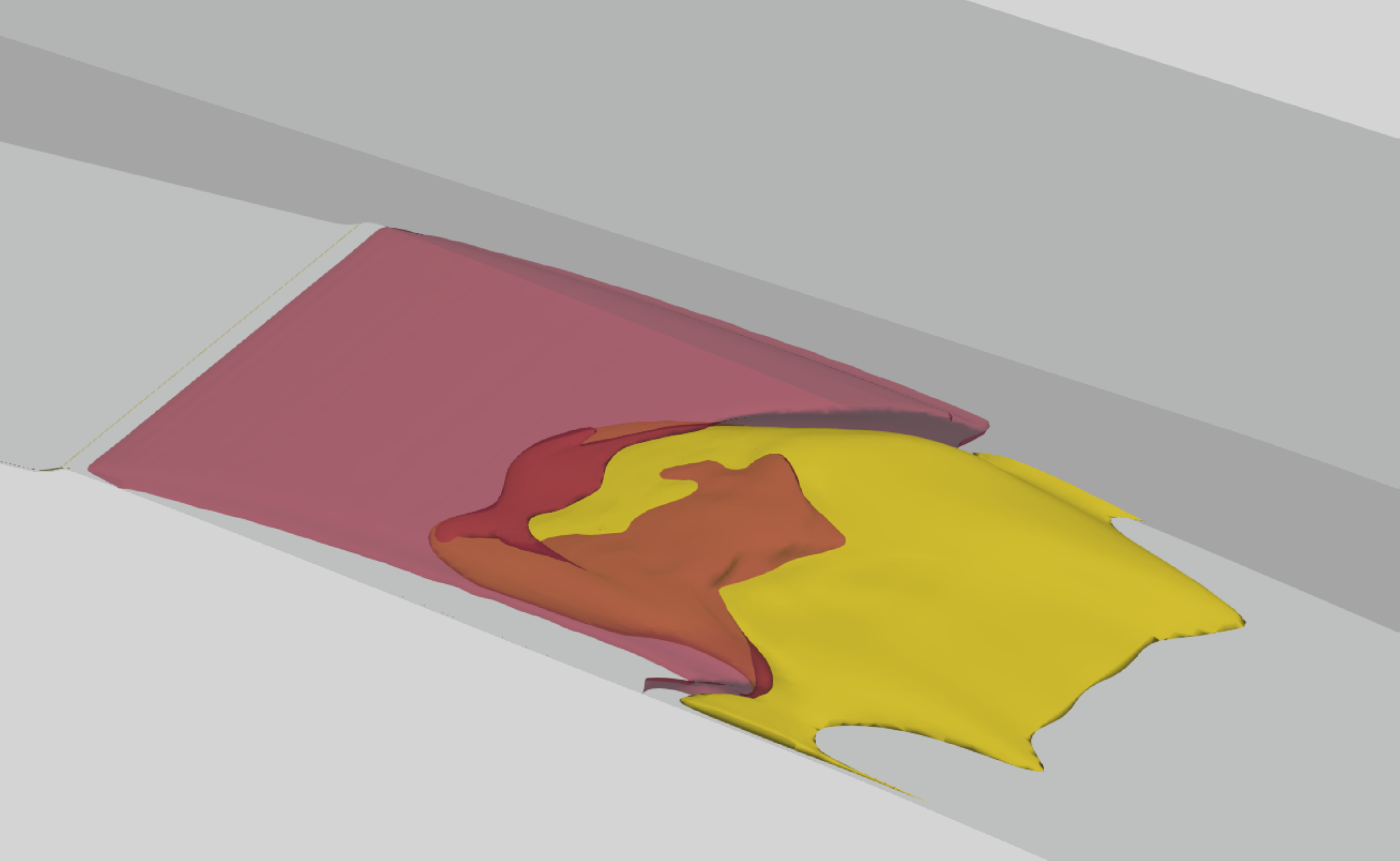}
    \caption{$t_1$}
    \end{subfigure}
    \hfill
    \begin{subfigure}{0.31\textwidth}
    \vspace{0.3cm}
    \includegraphics[width=\textwidth]{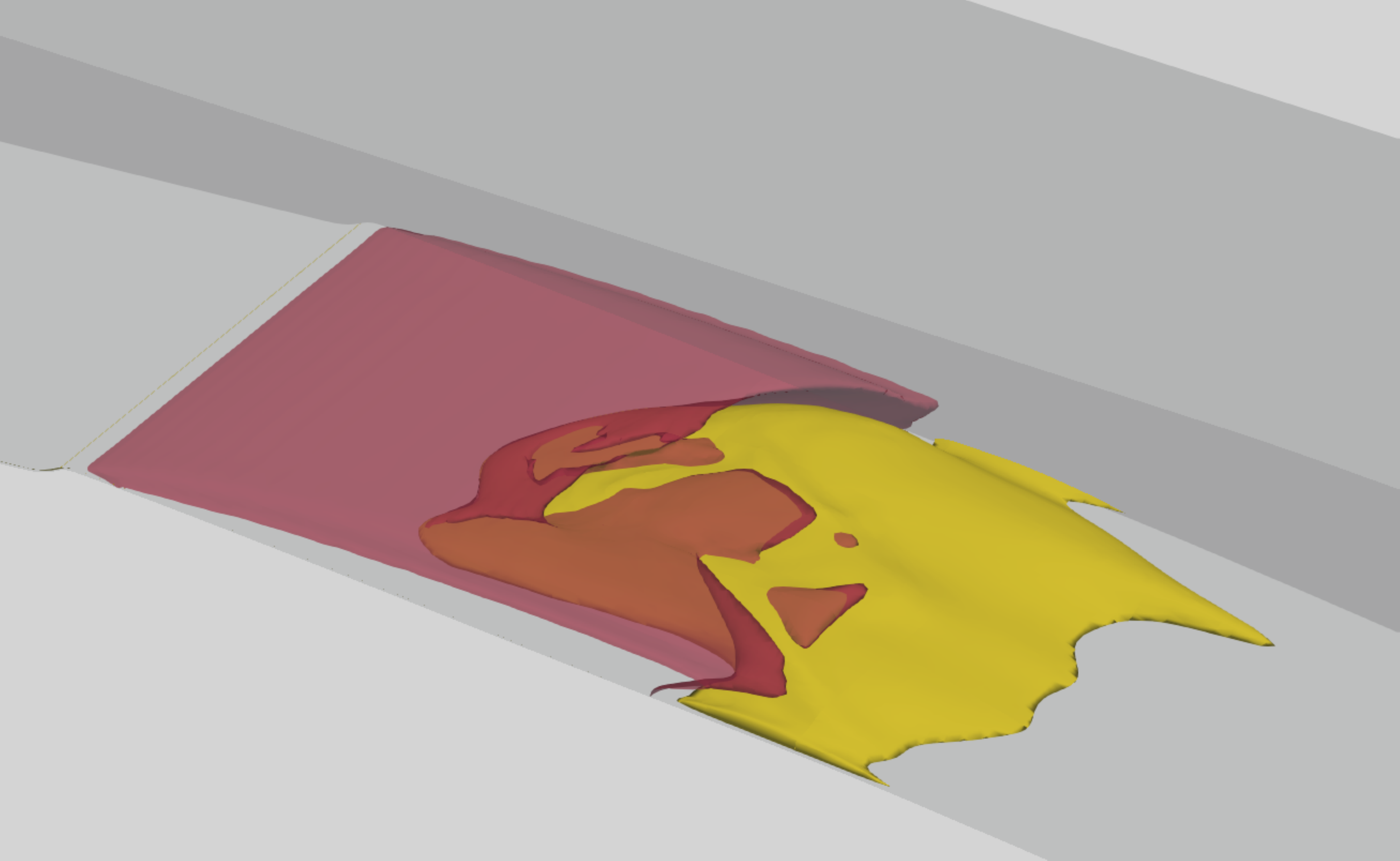}
    \caption{$t_2$}
    \end{subfigure}
    \hfill
    \begin{subfigure}{0.31\textwidth}
    \vspace{0.3cm}
    \includegraphics[width=\textwidth]{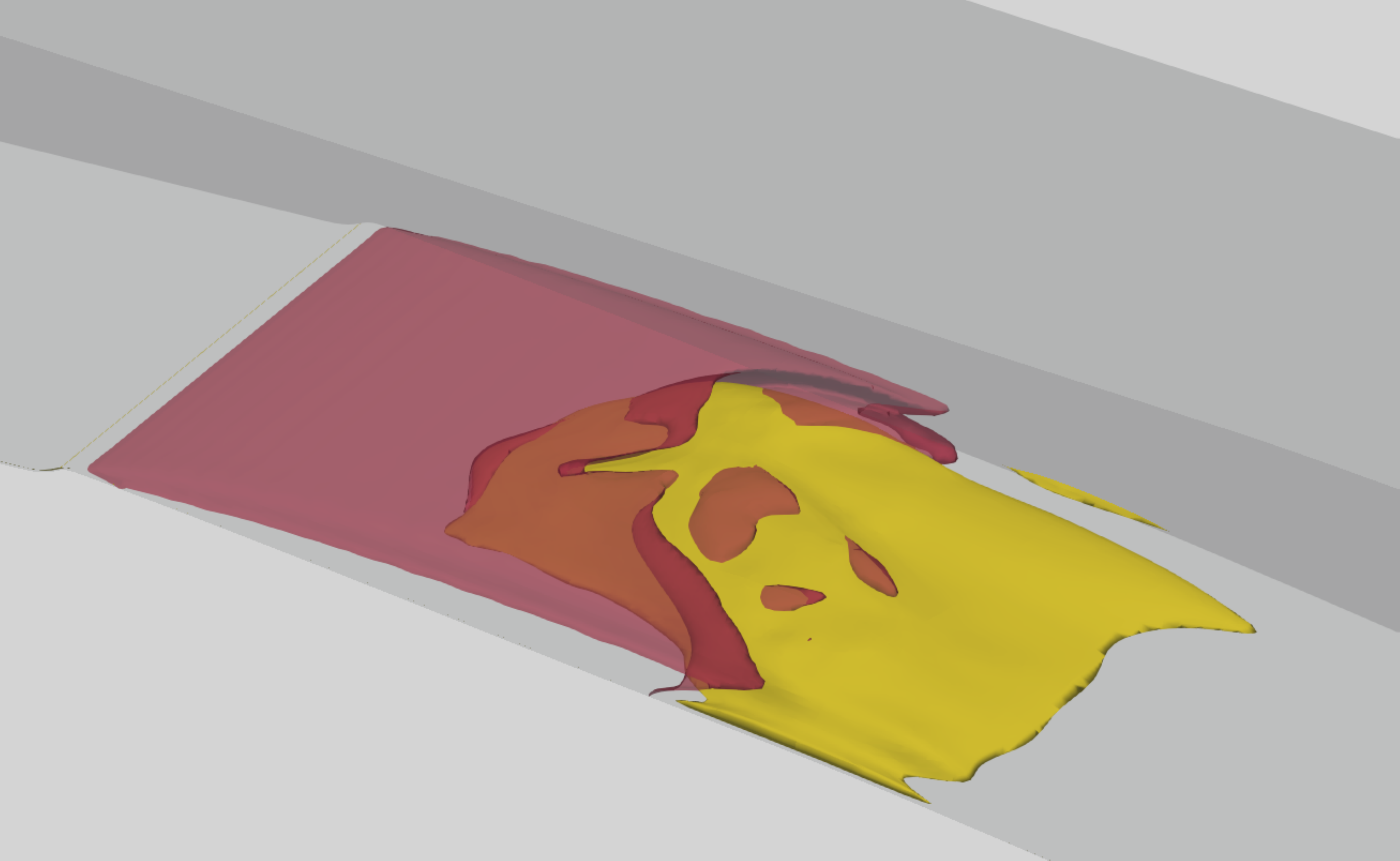}
    \caption{$t_3$}
    \end{subfigure}
    \hfill
    \begin{subfigure}{0.31\textwidth}
    \vspace{0.3cm}
    \includegraphics[width=\textwidth]{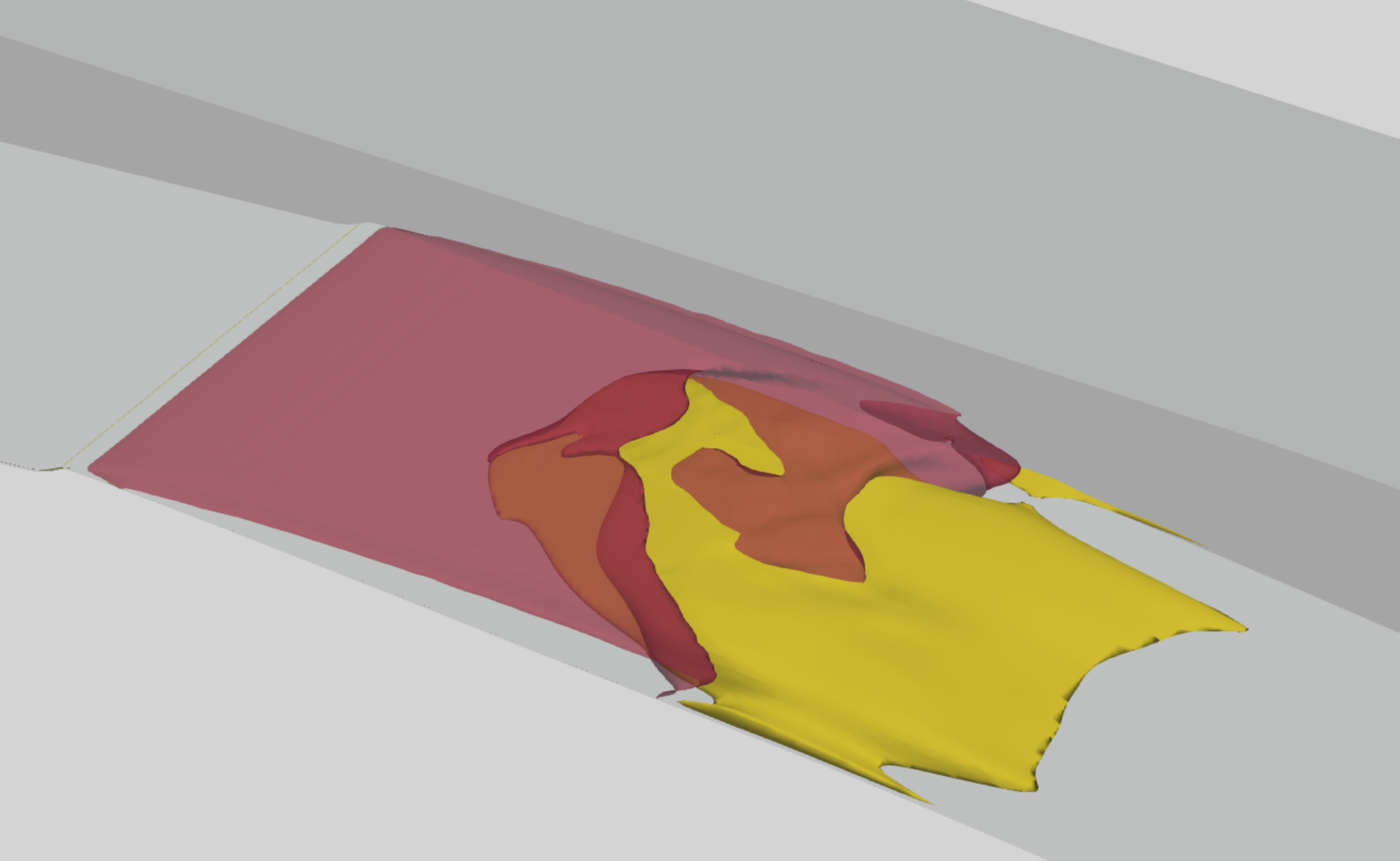}
    \caption{$t_4$}
    \end{subfigure}
    \hfill
    \begin{subfigure}{0.31\textwidth}
    \vspace{0.3cm}
    \includegraphics[width=\textwidth]{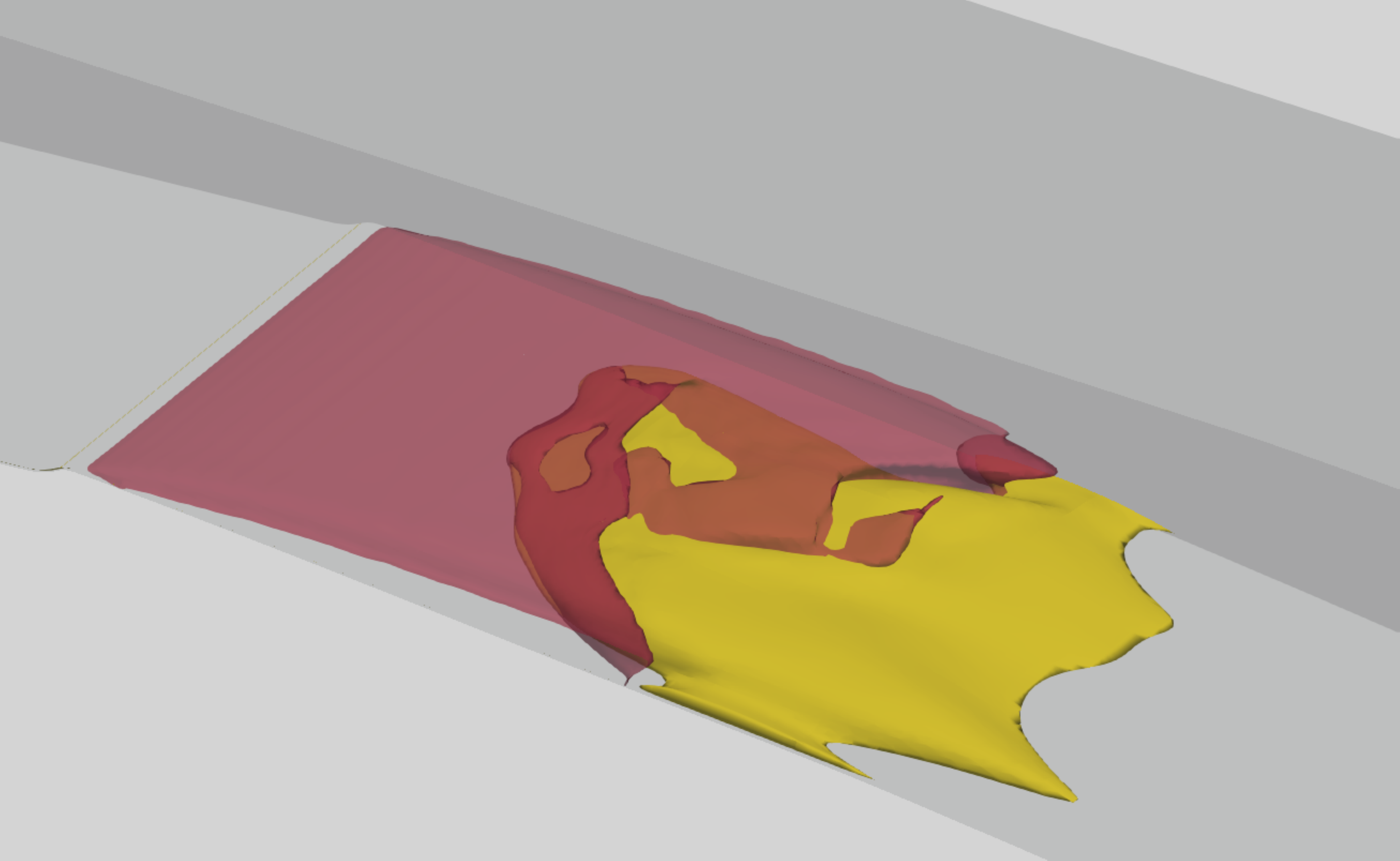}
    \caption{$t_5$}
    \end{subfigure}
    \hfill
    \begin{subfigure}{0.31\textwidth}
    \vspace{0.3cm}
    \includegraphics[width=\textwidth]{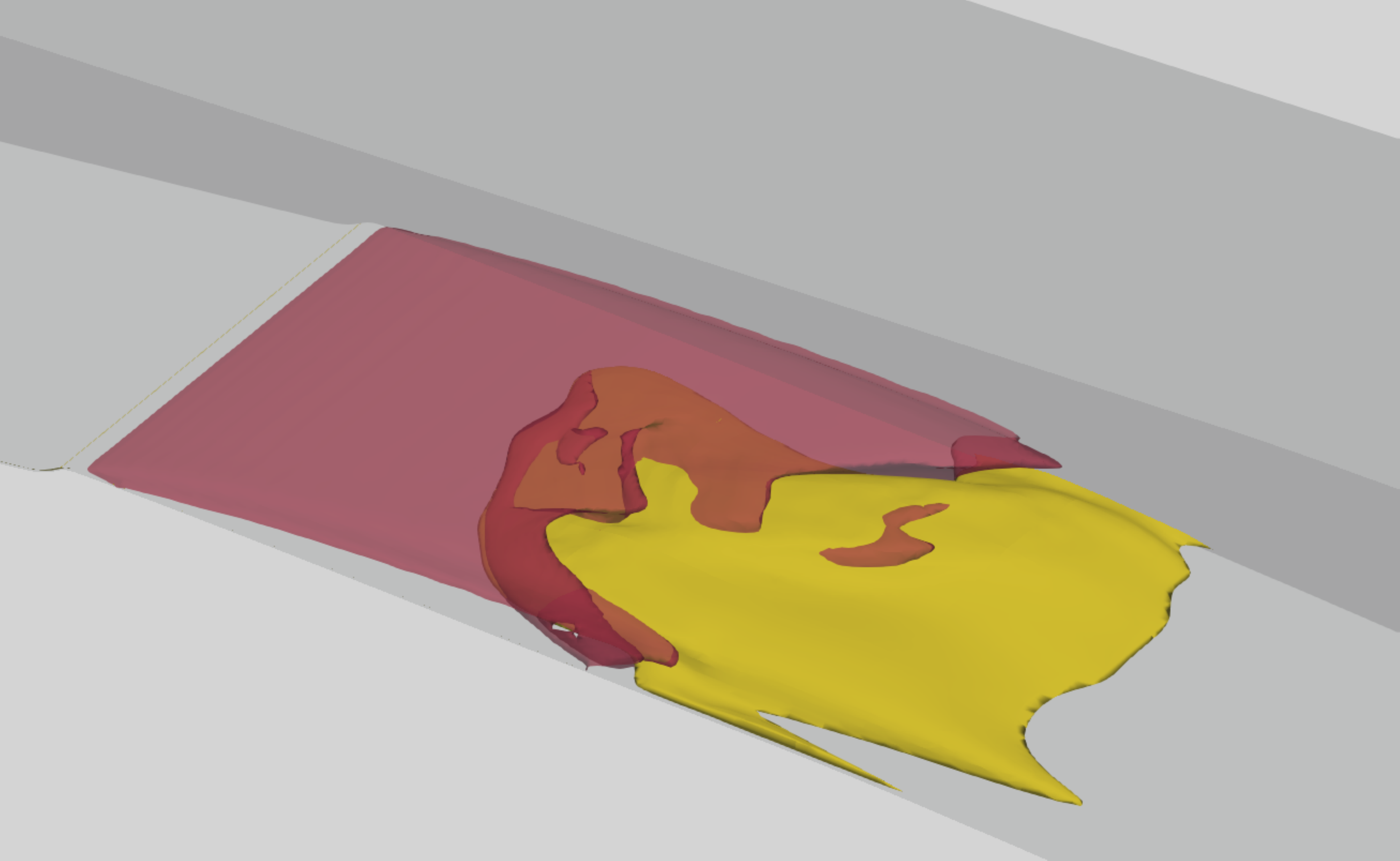}
    \caption{$t_6$}
    \end{subfigure}
    \end{center}
    \caption{Dynamics of the re-entrant jet regarding the vapor cavity : (a) void ratio over time at $x^*=0.8$ and $y^*=3L_y^*/4$ for the case with sidewalls; (a)-(f): Snapshots with timestep $\Delta t = 4.58\times 10^{-3} s$ of $\alpha=0.5$ (purple) and $u^*=-0.1 \ m.s^{-1}$ (yellow) contours.\label{fig:screenshot_pocket_jet}}
\end{figure}
\begin{figure}[H]
    \centering
    \includegraphics[width=\textwidth]{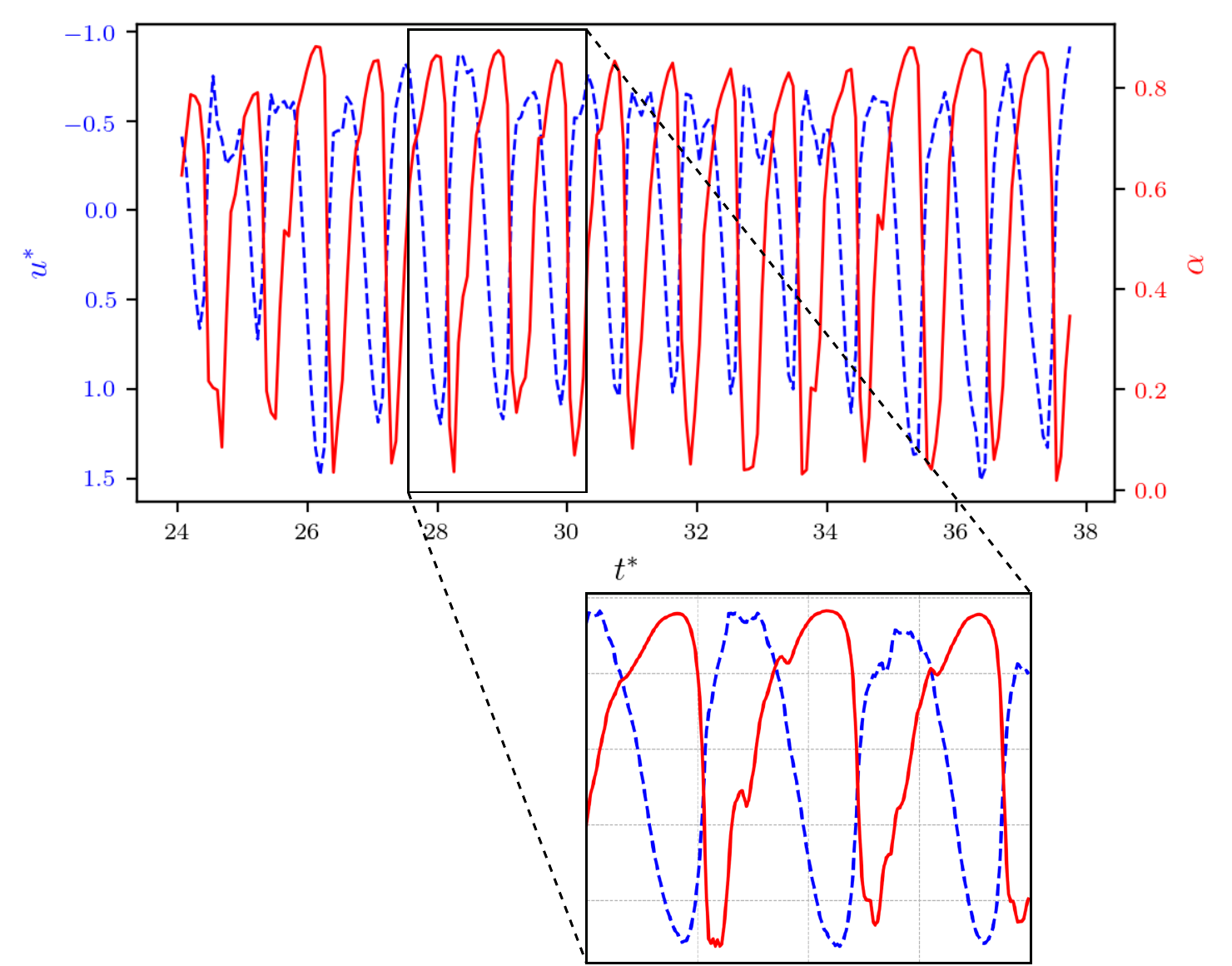}
    \caption{Time evolution of the flow direction velocity $u$ and the void ratio $\alpha$ at $x^*=0.64$ and $y^*=L_y/4$ for the case with sidewalls. }
    \label{fig:alpha_u_x02}
\end{figure}

In this section, the behavior of the re-entrant jet is studied in regards to the cavitation pocket oscillation for the case with sidewalls. Figure \ref{fig:screenshot_pocket_jet} describes the re-entrant jet position in relation to the cavity position at different times over an oscillation period. As expected, the re-entrant jet is located nearby the cavity closure and is time-dependent. Moreover, PSD maps of the streamwise velocity highlight a dynamics around the cavity closure based on the Strouhal number $St_0$. Thus, the re-entrant jet oscillates at the same frequency as the cavitation pocket from a spanwise wall to another. However, the position of the re-entrant jet compared to the cavity has to be determined. Figure \ref{fig:alpha_u_x02} presents the time evolution of the void ratio $\alpha$ and the flow direction velocity $u$ around the cavity closure at the quarter width. The negative values of the streamwise velocity component illustrate the re-entrant jet position and the highest value of the void ratio represents the cavity position. The time evolution of $u$ and $\alpha$ indicates an opposition in both oscillations of the cavitation pocket and the re-entrant jet. When the pocket is asymmetric in the spanwise direction, the re-entrant jet presents an opposite asymmetry. The cavity growth near side walls is smoother when compared to the cavity disappearance at the same position. For the re-entrant jet signal, it is the opposite behavior, high growth and a smoother decrease. Therefore, the maximum peaks of void ratio exactly correspond, in time, to the maximum ones of flow direction velocity, while both minimum ones are time-shifted. Hence, the spanwise movement is not uniform. When the cavity moves nearby sidewalls, it is pushed back with acceleration and, at the same time, the re-entrant jet motion changes its spanwise direction with an acceleration. Such behavior indicates a possible causality effect between both phenomena.




\section{Modal decomposition analysis}
\label{sec:modal}

Early results highlighted a dominant flow component at the Strouhal number $St_0$ probably linked to a cavitation pocket oscillation. Modal decomposition analysis are performed to corroborate previous observations. The Spectral Proper Orthogonal Decomposition (SPOD) is computed from the computational data to identify spatiotemporal mechanisms. The choice of the SPOD is motivated by the extraction of spatiotemporal modes which is the most coherent method to study unsteady flow as presented by Towne et al.\@ (2018)\cite{Towne2018}. The SPOD methodology employed, based on Schmidt and Colonius (2020)\cite{Schmidt2020} work, is described below.\\

Given a snapshot $\bm{q}_i'=\bm{q}_i-\bm{\bar q}$ which represents the fluctuation of the flow result for the density and the velocity components at a time $t_i$, the data matrix $\bm{Q}$ is defined as:
\begin{equation}
\bm{Q}=[\bm{q}_1' \ \bm{q}_2' \ \bm{q}_3' \ ... \ \bm{q}_N'],
\end{equation}
with $\bm{\bar q}$ the temporal mean. The cavitating flow is modeled using a compressible formulation, therefore, the instantaneous energy is defined based on the Chu norm \citep{Chu1965} with the temperature fluctuation neglected. It is expressed with a spatial inner product:
\begin{equation}
\| \bm{q}_i'\|_E = \langle \bm{q}_i',\bm{q}_i'\rangle_E = \bm{q}_i'^T \bm{M q}_i' = \bm{q}_i'^T \bm{R}^T\bm{R q}_i' = \| \bm{Rq}_i'\|_2 ,
\end{equation}

\begin{center}
\begin{tabular}{l}
$\bm{M} = \bm{R}^T\bm{R} = \left(\begin{array}{cccc}
            \frac{\bar T}{2\bar\rho\gamma M_{\infty}^2}\bm{A} & 0 & 0 & 0\\
            0 & \frac{\bar\rho}{2}\bm{A} & 0 & 0\\
            0 & 0 & \frac{\bar\rho}{2}\bm{A} & 0\\
            0 & 0 & 0 & \frac{\bar\rho}{2}\bm{A}
          \end{array}
          \right)$.
\end{tabular}
\end{center}
Where $\bm{A}$ stands for the diagonal cell volume matrix, $\bar T$ the temporal mean temperature, $\bar \rho$ the temporal mean density and $M_\infty$ the far-field Mach number. The first step of the SPOD decomposition is to apply Welch's method to the data matrix. It consists of separating the data in $N_{blk}$ overlapping blocks of snapshots and then performed the discrete Fourier transform on each block. Thus, $N_{blk}$ matrix of $N_{freq}$ discrete frequency realisations are obtained. Then, the Fourier realisations of all blocks are grouped by frequency to obtain $N_{freq}$ matrix $\bm{\hat Q}_{fr}$:
\begin{equation}
    \bm{\hat Q}_{fr}=[\bm{\hat q}^1_{fr} \ \bm{\hat q}^2_{fr} \ \bm{\hat q}^3_{fr} \ ... \ \bm{\hat q}^{N_{blk}}_{fr}],
\end{equation}
where $\bm{\hat q}^i_{fr}$ denotes the Fourier realisation of the $i^{th}$ block at the frequency $fr$. The second part of the SPOD decomposition is to compute the cross-spectral density matrix at each frequency:
\begin{equation}
    \bm{S}_{fr}=\bm{\hat Q}_{fr} \bm{\hat Q}_{fr}^*,
\end{equation}
$(.)^*$ denotes the complex conjugate. Thus, the SPOD modes $\bm{\Psi}_{fr}$ are generated by the eigenvalue decomposition of the cross-spectral density matrix:
\begin{equation}
     \bm{S}_{fr}\bm{M} \bm{\Psi}_{fr}^* = \bm{\Psi}_{fr} \bm{\Lambda}_{fr},
\end{equation}
with $\bm{\Lambda}_{fr}$ the diagonal matrix of eigenvalues representing the mode energy from the most energetic, corresponding to the leading SPOD mode, to the less one.\\

\begin{figure}[H]
    \centering
    \begin{subfigure}{\textwidth}
    \includegraphics[width=0.9\textwidth]{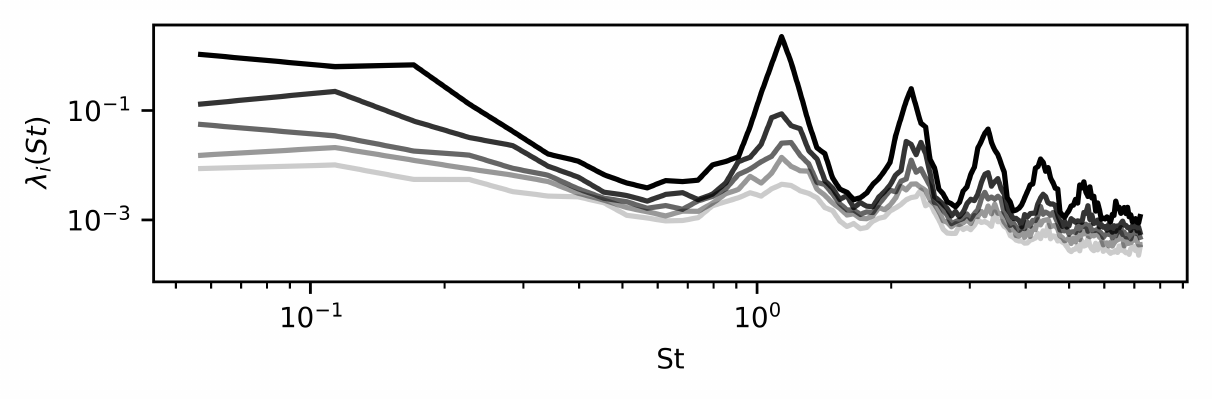}
    \caption{}
    \end{subfigure}
    \begin{subfigure}{\textwidth}
    \includegraphics[width=0.9\textwidth]{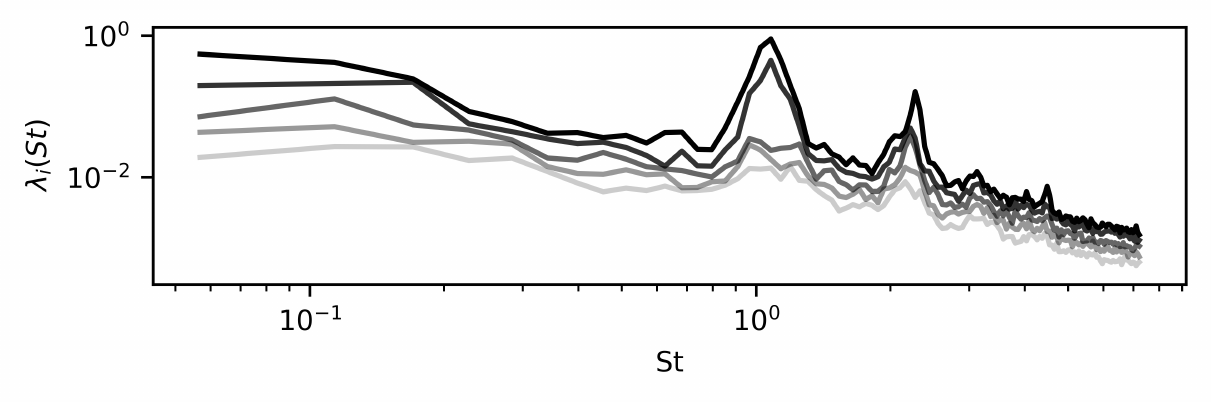}
    \caption{}
    \end{subfigure}
    \caption{SPOD spectrums representing energy gain over Strouhal number : (a) for the sidewalls case; (b) for the periodic case. The black to grey scale represents the most energetic mode to the less energetic one for each Strouhal number.}
    \label{fig:SPOD_spectrum}
\end{figure}

Figure \ref{fig:SPOD_spectrum} presents the SPOD spectrum for both studied cases. The Strouhal number $St_0=1.09$ and its harmonics emerge from the spectrum for the simulation with sidewalls. Similarly, for the periodic sides simulation, the Strouhal number $St_0$ is extracted from the spectrum but with only the first harmonic. Hence, the dominant flow mechanism seems to be characterised by the Strouhal number $St_0$. Moreover, for the sidewalls case, the observation of the energy gain gap between the first and the second SPOD mode at the corresponding frequency proves that the associated mechanism is mostly led by the first mode. Nevertheless, for the periodic sides case, the energy gap is located between the second and the third mode. The associated mechanism is then mostly driven by the two first modes. \\

\begin{figure*}
    \centering
    \begin{subfigure}{0.49\textwidth}
    \includegraphics[width=\textwidth]{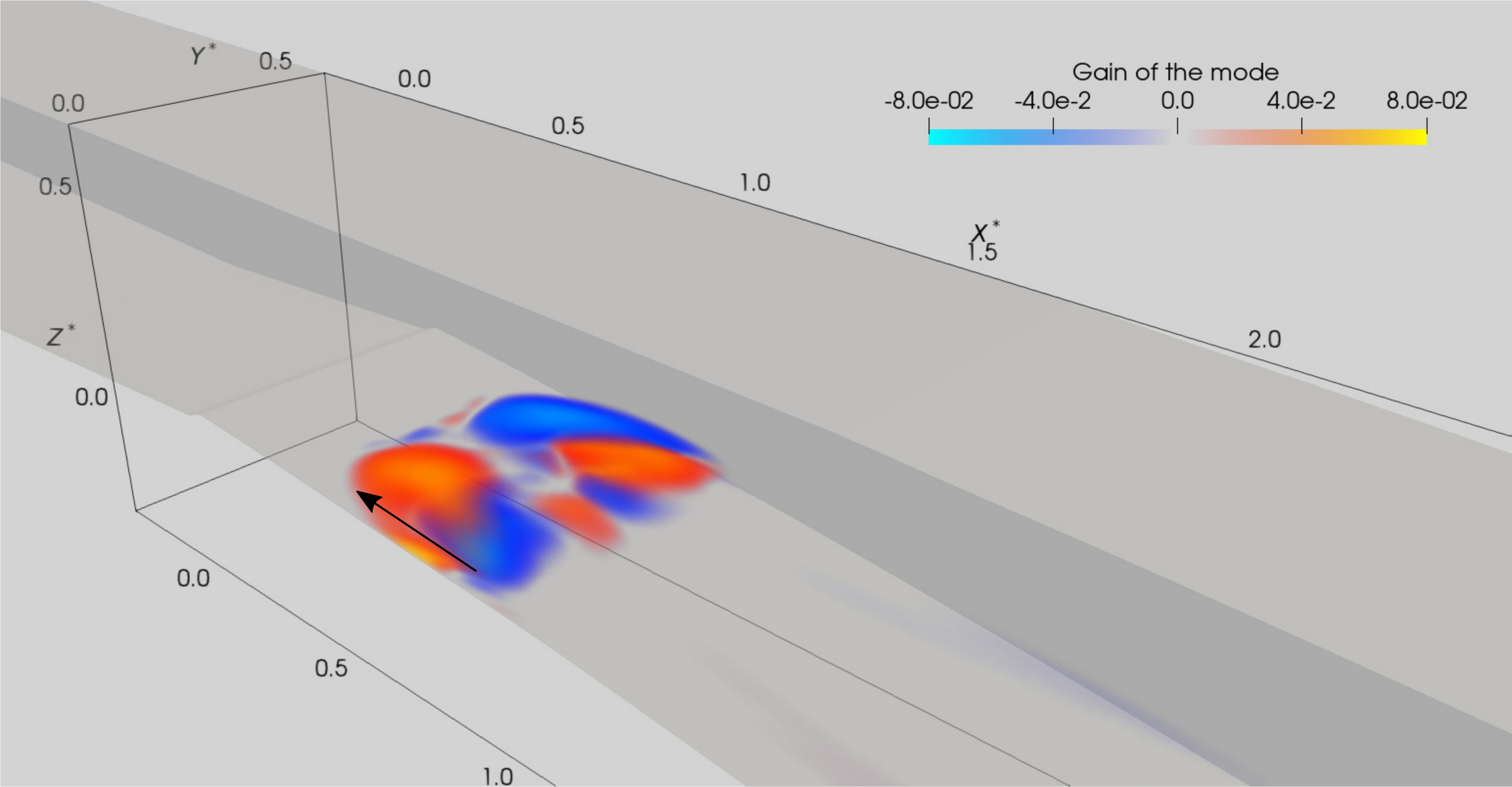}
    \caption{}
    \label{fig:rho_r_SPOD_walls}
    \end{subfigure}
    \begin{subfigure}{0.49\textwidth}
    \includegraphics[width=\textwidth]{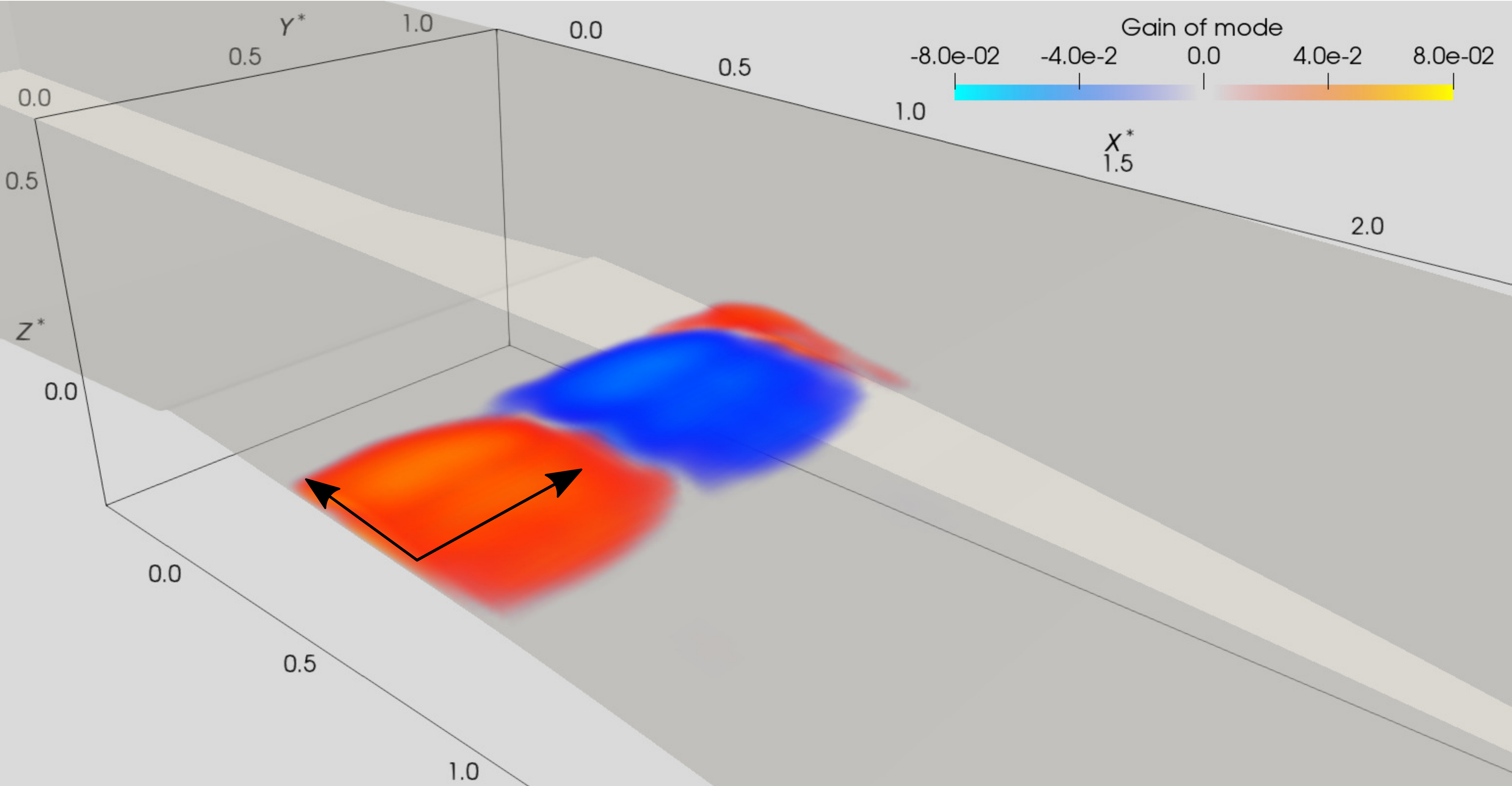}
    \caption{}
    \label{fig:rho_r_SPOD_perio}
    \end{subfigure}
    \caption{Volume rendering of the dominant mode of the SPOD for $\rho$ at $St=1.09$; $\rightarrowtriangle$ : Time evolution behavior; (a) for the case with sidewalls; (b) for the case with periodic side boundaries.}
    \label{fig:rho_r_SPOD}
\end{figure*}

\begin{figure*}
    \centering
     \begin{subfigure}{0.49\textwidth}
    \includegraphics[width=\textwidth]{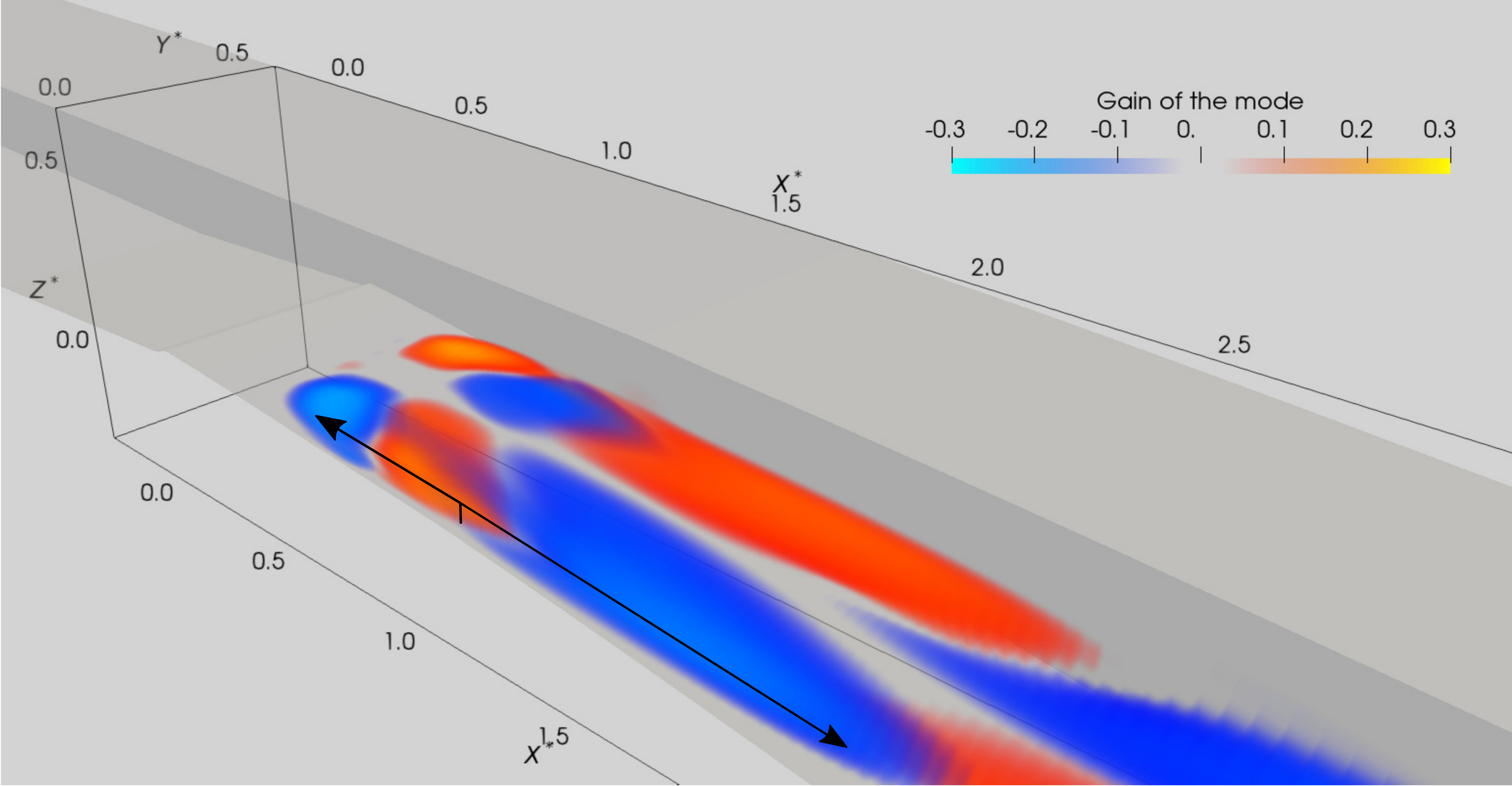}
    \caption{}
    \label{fig:u_r_SPOD_walls}
    \end{subfigure}
    \begin{subfigure}{0.49\textwidth}
    \includegraphics[width=\textwidth]{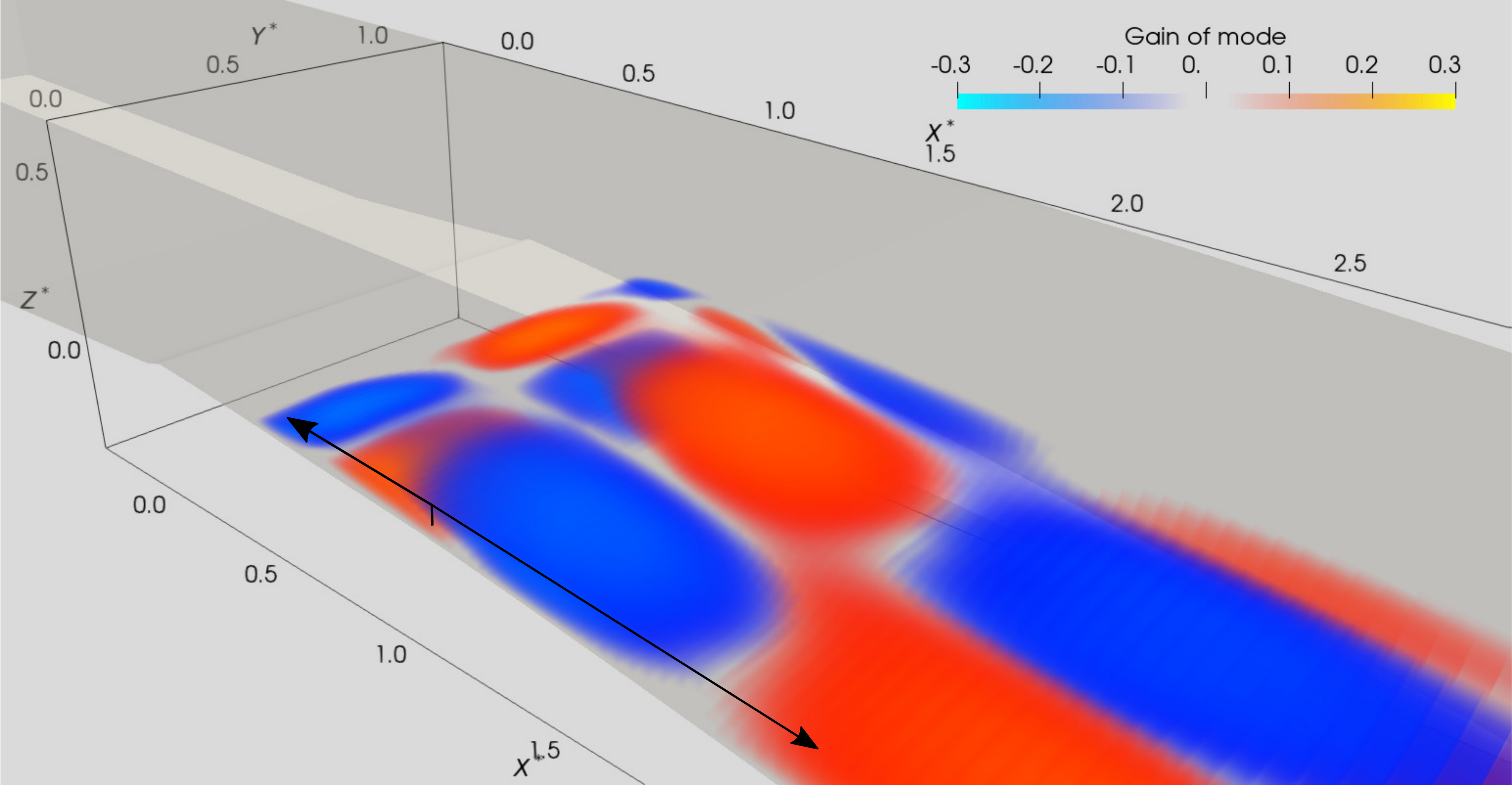}
    \caption{}
    \label{fig:u_r_SPOD_perio}
    \end{subfigure}
    \caption{Volume rendering of the dominant mode of the SPOD for $u$ at $St=1.09$; $\rightarrowtriangle$ : Time evolution behavior; (a) for the case with sidewalls; (b) for the case with periodic side boundaries.}
    \label{fig:u_r_SPOD}
\end{figure*}

\begin{figure*}
    \centering
    \begin{subfigure}{0.49\textwidth}
    \includegraphics[width=\textwidth]{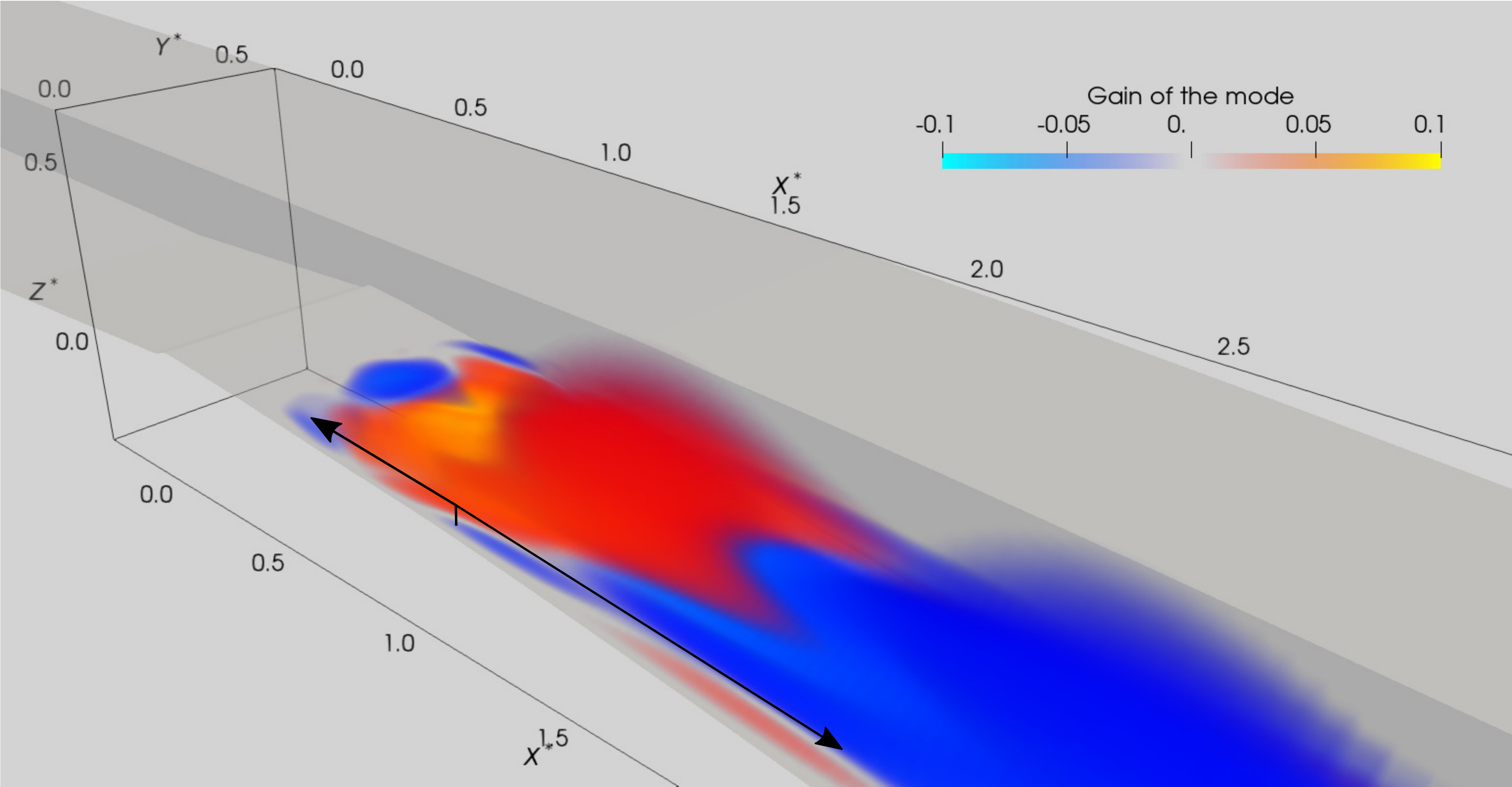}
    \caption{}
    \label{fig:v_r_SPOD_walls}
    \end{subfigure}
    \begin{subfigure}{0.49\textwidth}
    \includegraphics[width=\textwidth]{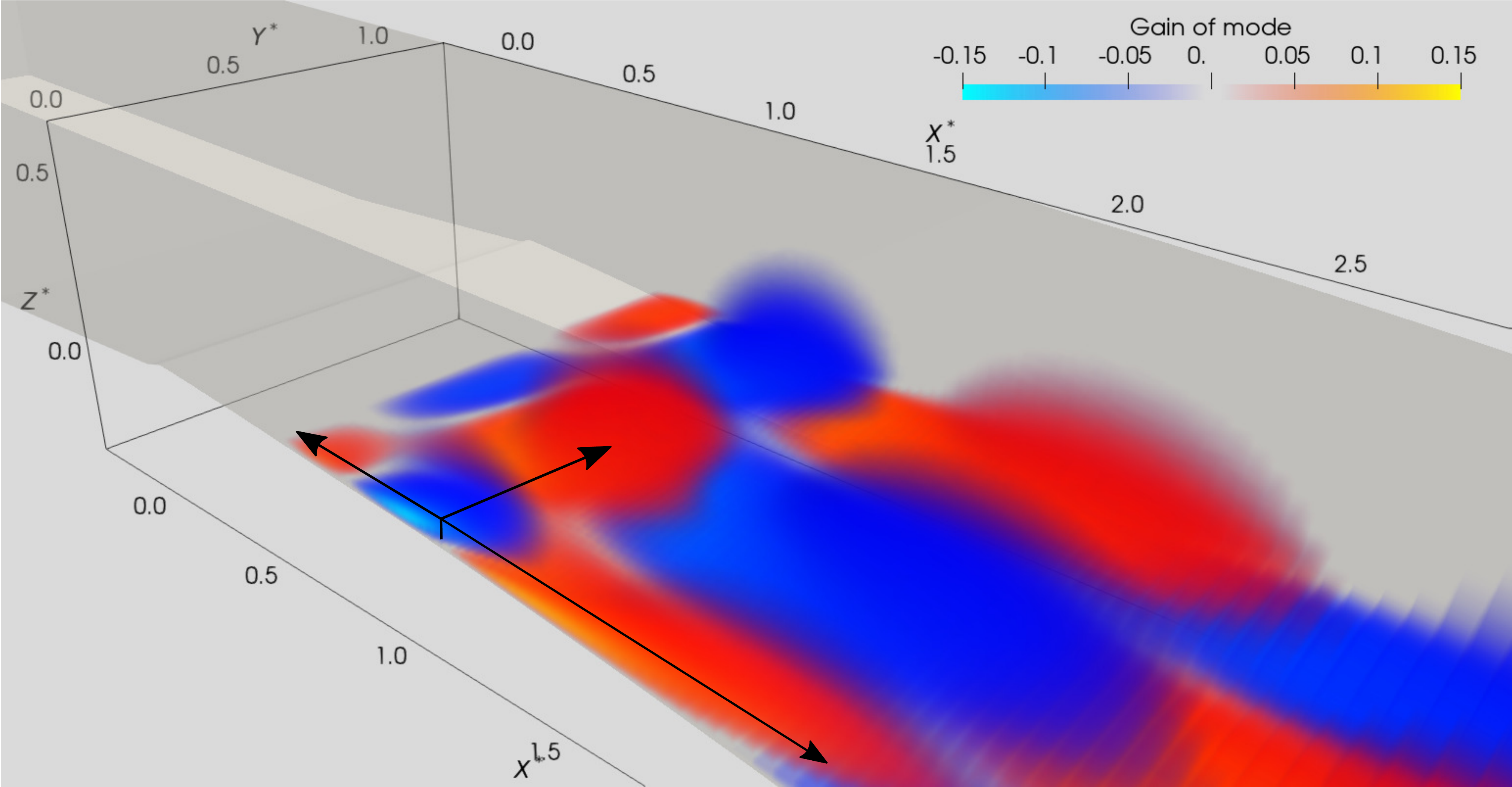}
    \caption{}
    \label{fig:v_r_SPOD_perio}
    \end{subfigure}
    \caption{Volume rendering of the dominant mode of the SPOD for $v$ at $St=1.09$; $\rightarrowtriangle$ : Time evolution behavior; (a) for the case with sidewalls; (b) for the case with periodic side boundaries.}
    \label{fig:v_r_SPOD}
\end{figure*}

\begin{figure*}
    \centering
    \begin{subfigure}{0.49\textwidth}
    \includegraphics[width=\textwidth]{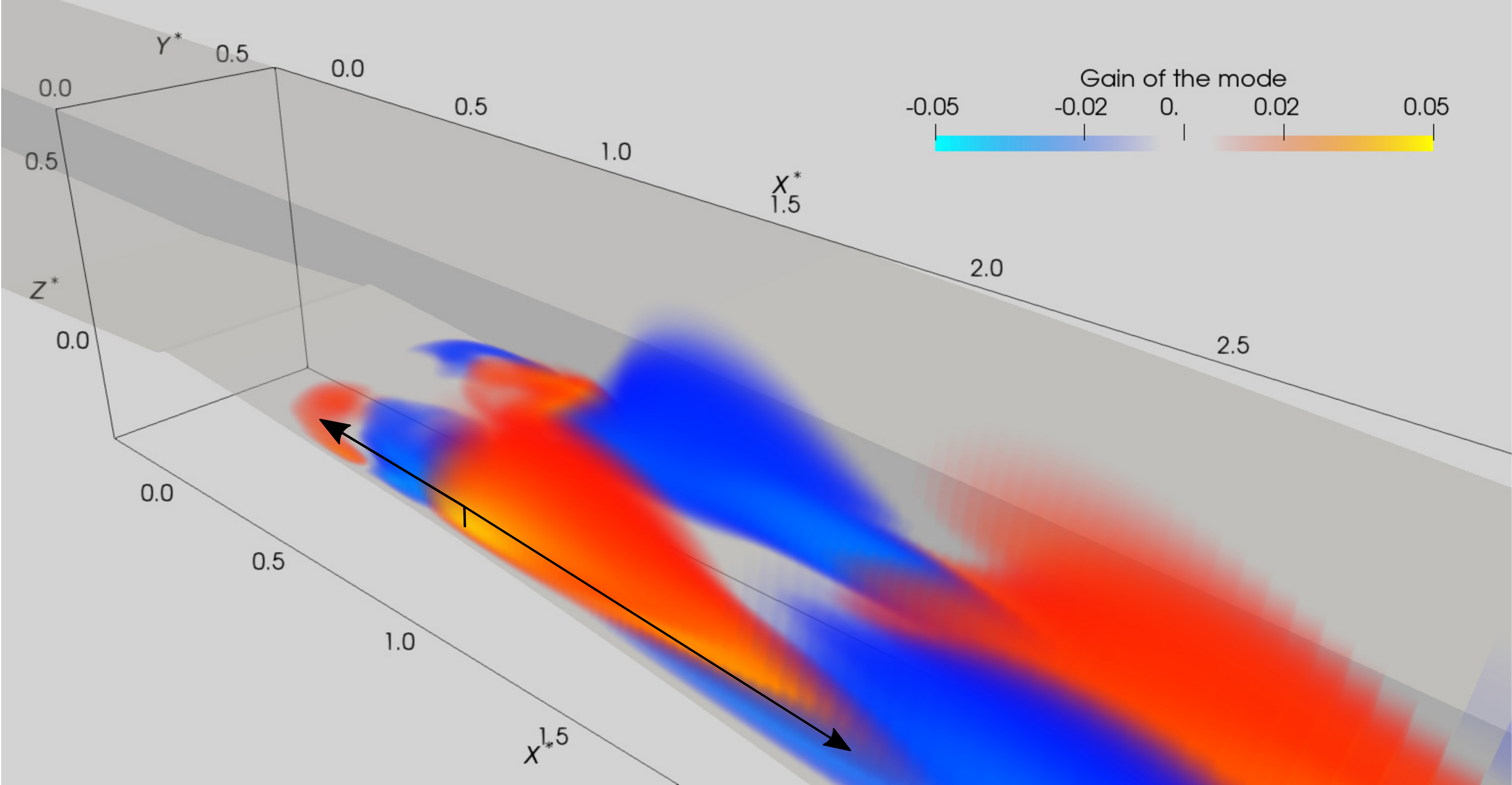}
    \caption{}
    \label{fig:w_r_SPOD_walls}
    \end{subfigure}
    \begin{subfigure}{0.49\textwidth}
    \includegraphics[width=\textwidth]{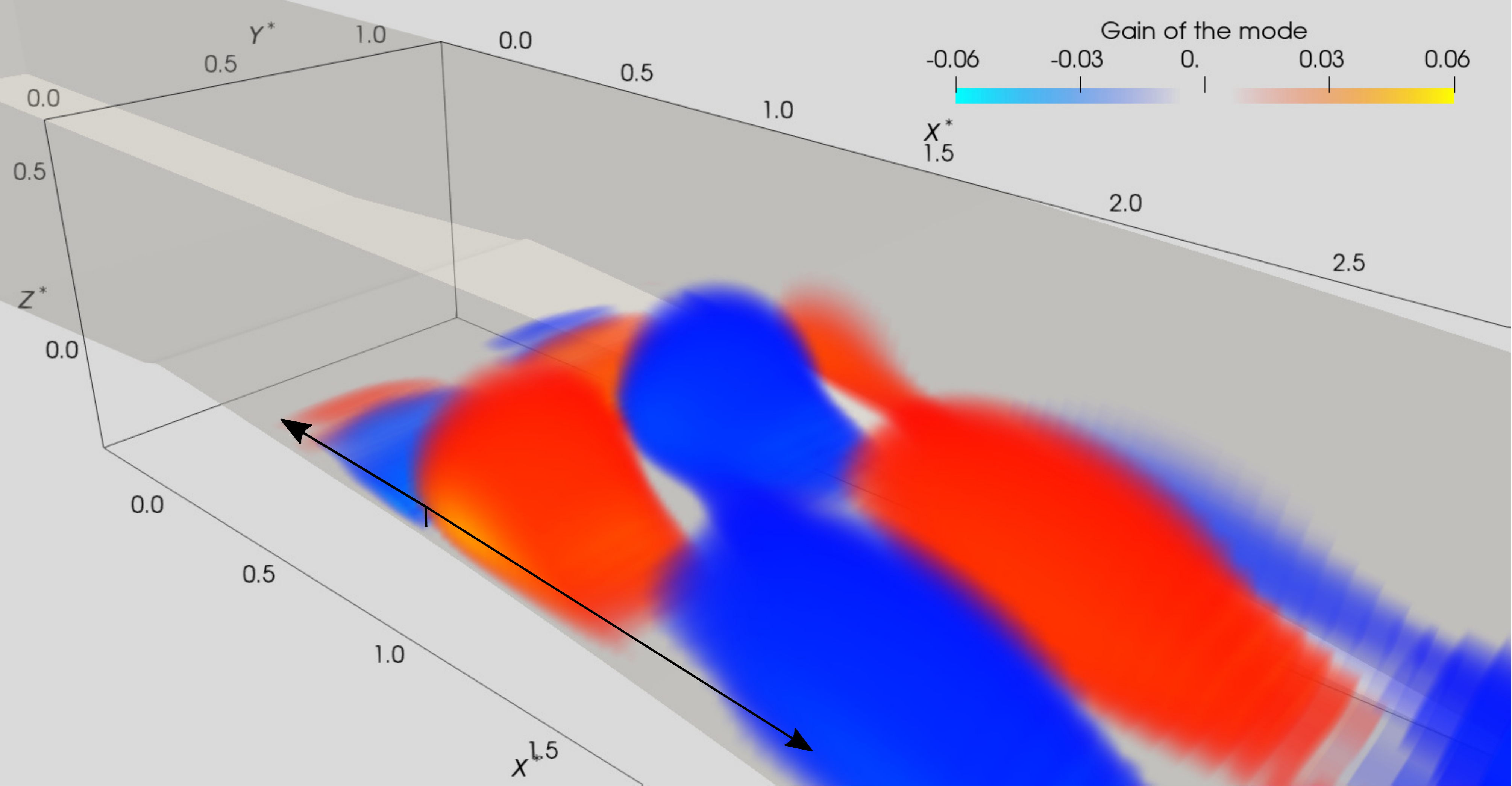}
    \caption{}
    \label{fig:w_r_SPOD_perio}
    \end{subfigure}
    \caption{Volume rendering of the dominant mode of the SPOD for $w$ at $St=1.09$; $\rightarrowtriangle$ : Time evolution behavior; (a) for the case with sidewalls; (b) for the case with periodic side boundaries.}
    \label{fig:w_r_SPOD}
\end{figure*}

Figures \ref{fig:rho_r_SPOD} to \ref{fig:w_r_SPOD} show the real part of the dominant mode for density and velocity components for both cases. Arrows underline the time evolution of the mode. For the density in the simulation with sidewalls in Fig.\@\ref{fig:rho_r_SPOD_walls}, the mode is antisymmetric and corresponds to a spanwise oscillation coupled with a non-uniform upstream flow. It is noticed that the mid-width plane is not affected by the fluctuations. Similarly, the spanwise oscillations are observed by the antisymmetric mode for the streamwise velocity in Fig.\@\ref{fig:u_r_SPOD_walls} and the vertical velocity in Fig.\@\ref{fig:w_r_SPOD_walls} with the upstream behavior. Conversely, the mode of the spanwise velocity in Fig.\@\ref{fig:v_r_SPOD_walls} is symmetric but also corresponds to the spanwise oscillation between sidewalls. However, in contrast with the density oscillation, the dominant mode for velocity components is also propagated downstream by the flow with a higher speed. These results substantiate the previous ones observed in the PSD analysis. By examining the harmonic modes, it is determined that, contrary to the dominant one, the first presents a symmetric behavior for density, longitudinal velocity and vertical velocity while it illustrates an antisymmetric behavior for the spanwise velocity. Nevertheless, the second harmonic mode shows the same symmetrical and antisymmetrical characteristics as the dominant mode. Hence, an alternation of symmetry and antisymmetry is observed in harmonic modes.
The dominant SPOD mode for the case with periodic sides mainly differs from the sidewalls case for the density and the spanwise velocity component. In the first one, as observed in Fig.\@\ref{fig:rho_r_SPOD_perio}, a spanwise alternation of the cavitation pocket and an upstream flow are depicted. Then, a non-homogeneous spanwise motion is also captured around the cavity closure. It has to be noticed that the second SPOD mode, at the Strouhal number $St_0$, presents a similar behavior with an opposite direction for the spanwise motion. The first two modes are theoretically equiprobable. In the dominant mode for the spanwise velocity component, Fig.\@\ref{fig:v_r_SPOD_perio}, both upstream and downstream motions are detected but with also a non-homogeneous spanwise movement at the cavity closure. Unlike the case with sidewalls, this mode is not symmetric and it highlights an alternation of the spanwise velocity component along the spanwise axis. Furthermore, the second mode has an opposite movement along the spanwise axis. The SPOD mode behavior, for the streamwise and the vertical velocity components, is similar to the ones for the case with sidewalls.




\section{Discussion}

A dominant dynamics has been highlighted by diverse analysis of the cavitating flow. It has to be noticed that a similar behavior has been detected in computations with the turbulence model of Spalart Allmaras. The current section offers the authors interpretation of the phenomenon.




Oblique-shape behaviors of the cavitation pocket have already been observed in two different sheet cavitation experiments. The first one is a venturi flow experiment with 8$^o$ divergent angle carried out by Dular et al.\@ (2012)\cite{Dular2012}. In this case, vapor release into the flow appears under certain conditions and, for one of the studied geometry, the cavity presents a non-symmetrical shape. The authors suggest that it is caused by the re-entrant jet, which, besides going upstream, turns to the side. Nevertheless, due to the higher divergent angle, the pocket is cut by the re-entrant jet and leads to a vapor release. The second experiment is a flow around a guide vane profile carried out by Timoshevskiy et al.\@ (2016)\cite{Timoshevskiy2016}. Under the vapor release regime, non-symmetrical cavitation pocket behavior appears, as for the previous experiment. However, this regime is defined as non-persistent by the authors. Both experiments present a vapor release regime which could explain that the re-entrant jet bifurcation and the cavity shape do not lead to an identical spanwise oscillation. Nonetheless, it is suggested that the leading mechanism of those phenomena could be the same. The behavior of the re-entrant jet against the cavitation pocket in both experiments and the current work is identical: when the cavity is expanded on one side, the re-entrant jet is more developed on the other side.\\

The SPOD analysis validates the presence of a dominant dynamics of the flow at a Strouhal number of $1.09$. First, for the case with sidewalls, such a phenomenon is associated to the spanwise oscillation of the flow. The re-entrant jet is then captured as observed in the mode dynamics for density and velocity components. For this reason, it presents a significant role in the spanwise oscillation. The present study does not give enough information to ensure that the re-entrant jet is the mechanism which leads to the oscillation. Nevertheless, experiments describing the vapor release regime considered that the shedding is triggered by the re-entrant jet which "cuts" the cavity by going upward. Therefore, given the promiscuity of the physical phenomenon, it can be presumed that the re-entrant jet is the leading mechanism of the oscillation for the present case.

The spanwise oscillation is easily observed through the dominant mode dynamics for the spanwise velocity which highlights an alternation of positive and negative velocity around the cavity closure. By analysing the predominant mode dynamics for the velocity components, it is noticed that the oscillation pattern is simultaneously propagated upstream, by the re-entrant jet, and downstream by the main flow. The downstream flow is thus highly influenced by the pocket and the re-entrant jet dynamics while the oscillation seems to be self-sustained by the upstream flow suggesting the possible existence of a global mode driving this dynamic.

Fluctuations observed just downstream the cavitation pocket are interpreted as an oscillation of magnitude for the vertical and the streamwise velocity components. When the pocket is expanded near a sidewall, the magnitude of these two components is increased on the same side and decreased on the other. This information involves that the longer the cavity is, the more accelerated the downward and the downstream speeds are. When the pocket moves to the other side, the velocity effects are reversed.  Hence, the cavitation pocket interferes with the flow can be seen as a dynamic fluidic obstacle.\\

The results of the simulation with periodic sides boundaries give relevant information about the sidewalls impact on the flow. For this case, the cavity and the re-entrant jet shapes are unchanged along the spanwise axis. Therefore, the U-shape of the cavitation pocket and the re-entrant jet shape are linked to the presence of sidewalls. However, the extraction of a dominant mode at the same Strouhal number suggests that the sidewalls do not trigger the spanwise oscillation but only amplify a phenomenon. The dominant dynamics is then led by a mechanism specific to the cavitating flow. The Strouhal number of $1.09$ linked to the dominant mode is obtained by using two characteristic quantities: the length of the cavitation pocket $L_c=0.0788 \ m$ and the maximum time-averaged velocity of the re-entrant jet at the midspan $u^{max}_{jet}=2.38 \ m.s^{-1}$. Hence, the mechanism highlighted by the current study seems to be inherent to the pocket development.\\

For the periodic case, the associated energy of the two first SPOD modes extracted at the Strouhal number $St_0$ are close. The impact of these modes on the flow is then almost identical. Furthermore, both present an opposite spanwise movement of the fluctuations for the density and the spanwise velocity component. The two modes could be described as a bifurcation of the flow with an equal probability of appearance. This hypothesis could be investigated by simulating the case on a much longer time. 




\section{Conclusion}

The 3-D effects of cavitating flow, with a partial cavity, are studied in the case of a 4$^o$ divergent angle venturi. Two configurations, with sidewalls and with periodic side boundaries, are computed for an identical physical time. Time-averaged data and dynamic analysis highlight flow structure differences, in particular in the cavitation pocket and re-entrant-jet shapes. Nonetheless, an identic Strouhal number of $1.09$ linked to flow fluctuations is extracted in both cases. The SPOD analysis provides details about this phenomenon. A dominant mode at this Strouhal number is captured for both cases. For the simulation with sidewalls, it represents a spanwise oscillation of the flow observed through the cavitation pocket and re-entrant jet oscillation. It is noticed that the cavity and the re-entrant jet oscillations are in opposition of phase. Moreover, the flow bypassing the cavitation pocket accelerates and follows the pocket oscillation. Thus, the cavitation pocket acts as a dynamic obstacle. For the case with periodic spanwise boundaries, the dominant SPOD mode is energetically close to the second one at the same Strouhal number. Both show similar dynamics but with an opposite spanwise motion. These two modes could represent a flow bifurcation concerning spanwise fluctuations and are ideally equiprobable. Finally, the analysis of both cases involves that 3-D effects, non-related to the presence of sidewalls, appeares in this cavitating flow configuration. The results also suggests that the phenomenon is linked to two characteristic variables of the flow: the cavity length and the maximum of the time-averaged re-entrant jet velocity at midspan. The link between the extracted Strouhal number and the two characteristic variables could be investigated. Other flow configurations presenting distinct cavity length could be used to attest the robustness of this conclusion. 


\begin{acknowledgments}
This research was supported by the French National Research Agency ANR (project 18-CE46-009). This work was granted access to the HPC resources of IDRIS under the allocation 2020-A0072A06362 made by GENCI. The authors thank Jean-Christophe Loiseau for providing the SPOD code.
\end{acknowledgments}



\appendix

\section{Modified third order Jameson-Schmidt-Turkel scheme}
\label{ap:JST3}

For stiff problems like sheet cavitation modeled with a 1-fluid method, centered schemes with artificial dissipation has been selected. The Jameson-Schmidt-Turkel (JST) scheme is a second-order scheme proposed by Jameson et al.\@ (1981)\cite{Jameson1981}. It is composed of a second-order centered scheme and a dissipation term:
\begin{equation}
\Ipdemi{\mathbf{F}_c}= \dfrac{1}{2}(\mathbf{F}_{c_{i+1,j}}+\mathbf{F}_{c_{i,j}}) - \Ipdemi{D_1}(\mathbf{w}_{i+1,j}-\mathbf{w}_{i,j}).
\end{equation}
In this 2-D example, the flux is computed on the interface $i+\dfrac{1}{2}$ in the x direction. The term $D_1$ represents the dissipation separated in two parts. The first part is a second-order dissipation based on a pressure sensor $\eta_i$ and the second part is a four-order dissipation. It can be written, for the current example, as:
\begin{multline}
 D_{i+1/2,j}(\mathbf{w}_{i+1,j}-\mathbf{w}_{i,j}) \ = \ \epsilon^{(2)}_{i+1/2,j} \rho(A_{i+1/2,j})(\mathbf{w}_{i+1,j}-\mathbf{w}_{i,j}) \\ - \epsilon^{(4)}_{i+1/2,j}  \rho(\mathbf{A}_{i+1/2,j})(\mathbf{w}_{i+2,j}-3\mathbf{w}_{i+1,j}+3\mathbf{w}_{i,j}-\mathbf{w}_{i-1,j}),
 \label{dissip_JST}
\end{multline}
with $\rho(\mathbf{A}_{i+1/2,j})$ the spectral radius of the jacobian matrix $\mathbf{A}$. The term $\epsilon^{(2)}_{i+1/2,j}$ is defined with a parameter $k^{(2)} \in [0,1]$ and the pressure sensor $\eta_i$:
\begin{eqnarray*}
\Ipdemi{\epsilon^{(2)}} &=& k^{(2)}\text{max}[\eta_i;\eta_{i+1}],\\
\eta_i &=& \dfrac{|p_{i+1} - 2p_i + p_{i-1}|}{ p_{i+1} + 2p_i + p_{i-1} }.
\end{eqnarray*}
This sensor allows triggering the second-order dissipation only around high pressure gradients. The term $\epsilon^{(4)}_{i+1/2,j}$ is defined with a parameter $k^{(4)} \in [0.008,0.064]$ and allows to damp small oscillation far from shocks:
$$
\Ipdemi{\epsilon^{(4)}} = \text{max}\left[0,k^{(4)}-\Ipdemi{\epsilon}^2\right]
$$
Nevertheless, in the 1-fluid flow, high density gradients appear along with the interface between liquid and vapor. To prevent computational problems, a third term is added to the dissipation with the same formula that the $\epsilon^{(2)}_{i+1/2,j}$ but with another sensor $\eta_i^{(I)}$ based on the density:
\begin{align}
 \eta_i^{(I)} \ = \ \frac{|\rho_{i+1}-2\rho_i+\rho_{i-1}|}{\rho_{i+1}+2\rho_i+\rho_{i-1}}.
\end{align}
Hence, the dissipation term $\epsilon^{(2)}_{i+1/2,j}$ could be written as a sum of a dissipation around shocks and another around interface with their own constant $k^{(2)}$ and $k_I^{(2)}$:
\begin{align*}
    \epsilon^{(2)}_{i+1/2,j} \ = \ \epsilon^{(2) \ shock}_{i+1/2,j}+ \epsilon^{(2) \ interface}_{i+1/2,j}.
\end{align*}
Finally, the second-order centered scheme could be updated to a fourth-order one:
\begin{equation}
\Ipdemi{\mathbf{F}_c}= \dfrac{1}{12}(-\mathbf{F}_{c_{i+2,j}}+7 \mathbf{F}_{c_{i+1,j}}+7 \mathbf{F}_{c_{i,j}}-\mathbf{F}_{c_{i-1,j}}).
\end{equation}
The global numerical scheme (centered part plus dissipation) is, therefore, a third-order one.


\section*{Data Availability}
The data that support the findings of this study are available from the corresponding author upon reasonable request

\bibliography{Biblio}

\providecommand{\noopsort}[1]{}\providecommand{\singleletter}[1]{#1}%
\begin{thebibliography}{80}%
\makeatletter
\providecommand \@ifxundefined [1]{%
 \@ifx{#1\undefined}
}%
\providecommand \@ifnum [1]{%
 \ifnum #1\expandafter \@firstoftwo
 \else \expandafter \@secondoftwo
 \fi
}%
\providecommand \@ifx [1]{%
 \ifx #1\expandafter \@firstoftwo
 \else \expandafter \@secondoftwo
 \fi
}%
\providecommand \natexlab [1]{#1}%
\providecommand \enquote  [1]{``#1''}%
\providecommand \bibnamefont  [1]{#1}%
\providecommand \bibfnamefont [1]{#1}%
\providecommand \citenamefont [1]{#1}%
\providecommand \href@noop [0]{\@secondoftwo}%
\providecommand \href [0]{\begingroup \@sanitize@url \@href}%
\providecommand \@href[1]{\@@startlink{#1}\@@href}%
\providecommand \@@href[1]{\endgroup#1\@@endlink}%
\providecommand \@sanitize@url [0]{\catcode `\\12\catcode `\$12\catcode
  `\&12\catcode `\#12\catcode `\^12\catcode `\_12\catcode `\%12\relax}%
\providecommand \@@startlink[1]{}%
\providecommand \@@endlink[0]{}%
\providecommand \url  [0]{\begingroup\@sanitize@url \@url }%
\providecommand \@url [1]{\endgroup\@href {#1}{\urlprefix }}%
\providecommand \urlprefix  [0]{URL }%
\providecommand \Eprint [0]{\href }%
\providecommand \doibase [0]{http://dx.doi.org/}%
\providecommand \selectlanguage [0]{\@gobble}%
\providecommand \bibinfo  [0]{\@secondoftwo}%
\providecommand \bibfield  [0]{\@secondoftwo}%
\providecommand \translation [1]{[#1]}%
\providecommand \BibitemOpen [0]{}%
\providecommand \bibitemStop [0]{}%
\providecommand \bibitemNoStop [0]{.\EOS\space}%
\providecommand \EOS [0]{\spacefactor3000\relax}%
\providecommand \BibitemShut  [1]{\csname bibitem#1\endcsname}%
\let\auto@bib@innerbib\@empty
\bibitem [{\citenamefont {Laberteaux}\ and\ \citenamefont
  {Ceccio}(2001)}]{Laberteaux2001}%
  \BibitemOpen
  \bibfield  {author} {\bibinfo {author} {\bibfnamefont {K.~R.}\ \bibnamefont
  {Laberteaux}}\ and\ \bibinfo {author} {\bibfnamefont {S.~L.}\ \bibnamefont
  {Ceccio}},\ }\bibfield  {title} {\enquote {\bibinfo {title} {Partial cavity
  flows. {Part 1. C}avities forming on models without spanwise variation},}\
  }\href {\doibase 10.1017/S0022112000002925} {\bibfield  {journal} {\bibinfo
  {journal} {Journal of Fluid Mechanics}\ }\textbf {\bibinfo {volume} {431}},\
  \bibinfo {pages} {1--41} (\bibinfo {year} {2001})}\BibitemShut {NoStop}%
\bibitem [{\citenamefont {{de Lange}}, \citenamefont {{de Bruin}},\ and\
  \citenamefont {{van Wijngaarden}}(1994)}]{Lange1994}%
  \BibitemOpen
  \bibfield  {author} {\bibinfo {author} {\bibfnamefont {D.~F.}\ \bibnamefont
  {{de Lange}}}, \bibinfo {author} {\bibfnamefont {G.~J.}\ \bibnamefont {{de
  Bruin}}}, \ and\ \bibinfo {author} {\bibfnamefont {L.}~\bibnamefont {{van
  Wijngaarden}}},\ }\bibfield  {title} {\enquote {\bibinfo {title} {On the
  mechanism of cloud cavitation - experiment and modelling.}}\ }in\ \href@noop
  {} {\emph {\bibinfo {booktitle} {The Second International Symposium on
  Cavitation, CAV1994}}}\ (\bibinfo  {publisher} {H. Kato},\ \bibinfo {address}
  {Tokyo, Japan},\ \bibinfo {year} {1994})\ pp.\ \bibinfo {pages}
  {45--49}\BibitemShut {NoStop}%
\bibitem [{\citenamefont {Kawanami}\ \emph {et~al.}(1997)\citenamefont
  {Kawanami}, \citenamefont {Kato}, \citenamefont {Yamaguchi}, \citenamefont
  {Tanimura},\ and\ \citenamefont {Tagaya}}]{Kawanami1997}%
  \BibitemOpen
  \bibfield  {author} {\bibinfo {author} {\bibfnamefont {Y.}~\bibnamefont
  {Kawanami}}, \bibinfo {author} {\bibfnamefont {H.}~\bibnamefont {Kato}},
  \bibinfo {author} {\bibfnamefont {H.}~\bibnamefont {Yamaguchi}}, \bibinfo
  {author} {\bibfnamefont {M.}~\bibnamefont {Tanimura}}, \ and\ \bibinfo
  {author} {\bibfnamefont {Y.}~\bibnamefont {Tagaya}},\ }\bibfield  {title}
  {\enquote {\bibinfo {title} {Mechanism and control of cloud cavitation},}\
  }\href {\doibase 10.1115/1.2819499} {\bibfield  {journal} {\bibinfo
  {journal} {Journal of Fluids Engineering}\ }\textbf {\bibinfo {volume}
  {119}},\ \bibinfo {pages} {788--794} (\bibinfo {year} {1997})}\BibitemShut
  {NoStop}%
\bibitem [{\citenamefont {Reisman}, \citenamefont {Wang},\ and\ \citenamefont
  {Brennen}(1998)}]{Reisman98}%
  \BibitemOpen
  \bibfield  {author} {\bibinfo {author} {\bibfnamefont {G.}~\bibnamefont
  {Reisman}}, \bibinfo {author} {\bibfnamefont {Y.-C.}\ \bibnamefont {Wang}}, \
  and\ \bibinfo {author} {\bibfnamefont {C.}~\bibnamefont {Brennen}},\
  }\bibfield  {title} {\enquote {\bibinfo {title} {Observations of shock waves
  in cloud cavitation},}\ }\href {\doibase 10.1017/S0022112097007830}
  {\bibfield  {journal} {\bibinfo  {journal} {Journal of Fluid Mechanics}\
  }\textbf {\bibinfo {volume} {355}},\ \bibinfo {pages} {255--283} (\bibinfo
  {year} {1998})}\BibitemShut {NoStop}%
\bibitem [{\citenamefont {Gopalan}\ and\ \citenamefont
  {Katz}(2000)}]{Gopalan00}%
  \BibitemOpen
  \bibfield  {author} {\bibinfo {author} {\bibfnamefont {S.}~\bibnamefont
  {Gopalan}}\ and\ \bibinfo {author} {\bibfnamefont {J.}~\bibnamefont {Katz}},\
  }\bibfield  {title} {\enquote {\bibinfo {title} {Flow structure and modeling
  issues in the closure region of attached cavitation},}\ }\href {\doibase
  10.1063/1.870344} {\bibfield  {journal} {\bibinfo  {journal} {Physics of
  Fluids}\ }\textbf {\bibinfo {volume} {12}},\ \bibinfo {pages} {895--911}
  (\bibinfo {year} {2000})}\BibitemShut {NoStop}%
\bibitem [{\citenamefont {Callenaere}\ \emph {et~al.}(2001)\citenamefont
  {Callenaere}, \citenamefont {Franc}, \citenamefont {Michel},\ and\
  \citenamefont {Riondet}}]{Callenaere2001}%
  \BibitemOpen
  \bibfield  {author} {\bibinfo {author} {\bibfnamefont {M.}~\bibnamefont
  {Callenaere}}, \bibinfo {author} {\bibfnamefont {J.-P.}\ \bibnamefont
  {Franc}}, \bibinfo {author} {\bibfnamefont {J.-M.}\ \bibnamefont {Michel}}, \
  and\ \bibinfo {author} {\bibfnamefont {M.}~\bibnamefont {Riondet}},\
  }\bibfield  {title} {\enquote {\bibinfo {title} {The cavitation instability
  induced by the development of a re-entrant jet},}\ }\href {\doibase
  10.1017/s0022112001005420} {\bibfield  {journal} {\bibinfo  {journal}
  {Journal of Fluid Mechanics}\ }\textbf {\bibinfo {volume} {444}},\ \bibinfo
  {pages} {223--256} (\bibinfo {year} {2001})}\BibitemShut {NoStop}%
\bibitem [{\citenamefont {Coutier-Delgosha}, \citenamefont {Devillers},\ and\
  \citenamefont {Pichon}(2006)}]{Coutier-Delgosha2006}%
  \BibitemOpen
  \bibfield  {author} {\bibinfo {author} {\bibfnamefont {O.}~\bibnamefont
  {Coutier-Delgosha}}, \bibinfo {author} {\bibfnamefont {J.-F.}\ \bibnamefont
  {Devillers}}, \ and\ \bibinfo {author} {\bibfnamefont {T.}~\bibnamefont
  {Pichon}},\ }\bibfield  {title} {\enquote {\bibinfo {title} {Internal
  structure and dynamics of sheet cavitation},}\ }\href {\doibase
  10.1063/1.2149882} {\bibfield  {journal} {\bibinfo  {journal} {Physics of
  Fluids}\ }\textbf {\bibinfo {volume} {18}},\ \bibinfo {pages} {017103}
  (\bibinfo {year} {2006})}\BibitemShut {NoStop}%
\bibitem [{\citenamefont {Hayashi}\ and\ \citenamefont
  {Sato}(2014)}]{Hayashi2014}%
  \BibitemOpen
  \bibfield  {author} {\bibinfo {author} {\bibfnamefont {S.}~\bibnamefont
  {Hayashi}}\ and\ \bibinfo {author} {\bibfnamefont {K.}~\bibnamefont {Sato}},\
  }\bibfield  {title} {\enquote {\bibinfo {title} {Unsteady behavior of
  cavitating waterjet in an axisymmetric convergent-divergent nozzle: High
  speed observation and image analysis based on frame difference method},}\
  }\href {\doibase 10.4236/jfcmv.2014.23011} {\bibfield  {journal} {\bibinfo
  {journal} {Journal of Flow Control, Measurement \& Visualization}\ }\textbf
  {\bibinfo {volume} {2}},\ \bibinfo {pages} {94--104} (\bibinfo {year}
  {2014})}\BibitemShut {NoStop}%
\bibitem [{\citenamefont {Kravtsova}\ \emph {et~al.}(2014)\citenamefont
  {Kravtsova}, \citenamefont {Markovich}, \citenamefont {Pervunin},
  \citenamefont {Timoshevskii},\ and\ \citenamefont
  {Hanjalic}}]{Kravtsova2014}%
  \BibitemOpen
  \bibfield  {author} {\bibinfo {author} {\bibfnamefont {A.~Y.}\ \bibnamefont
  {Kravtsova}}, \bibinfo {author} {\bibfnamefont {D.~M.}\ \bibnamefont
  {Markovich}}, \bibinfo {author} {\bibfnamefont {K.~S.}\ \bibnamefont
  {Pervunin}}, \bibinfo {author} {\bibfnamefont {M.~V.}\ \bibnamefont
  {Timoshevskii}}, \ and\ \bibinfo {author} {\bibfnamefont {K.}~\bibnamefont
  {Hanjalic}},\ }\bibfield  {title} {\enquote {\bibinfo {title} {Cavitation on
  a semicircular leading-edge plate and {NACA0015} hydrofoil: Visualization and
  velocity measurement},}\ }\href {\doibase 10.1134/S0040601514140055}
  {\bibfield  {journal} {\bibinfo  {journal} {Thermal Engineering volume}\
  }\textbf {\bibinfo {volume} {61}},\ \bibinfo {pages} {1007--1014} (\bibinfo
  {year} {2014})}\BibitemShut {NoStop}%
\bibitem [{\citenamefont {Jahangir}, \citenamefont {Hogendoorn},\ and\
  \citenamefont {Poelma}(2018)}]{Jahangir2018}%
  \BibitemOpen
  \bibfield  {author} {\bibinfo {author} {\bibfnamefont {S.}~\bibnamefont
  {Jahangir}}, \bibinfo {author} {\bibfnamefont {W.}~\bibnamefont
  {Hogendoorn}}, \ and\ \bibinfo {author} {\bibfnamefont {C.}~\bibnamefont
  {Poelma}},\ }\bibfield  {title} {\enquote {\bibinfo {title} {Dynamics of
  partial cavitation in an axisymmetric converging-diverging nozzle},}\ }\href
  {\doibase 10.1016/j.ijmultiphaseflow.2018.04.019} {\bibfield  {journal}
  {\bibinfo  {journal} {International Journal of Multiphase Flow}\ }\textbf
  {\bibinfo {volume} {106}},\ \bibinfo {pages} {34--45} (\bibinfo {year}
  {2018})}\BibitemShut {NoStop}%
\bibitem [{\citenamefont {Arndt}\ \emph {et~al.}(2000)\citenamefont {Arndt},
  \citenamefont {Song}, \citenamefont {He},\ and\ \citenamefont
  {Keller}}]{Arndt2000}%
  \BibitemOpen
  \bibfield  {author} {\bibinfo {author} {\bibfnamefont {R.~A.~E.}\
  \bibnamefont {Arndt}}, \bibinfo {author} {\bibfnamefont {C.~C.~S.}\
  \bibnamefont {Song}}, \bibinfo {author} {\bibfnamefont {M.~K.~J.}\
  \bibnamefont {He}}, \ and\ \bibinfo {author} {\bibfnamefont {A.}~\bibnamefont
  {Keller}},\ }\bibfield  {title} {\enquote {\bibinfo {title} {Instability of
  partial cavitation: A numerical/experimental approach},}\ }in\ \href@noop {}
  {\emph {\bibinfo {booktitle} {Twenty-Third Symposium on Naval
  Hydrodynamics}}}\ (\bibinfo  {publisher} {National Academies Press},\
  \bibinfo {year} {2000})\BibitemShut {NoStop}%
\bibitem [{\citenamefont {Stanley}\ \emph {et~al.}(2011)\citenamefont
  {Stanley}, \citenamefont {Barber}, \citenamefont {Milton},\ and\
  \citenamefont {Rosengarten}}]{Stanley2011}%
  \BibitemOpen
  \bibfield  {author} {\bibinfo {author} {\bibfnamefont {C.}~\bibnamefont
  {Stanley}}, \bibinfo {author} {\bibfnamefont {T.}~\bibnamefont {Barber}},
  \bibinfo {author} {\bibfnamefont {B.}~\bibnamefont {Milton}}, \ and\ \bibinfo
  {author} {\bibfnamefont {G.}~\bibnamefont {Rosengarten}},\ }\bibfield
  {title} {\enquote {\bibinfo {title} {Periodic cavitation shedding in a
  cylindrical orifice},}\ }\href {\doibase 10.1007/s00348-011-1138-7}
  {\bibfield  {journal} {\bibinfo  {journal} {Experiments in Fluids}\ }\textbf
  {\bibinfo {volume} {51}},\ \bibinfo {pages} {1189--1200} (\bibinfo {year}
  {2011})}\BibitemShut {NoStop}%
\bibitem [{\citenamefont {Stanley}, \citenamefont {Barber},\ and\ \citenamefont
  {Rosengarten}(2014)}]{Stanley2014}%
  \BibitemOpen
  \bibfield  {author} {\bibinfo {author} {\bibfnamefont {C.}~\bibnamefont
  {Stanley}}, \bibinfo {author} {\bibfnamefont {T.}~\bibnamefont {Barber}}, \
  and\ \bibinfo {author} {\bibfnamefont {G.}~\bibnamefont {Rosengarten}},\
  }\bibfield  {title} {\enquote {\bibinfo {title} {Re-entrant jet mechanism for
  periodic cavitation shedding in a cylindrical orifice},}\ }\href {\doibase
  10.1016/j.ijheatfluidflow.2014.07.004} {\bibfield  {journal} {\bibinfo
  {journal} {International Journal of Heat and Fluid Flow}\ }\textbf {\bibinfo
  {volume} {50}},\ \bibinfo {pages} {169--176} (\bibinfo {year}
  {2014})}\BibitemShut {NoStop}%
\bibitem [{\citenamefont {Ganesh}, \citenamefont {Mäkiharju},\ and\
  \citenamefont {Ceccio}(2016)}]{Ganesh2016}%
  \BibitemOpen
  \bibfield  {author} {\bibinfo {author} {\bibfnamefont {H.}~\bibnamefont
  {Ganesh}}, \bibinfo {author} {\bibfnamefont {S.~A.}\ \bibnamefont
  {Mäkiharju}}, \ and\ \bibinfo {author} {\bibfnamefont {S.~L.}\ \bibnamefont
  {Ceccio}},\ }\bibfield  {title} {\enquote {\bibinfo {title} {Bubbly shock
  propagation as a mechanism for sheet-to-cloud transition of partial
  cavities},}\ }\href {\doibase 10.1017/jfm.2016.425} {\bibfield  {journal}
  {\bibinfo  {journal} {Journal of Fluid Mechanics}\ }\textbf {\bibinfo
  {volume} {802}},\ \bibinfo {pages} {37--78} (\bibinfo {year}
  {2016})}\BibitemShut {NoStop}%
\bibitem [{\citenamefont {Charrière}\ and\ \citenamefont
  {Goncalvès}(2017)}]{Charriere2017}%
  \BibitemOpen
  \bibfield  {author} {\bibinfo {author} {\bibfnamefont {B.}~\bibnamefont
  {Charrière}}\ and\ \bibinfo {author} {\bibfnamefont {E.}~\bibnamefont
  {Goncalvès}},\ }\bibfield  {title} {\enquote {\bibinfo {title} {Numerical
  investigation of periodic cavitation shedding in a {V}enturi},}\ }\href
  {\doibase 10.1016/j.ijheatfluidflow.2017.01.011} {\bibfield  {journal}
  {\bibinfo  {journal} {International Journal of Heat and Fluid Flow}\ }\textbf
  {\bibinfo {volume} {64}},\ \bibinfo {pages} {41--54} (\bibinfo {year}
  {2017})}\BibitemShut {NoStop}%
\bibitem [{\citenamefont {{de Lange}}\ and\ \citenamefont {{de
  Bruin}}(1997)}]{Lange1997}%
  \BibitemOpen
  \bibfield  {author} {\bibinfo {author} {\bibfnamefont {D.}~\bibnamefont {{de
  Lange}}}\ and\ \bibinfo {author} {\bibfnamefont {G.}~\bibnamefont {{de
  Bruin}}},\ }\bibfield  {title} {\enquote {\bibinfo {title} {Sheet cavitation
  and cloud cavitation},}\ }\href {\doibase 10.1023/A:1000763130780} {\bibfield
   {journal} {\bibinfo  {journal} {Flow, Turbulence and Combustion}\ }\textbf
  {\bibinfo {volume} {58}},\ \bibinfo {pages} {91--114} (\bibinfo {year}
  {1997})}\BibitemShut {NoStop}%
\bibitem [{\citenamefont {Foeth}\ \emph {et~al.}(2006)\citenamefont {Foeth},
  \citenamefont {van Doorne}, \citenamefont {van Terwisga},\ and\ \citenamefont
  {Wieneke}}]{Foeth2006}%
  \BibitemOpen
  \bibfield  {author} {\bibinfo {author} {\bibfnamefont {E.~J.}\ \bibnamefont
  {Foeth}}, \bibinfo {author} {\bibfnamefont {C.~W.~H.}\ \bibnamefont {van
  Doorne}}, \bibinfo {author} {\bibfnamefont {T.}~\bibnamefont {van Terwisga}},
  \ and\ \bibinfo {author} {\bibfnamefont {B.}~\bibnamefont {Wieneke}},\
  }\bibfield  {title} {\enquote {\bibinfo {title} {Time resolved {PIV} and flow
  visualization of {3D} sheet cavitation},}\ }\href {\doibase
  10.1007/s00348-005-0082-9} {\bibfield  {journal} {\bibinfo  {journal}
  {Experiments in Fluids Volume}\ }\textbf {\bibinfo {volume} {40}},\ \bibinfo
  {pages} {503--513} (\bibinfo {year} {2006})}\BibitemShut {NoStop}%
\bibitem [{\citenamefont {Dular}\ \emph {et~al.}(2007)\citenamefont {Dular},
  \citenamefont {Bachert}, \citenamefont {Schaad},\ and\ \citenamefont
  {Stoffel}}]{Dular2007}%
  \BibitemOpen
  \bibfield  {author} {\bibinfo {author} {\bibfnamefont {M.}~\bibnamefont
  {Dular}}, \bibinfo {author} {\bibfnamefont {R.}~\bibnamefont {Bachert}},
  \bibinfo {author} {\bibfnamefont {C.}~\bibnamefont {Schaad}}, \ and\ \bibinfo
  {author} {\bibfnamefont {B.}~\bibnamefont {Stoffel}},\ }\bibfield  {title}
  {\enquote {\bibinfo {title} {Investigation of a re-entrant jet reflection at
  an inclined cavity closure line},}\ }\href {\doibase
  10.1016/j.euromechflu.2007.01.001} {\bibfield  {journal} {\bibinfo  {journal}
  {European Journal of Mechanics - B/Fluids}\ }\textbf {\bibinfo {volume}
  {26}},\ \bibinfo {pages} {688--705} (\bibinfo {year} {2007})}\BibitemShut
  {NoStop}%
\bibitem [{\citenamefont {Schnerr}, \citenamefont {Sezal},\ and\ \citenamefont
  {Schmid}(2008)}]{Schnerr2008}%
  \BibitemOpen
  \bibfield  {author} {\bibinfo {author} {\bibfnamefont {G.~H.}\ \bibnamefont
  {Schnerr}}, \bibinfo {author} {\bibfnamefont {I.~H.}\ \bibnamefont {Sezal}},
  \ and\ \bibinfo {author} {\bibfnamefont {S.~J.}\ \bibnamefont {Schmid}},\
  }\bibfield  {title} {\enquote {\bibinfo {title} {Numerical investigation of
  three-dimensional cloud cavitationwith special emphasis on collapse induced
  shock dynamics},}\ }\href {\doibase 10.1063/1.2911039} {\bibfield  {journal}
  {\bibinfo  {journal} {Physics of Fluids}\ }\textbf {\bibinfo {volume} {20}},\
  \bibinfo {pages} {040703} (\bibinfo {year} {2008})}\BibitemShut {NoStop}%
\bibitem [{\citenamefont {Kubota}\ \emph {et~al.}(1989)\citenamefont {Kubota},
  \citenamefont {Kato}, \citenamefont {Yamaguchi},\ and\ \citenamefont
  {Maeda}}]{Kubota1989}%
  \BibitemOpen
  \bibfield  {author} {\bibinfo {author} {\bibfnamefont {A.}~\bibnamefont
  {Kubota}}, \bibinfo {author} {\bibfnamefont {H.}~\bibnamefont {Kato}},
  \bibinfo {author} {\bibfnamefont {H.}~\bibnamefont {Yamaguchi}}, \ and\
  \bibinfo {author} {\bibfnamefont {M.}~\bibnamefont {Maeda}},\ }\bibfield
  {title} {\enquote {\bibinfo {title} {Unsteady structure measurement of cloud
  cavitation on a foil section using conditional sampling technique},}\ }\href
  {\doibase 10.1115/1.3243624} {\bibfield  {journal} {\bibinfo  {journal}
  {Journal of Fluids Engineering}\ }\textbf {\bibinfo {volume} {111}},\
  \bibinfo {pages} {204--210} (\bibinfo {year} {1989})}\BibitemShut {NoStop}%
\bibitem [{\citenamefont {Peng}\ \emph {et~al.}(2016)\citenamefont {Peng},
  \citenamefont {Ji}, \citenamefont {Cao}, \citenamefont {Xu}, \citenamefont
  {Zhang}, \citenamefont {Luo},\ and\ \citenamefont {Long}}]{Peng2016}%
  \BibitemOpen
  \bibfield  {author} {\bibinfo {author} {\bibfnamefont {X.}~\bibnamefont
  {Peng}}, \bibinfo {author} {\bibfnamefont {B.}~\bibnamefont {Ji}}, \bibinfo
  {author} {\bibfnamefont {Y.}~\bibnamefont {Cao}}, \bibinfo {author}
  {\bibfnamefont {L.}~\bibnamefont {Xu}}, \bibinfo {author} {\bibfnamefont
  {G.}~\bibnamefont {Zhang}}, \bibinfo {author} {\bibfnamefont
  {X.}~\bibnamefont {Luo}}, \ and\ \bibinfo {author} {\bibfnamefont
  {X.}~\bibnamefont {Long}},\ }\bibfield  {title} {\enquote {\bibinfo {title}
  {Combined experimental observation and numerical simulation of the cloud
  cavitation with {U}-type flow structures on hydrofoils},}\ }\href {\doibase
  10.1016/j.ijmultiphaseflow.2015.10.006} {\bibfield  {journal} {\bibinfo
  {journal} {International Journal of Multiphase Flow}\ }\textbf {\bibinfo
  {volume} {79}},\ \bibinfo {pages} {10--22} (\bibinfo {year}
  {2016})}\BibitemShut {NoStop}%
\bibitem [{\citenamefont {Kadivar}, \citenamefont {el~Moctar},\ and\
  \citenamefont {Javadi}(2019)}]{Kadivar2019}%
  \BibitemOpen
  \bibfield  {author} {\bibinfo {author} {\bibfnamefont {E.}~\bibnamefont
  {Kadivar}}, \bibinfo {author} {\bibfnamefont {O.}~\bibnamefont {el~Moctar}},
  \ and\ \bibinfo {author} {\bibfnamefont {K.}~\bibnamefont {Javadi}},\
  }\bibfield  {title} {\enquote {\bibinfo {title} {Stabilization of cloud
  cavitation instabilities using cylindrical cavitating-bubble generators
  {(CCGs)}},}\ }\href {\doibase 10.1016/j.ijmultiphaseflow.2019.03.019}
  {\bibfield  {journal} {\bibinfo  {journal} {International Journal of
  Multiphase Flow}\ }\textbf {\bibinfo {volume} {115}},\ \bibinfo {pages}
  {108--125} (\bibinfo {year} {2019})}\BibitemShut {NoStop}%
\bibitem [{\citenamefont {Che}\ \emph {et~al.}(2019)\citenamefont {Che},
  \citenamefont {Cao}, \citenamefont {Chu}, \citenamefont {Likhachev},\ and\
  \citenamefont {Wu}}]{Che2019}%
  \BibitemOpen
  \bibfield  {author} {\bibinfo {author} {\bibfnamefont {B.}~\bibnamefont
  {Che}}, \bibinfo {author} {\bibfnamefont {L.}~\bibnamefont {Cao}}, \bibinfo
  {author} {\bibfnamefont {N.}~\bibnamefont {Chu}}, \bibinfo {author}
  {\bibfnamefont {D.}~\bibnamefont {Likhachev}}, \ and\ \bibinfo {author}
  {\bibfnamefont {D.}~\bibnamefont {Wu}},\ }\bibfield  {title} {\enquote
  {\bibinfo {title} {Dynamic behaviors of re-entrant jet and cavity shedding
  during transitional cavity oscillation on {NACA0015} hydrofoil},}\ }\href
  {\doibase 10.1115/1.4041716} {\bibfield  {journal} {\bibinfo  {journal}
  {Journal of Fluids Engineering}\ }\textbf {\bibinfo {volume} {141}},\
  \bibinfo {pages} {061101} (\bibinfo {year} {2019})}\BibitemShut {NoStop}%
\bibitem [{\citenamefont {Long}\ \emph {et~al.}(2018)\citenamefont {Long},
  \citenamefont {Cheng}, \citenamefont {BinJi}, \citenamefont {Arndt},\ and\
  \citenamefont {Peng}}]{Long2018}%
  \BibitemOpen
  \bibfield  {author} {\bibinfo {author} {\bibfnamefont {X.}~\bibnamefont
  {Long}}, \bibinfo {author} {\bibfnamefont {H.}~\bibnamefont {Cheng}},
  \bibinfo {author} {\bibnamefont {BinJi}}, \bibinfo {author} {\bibfnamefont
  {R.~E.}\ \bibnamefont {Arndt}}, \ and\ \bibinfo {author} {\bibfnamefont
  {X.}~\bibnamefont {Peng}},\ }\bibfield  {title} {\enquote {\bibinfo {title}
  {Large eddy simulation and {E}uler-{L}agrangian coupling investigation of the
  transient cavitating turbulent flow around a twisted hydrofoil},}\ }\href
  {\doibase 10.1016/j.ijmultiphaseflow.2017.12.002} {\bibfield  {journal}
  {\bibinfo  {journal} {International Journal of Multiphase Flow}\ }\textbf
  {\bibinfo {volume} {100}},\ \bibinfo {pages} {41--56} (\bibinfo {year}
  {2018})}\BibitemShut {NoStop}%
\bibitem [{\citenamefont {Decaix}\ and\ \citenamefont
  {Goncalves}(2013)}]{Decaix2013}%
  \BibitemOpen
  \bibfield  {author} {\bibinfo {author} {\bibfnamefont {J.}~\bibnamefont
  {Decaix}}\ and\ \bibinfo {author} {\bibfnamefont {E.}~\bibnamefont
  {Goncalves}},\ }\bibfield  {title} {\enquote {\bibinfo {title} {Investigation
  of three-dimensional effects on a cavitating {V}enturi flow},}\ }\href
  {\doibase 10.1016/j.ijheatfluidflow.2013.08.013} {\bibfield  {journal}
  {\bibinfo  {journal} {International Journal of Heat and Fluid Flow}\ }\textbf
  {\bibinfo {volume} {44}},\ \bibinfo {pages} {576--595} (\bibinfo {year}
  {2013})}\BibitemShut {NoStop}%
\bibitem [{\citenamefont {Timoshevskiy}\ \emph {et~al.}(2016)\citenamefont
  {Timoshevskiy}, \citenamefont {Churkin}, \citenamefont {Kravtsova},
  \citenamefont {Pervunin}, \citenamefont {Markovich},\ and\ \citenamefont
  {Hanjalic}}]{Timoshevskiy2016}%
  \BibitemOpen
  \bibfield  {author} {\bibinfo {author} {\bibfnamefont {M.~V.}\ \bibnamefont
  {Timoshevskiy}}, \bibinfo {author} {\bibfnamefont {S.~A.}\ \bibnamefont
  {Churkin}}, \bibinfo {author} {\bibfnamefont {A.~Y.}\ \bibnamefont
  {Kravtsova}}, \bibinfo {author} {\bibfnamefont {K.~S.}\ \bibnamefont
  {Pervunin}}, \bibinfo {author} {\bibfnamefont {D.~M.}\ \bibnamefont
  {Markovich}}, \ and\ \bibinfo {author} {\bibfnamefont {K.}~\bibnamefont
  {Hanjalic}},\ }\bibfield  {title} {\enquote {\bibinfo {title} {Cavitating
  flow around a scaled-down model of guide vanes of a high-pressure turbine},}\
  }\href {\doibase 10.1016/j.ijmultiphaseflow.2015.09.014} {\bibfield
  {journal} {\bibinfo  {journal} {International Journal of Multiphase Flow}\
  }\textbf {\bibinfo {volume} {78}},\ \bibinfo {pages} {75--87} (\bibinfo
  {year} {2016})}\BibitemShut {NoStop}%
\bibitem [{\citenamefont {Kawakami}\ \emph {et~al.}(2008)\citenamefont
  {Kawakami}, \citenamefont {Fuji}, \citenamefont {Tsujimoto},\ and\
  \citenamefont {Arndt}}]{Kawakami2008}%
  \BibitemOpen
  \bibfield  {author} {\bibinfo {author} {\bibfnamefont {D.~T.}\ \bibnamefont
  {Kawakami}}, \bibinfo {author} {\bibfnamefont {A.}~\bibnamefont {Fuji}},
  \bibinfo {author} {\bibfnamefont {Y.}~\bibnamefont {Tsujimoto}}, \ and\
  \bibinfo {author} {\bibfnamefont {R.~E.~A.}\ \bibnamefont {Arndt}},\
  }\bibfield  {title} {\enquote {\bibinfo {title} {An assessment of the
  influence of environmental factors on cavitation instabilities},}\ }\href
  {\doibase 10.1115/1.2842146} {\bibfield  {journal} {\bibinfo  {journal}
  {Journal of Fluids Engineering}\ }\textbf {\bibinfo {volume} {130}},\
  \bibinfo {pages} {031303} (\bibinfo {year} {2008})}\BibitemShut {NoStop}%
\bibitem [{\citenamefont {Dular}\ \emph {et~al.}(2012)\citenamefont {Dular},
  \citenamefont {Khlifa}, \citenamefont {Fuzier}, \citenamefont {Maiga},\ and\
  \citenamefont {Coutier-Delgosha}}]{Dular2012}%
  \BibitemOpen
  \bibfield  {author} {\bibinfo {author} {\bibfnamefont {M.}~\bibnamefont
  {Dular}}, \bibinfo {author} {\bibfnamefont {I.}~\bibnamefont {Khlifa}},
  \bibinfo {author} {\bibfnamefont {S.}~\bibnamefont {Fuzier}}, \bibinfo
  {author} {\bibfnamefont {M.~A.}\ \bibnamefont {Maiga}}, \ and\ \bibinfo
  {author} {\bibfnamefont {O.}~\bibnamefont {Coutier-Delgosha}},\ }\bibfield
  {title} {\enquote {\bibinfo {title} {Scale effect on unsteady cloud
  cavitation},}\ }\href {\doibase 10.1007/s00348-012-1356-7} {\bibfield
  {journal} {\bibinfo  {journal} {Experiments in Fluids}\ }\textbf {\bibinfo
  {volume} {53}},\ \bibinfo {pages} {1233--1250} (\bibinfo {year}
  {2012})}\BibitemShut {NoStop}%
\bibitem [{\citenamefont {Delannoy}\ and\ \citenamefont
  {Kueny}(1990)}]{Delannoy90}%
  \BibitemOpen
  \bibfield  {author} {\bibinfo {author} {\bibfnamefont {Y.}~\bibnamefont
  {Delannoy}}\ and\ \bibinfo {author} {\bibfnamefont {J.}~\bibnamefont
  {Kueny}},\ }\bibfield  {title} {\enquote {\bibinfo {title} {Two phase flow
  approach in unsteady cavitation modelling},}\ }in\ \href@noop {} {\emph
  {\bibinfo {booktitle} {Cavitation and Multiphase Flow Forum, American Society
  of Mechanical Engineers (ASME), Spring Meeting}}},\ Vol.~\bibinfo {volume}
  {98}\ (\bibinfo {address} {Toronto, Canada},\ \bibinfo {year} {1990})\ pp.\
  \bibinfo {pages} {153--158}\BibitemShut {NoStop}%
\bibitem [{\citenamefont {Clerc}(2000)}]{Clerc00}%
  \BibitemOpen
  \bibfield  {author} {\bibinfo {author} {\bibfnamefont {S.}~\bibnamefont
  {Clerc}},\ }\bibfield  {title} {\enquote {\bibinfo {title} {Numerical
  simulation of the homogeneous equilibrium model for two-phase flows},}\
  }\href {\doibase 10.1006/jcph.2000.6515} {\bibfield  {journal} {\bibinfo
  {journal} {Journal of Computational Physics}\ }\textbf {\bibinfo {volume}
  {161}},\ \bibinfo {pages} {354--375} (\bibinfo {year} {2000})}\BibitemShut
  {NoStop}%
\bibitem [{\citenamefont {Sinibaldi}, \citenamefont {Beux},\ and\ \citenamefont
  {Salvetti}(2006)}]{Sinibaldi06}%
  \BibitemOpen
  \bibfield  {author} {\bibinfo {author} {\bibfnamefont {E.}~\bibnamefont
  {Sinibaldi}}, \bibinfo {author} {\bibfnamefont {F.}~\bibnamefont {Beux}}, \
  and\ \bibinfo {author} {\bibfnamefont {M.}~\bibnamefont {Salvetti}},\
  }\bibfield  {title} {\enquote {\bibinfo {title} {A numerical method for 3{D}
  barotropic flows in turbomachinery},}\ }\href {\doibase
  10.1007/s10494-006-9025-7} {\bibfield  {journal} {\bibinfo  {journal} {Flow
  Turbulence Combustion}\ }\textbf {\bibinfo {volume} {76}},\ \bibinfo {pages}
  {371--381} (\bibinfo {year} {2006})}\BibitemShut {NoStop}%
\bibitem [{\citenamefont {Downar-Zapolski}\ \emph {et~al.}(1996)\citenamefont
  {Downar-Zapolski}, \citenamefont {Bilicki}, \citenamefont {Bolle},\ and\
  \citenamefont {Franco}}]{Downar-Zapolski1996}%
  \BibitemOpen
  \bibfield  {author} {\bibinfo {author} {\bibfnamefont {P.}~\bibnamefont
  {Downar-Zapolski}}, \bibinfo {author} {\bibfnamefont {Z.}~\bibnamefont
  {Bilicki}}, \bibinfo {author} {\bibfnamefont {L.}~\bibnamefont {Bolle}}, \
  and\ \bibinfo {author} {\bibfnamefont {J.}~\bibnamefont {Franco}},\
  }\bibfield  {title} {\enquote {\bibinfo {title} {The non-equilibrium
  relaxation model for one-dimensional flashing liquid flow},}\ }\href
  {\doibase 10.1016/0301-9322(95)00078-X} {\bibfield  {journal} {\bibinfo
  {journal} {International Journal of Multiphase Flow}\ }\textbf {\bibinfo
  {volume} {22}},\ \bibinfo {pages} {473--483} (\bibinfo {year}
  {1996})}\BibitemShut {NoStop}%
\bibitem [{\citenamefont {Kunz}\ \emph {et~al.}(2000)\citenamefont {Kunz},
  \citenamefont {Boger}, \citenamefont {Stinebring}, \citenamefont
  {Chyczewski}, \citenamefont {Lindau}, \citenamefont {Gibeling}, \citenamefont
  {Venkateswaran},\ and\ \citenamefont {Govindan}}]{Kunz2000}%
  \BibitemOpen
  \bibfield  {author} {\bibinfo {author} {\bibfnamefont {R.~F.}\ \bibnamefont
  {Kunz}}, \bibinfo {author} {\bibfnamefont {D.~A.}\ \bibnamefont {Boger}},
  \bibinfo {author} {\bibfnamefont {D.~R.}\ \bibnamefont {Stinebring}},
  \bibinfo {author} {\bibfnamefont {T.~S.}\ \bibnamefont {Chyczewski}},
  \bibinfo {author} {\bibfnamefont {J.~W.}\ \bibnamefont {Lindau}}, \bibinfo
  {author} {\bibfnamefont {H.~J.}\ \bibnamefont {Gibeling}}, \bibinfo {author}
  {\bibfnamefont {S.}~\bibnamefont {Venkateswaran}}, \ and\ \bibinfo {author}
  {\bibfnamefont {T.}~\bibnamefont {Govindan}},\ }\bibfield  {title} {\enquote
  {\bibinfo {title} {A preconditioned {Navier-Stokes} method for two-phase
  flows with application to cavitation prediction},}\ }\href {\doibase
  10.1016/S0045-7930(99)00039-0} {\bibfield  {journal} {\bibinfo  {journal}
  {Computers \& Fluids}\ }\textbf {\bibinfo {volume} {29}},\ \bibinfo {pages}
  {849--875} (\bibinfo {year} {2000})}\BibitemShut {NoStop}%
\bibitem [{\citenamefont {Zwart}, \citenamefont {Gerber},\ and\ \citenamefont
  {Belamri}(2004)}]{Zwart2004}%
  \BibitemOpen
  \bibfield  {author} {\bibinfo {author} {\bibfnamefont {P.~J.}\ \bibnamefont
  {Zwart}}, \bibinfo {author} {\bibfnamefont {A.~G.}\ \bibnamefont {Gerber}}, \
  and\ \bibinfo {author} {\bibfnamefont {T.}~\bibnamefont {Belamri}},\
  }\bibfield  {title} {\enquote {\bibinfo {title} {A two-phase flow model for
  predicting cavitation dynamics},}\ }in\ \href@noop {} {\emph {\bibinfo
  {booktitle} {ICMF 2004 International Conference on Multiphase Flow}}},\
  \bibinfo {series and number} {\bibinfo {number} {Paper {No.} 12}}\ (\bibinfo
  {address} {Yokohama, Japan},\ \bibinfo {year} {2004})\BibitemShut {NoStop}%
\bibitem [{\citenamefont {Helluy}\ and\ \citenamefont
  {Seguin}(2006)}]{Helluy06}%
  \BibitemOpen
  \bibfield  {author} {\bibinfo {author} {\bibfnamefont {P.}~\bibnamefont
  {Helluy}}\ and\ \bibinfo {author} {\bibfnamefont {N.}~\bibnamefont
  {Seguin}},\ }\bibfield  {title} {\enquote {\bibinfo {title} {Relaxation
  models of phase transition flows},}\ }\href {\doibase 10.1051/m2an:2006015}
  {\bibfield  {journal} {\bibinfo  {journal} {Mathematical Modelling and
  Numerical Analysis}\ }\textbf {\bibinfo {volume} {40}},\ \bibinfo {pages}
  {331--352} (\bibinfo {year} {2006})}\BibitemShut {NoStop}%
\bibitem [{\citenamefont {Saito}, \citenamefont {Takami},\ and\ \citenamefont
  {znd Toshiaki~Ikohagi}(2007)}]{Saito2007}%
  \BibitemOpen
  \bibfield  {author} {\bibinfo {author} {\bibfnamefont {Y.}~\bibnamefont
  {Saito}}, \bibinfo {author} {\bibfnamefont {R.}~\bibnamefont {Takami}}, \
  and\ \bibinfo {author} {\bibfnamefont {I.~N.}\ \bibnamefont {znd
  Toshiaki~Ikohagi}},\ }\bibfield  {title} {\enquote {\bibinfo {title}
  {Numerical analysis of unsteady behavior of cloud cavitation around a
  {NACA0015} foil},}\ }\href {\doibase 10.1007/s00466-006-0086-1} {\bibfield
  {journal} {\bibinfo  {journal} {Computational Mechanics Volume}\ }\textbf
  {\bibinfo {volume} {40}} (\bibinfo {year} {2007}),\
  10.1007/s00466-006-0086-1}\BibitemShut {NoStop}%
\bibitem [{\citenamefont {Park}\ and\ \citenamefont {Rhee}(2013)}]{Park2013}%
  \BibitemOpen
  \bibfield  {author} {\bibinfo {author} {\bibfnamefont {S.}~\bibnamefont
  {Park}}\ and\ \bibinfo {author} {\bibfnamefont {S.~H.}\ \bibnamefont
  {Rhee}},\ }\bibfield  {title} {\enquote {\bibinfo {title} {Numerical analysis
  of the three-dimensional cloud cavitating flow around a twisted hydrofoil},}\
  }\href {\doibase 10.1088/0169-5983/45/1/015502} {\bibfield  {journal}
  {\bibinfo  {journal} {Fluid Dynamics Research}\ }\textbf {\bibinfo {volume}
  {45}},\ \bibinfo {pages} {015502} (\bibinfo {year} {2013})}\BibitemShut
  {NoStop}%
\bibitem [{\citenamefont {Sedlar}\ \emph {et~al.}(2016)\citenamefont {Sedlar},
  \citenamefont {Ji}, \citenamefont {Kratky}, \citenamefont {Rebok},\ and\
  \citenamefont {Huzlik}}]{Sedlar2016}%
  \BibitemOpen
  \bibfield  {author} {\bibinfo {author} {\bibfnamefont {M.}~\bibnamefont
  {Sedlar}}, \bibinfo {author} {\bibfnamefont {B.}~\bibnamefont {Ji}}, \bibinfo
  {author} {\bibfnamefont {T.}~\bibnamefont {Kratky}}, \bibinfo {author}
  {\bibfnamefont {T.}~\bibnamefont {Rebok}}, \ and\ \bibinfo {author}
  {\bibfnamefont {R.}~\bibnamefont {Huzlik}},\ }\bibfield  {title} {\enquote
  {\bibinfo {title} {Numerical and experimental investigation of
  three-dimensional cavitating flow around the straight {NACA2412}
  hydrofoil},}\ }\href {\doibase 10.1016/j.oceaneng.2016.07.030} {\bibfield
  {journal} {\bibinfo  {journal} {Ocean Engineering}\ }\textbf {\bibinfo
  {volume} {123}},\ \bibinfo {pages} {357--382} (\bibinfo {year}
  {2016})}\BibitemShut {NoStop}%
\bibitem [{\citenamefont {Hidalgo}\ \emph {et~al.}(2019)\citenamefont
  {Hidalgo}, \citenamefont {Escaler}, \citenamefont {Valencia}, \citenamefont
  {Peng}, \citenamefont {Erazo}, \citenamefont {Puga},\ and\ \citenamefont
  {Luo}}]{Hidalgo2019}%
  \BibitemOpen
  \bibfield  {author} {\bibinfo {author} {\bibfnamefont {V.}~\bibnamefont
  {Hidalgo}}, \bibinfo {author} {\bibfnamefont {X.}~\bibnamefont {Escaler}},
  \bibinfo {author} {\bibfnamefont {E.}~\bibnamefont {Valencia}}, \bibinfo
  {author} {\bibfnamefont {X.}~\bibnamefont {Peng}}, \bibinfo {author}
  {\bibfnamefont {J.}~\bibnamefont {Erazo}}, \bibinfo {author} {\bibfnamefont
  {D.}~\bibnamefont {Puga}}, \ and\ \bibinfo {author} {\bibfnamefont
  {X.}~\bibnamefont {Luo}},\ }\bibfield  {title} {\enquote {\bibinfo {title}
  {Scale-adaptive simulation of unsteady cavitation around a {NACA66}
  hydrofoil},}\ }\href {\doibase 10.3390/app9183696} {\bibfield  {journal}
  {\bibinfo  {journal} {Applied Science}\ }\textbf {\bibinfo {volume} {9}},\
  \bibinfo {pages} {3696} (\bibinfo {year} {2019})}\BibitemShut {NoStop}%
\bibitem [{\citenamefont {Sun}\ \emph {et~al.}(2019)\citenamefont {Sun},
  \citenamefont {Zhang}, \citenamefont {Xu}, \citenamefont {Zhang},\ and\
  \citenamefont {Jiang}}]{Sun2019}%
  \BibitemOpen
  \bibfield  {author} {\bibinfo {author} {\bibfnamefont {T.}~\bibnamefont
  {Sun}}, \bibinfo {author} {\bibfnamefont {X.}~\bibnamefont {Zhang}}, \bibinfo
  {author} {\bibfnamefont {C.}~\bibnamefont {Xu}}, \bibinfo {author}
  {\bibfnamefont {G.}~\bibnamefont {Zhang}}, \ and\ \bibinfo {author}
  {\bibfnamefont {S.}~\bibnamefont {Jiang}},\ }\bibfield  {title} {\enquote
  {\bibinfo {title} {Numerical modeling and simulation of the shedding
  mechanism and vortex structures at the development stage of ventilated
  partial cavitating flows},}\ }\href {\doibase
  10.1016/j.euromechflu.2019.02.011} {\bibfield  {journal} {\bibinfo  {journal}
  {European Journal of Mechanics/B Fluids}\ }\textbf {\bibinfo {volume} {76}},\
  \bibinfo {pages} {223--232} (\bibinfo {year} {2019})}\BibitemShut {NoStop}%
\bibitem [{\citenamefont {Ji}\ \emph {et~al.}(2013{\natexlab{a}})\citenamefont
  {Ji}, \citenamefont {Luo}, \citenamefont {Wu}, \citenamefont {Peng},\ and\
  \citenamefont {Duan}}]{Ji2013}%
  \BibitemOpen
  \bibfield  {author} {\bibinfo {author} {\bibfnamefont {B.}~\bibnamefont
  {Ji}}, \bibinfo {author} {\bibfnamefont {X.}~\bibnamefont {Luo}}, \bibinfo
  {author} {\bibfnamefont {Y.}~\bibnamefont {Wu}}, \bibinfo {author}
  {\bibfnamefont {X.}~\bibnamefont {Peng}}, \ and\ \bibinfo {author}
  {\bibfnamefont {Y.}~\bibnamefont {Duan}},\ }\bibfield  {title} {\enquote
  {\bibinfo {title} {Numerical analysis of unsteady cavitating turbulent flow
  and shedding horse-shoe vortex structure around a twisted hydrofoil},}\
  }\href {\doibase 10.1016/j.ijmultiphaseflow.2012.11.008} {\bibfield
  {journal} {\bibinfo  {journal} {International Journal of Multiphase Flow}\
  }\textbf {\bibinfo {volume} {51}},\ \bibinfo {pages} {33--43} (\bibinfo
  {year} {2013}{\natexlab{a}})}\BibitemShut {NoStop}%
\bibitem [{\citenamefont {Huang}, \citenamefont {Zhao},\ and\ \citenamefont
  {Wang}(2014)}]{Huang2014}%
  \BibitemOpen
  \bibfield  {author} {\bibinfo {author} {\bibfnamefont {B.}~\bibnamefont
  {Huang}}, \bibinfo {author} {\bibfnamefont {Y.}~\bibnamefont {Zhao}}, \ and\
  \bibinfo {author} {\bibfnamefont {G.}~\bibnamefont {Wang}},\ }\bibfield
  {title} {\enquote {\bibinfo {title} {Large eddy simulation of turbulent
  vortex-cavitation interactionsin transient sheet/cloud cavitating flows},}\
  }\href {\doibase 10.1016/j.compfluid.2013.12.024} {\bibfield  {journal}
  {\bibinfo  {journal} {Computers \& Fluids}\ }\textbf {\bibinfo {volume}
  {92}},\ \bibinfo {pages} {113--124} (\bibinfo {year} {2014})}\BibitemShut
  {NoStop}%
\bibitem [{\citenamefont {Ji}\ \emph {et~al.}(2013{\natexlab{b}})\citenamefont
  {Ji}, \citenamefont {wu~Luo}, \citenamefont {xing Peng},\ and\ \citenamefont
  {lin Wu}}]{Ji2013a}%
  \BibitemOpen
  \bibfield  {author} {\bibinfo {author} {\bibfnamefont {B.}~\bibnamefont
  {Ji}}, \bibinfo {author} {\bibfnamefont {X.}~\bibnamefont {wu~Luo}}, \bibinfo
  {author} {\bibfnamefont {X.}~\bibnamefont {xing Peng}}, \ and\ \bibinfo
  {author} {\bibfnamefont {Y.}~\bibnamefont {lin Wu}},\ }\bibfield  {title}
  {\enquote {\bibinfo {title} {Three-dimensional large eddy simulation and
  vorticity analysis of unsteady cavitating flow around a twisted hydrofoil},}\
  }\href {\doibase 10.1016/S1001-6058(11)60390-X} {\bibfield  {journal}
  {\bibinfo  {journal} {Journal of Hydrodynamics}\ }\textbf {\bibinfo {volume}
  {25}},\ \bibinfo {pages} {510--519} (\bibinfo {year}
  {2013}{\natexlab{b}})}\BibitemShut {NoStop}%
\bibitem [{\citenamefont {Gnanaskandan}\ and\ \citenamefont
  {Mahesh}(2016{\natexlab{a}})}]{Gnanaskandan2016a}%
  \BibitemOpen
  \bibfield  {author} {\bibinfo {author} {\bibfnamefont {A.}~\bibnamefont
  {Gnanaskandan}}\ and\ \bibinfo {author} {\bibfnamefont {K.}~\bibnamefont
  {Mahesh}},\ }\bibfield  {title} {\enquote {\bibinfo {title} {Large eddy
  simulation of the transition from sheet to cloud cavitation over a wedge},}\
  }\href {\doibase 10.1016/j.ijmultiphaseflow.2016.03.015} {\bibfield
  {journal} {\bibinfo  {journal} {International Journal of Multiphase Flow}\
  }\textbf {\bibinfo {volume} {83}},\ \bibinfo {pages} {86--102} (\bibinfo
  {year} {2016}{\natexlab{a}})}\BibitemShut {NoStop}%
\bibitem [{\citenamefont {Chen}\ \emph {et~al.}(2019)\citenamefont {Chen},
  \citenamefont {Li}, \citenamefont {Gong}, \citenamefont {Chen},\ and\
  \citenamefont {Lu}}]{Chen2019}%
  \BibitemOpen
  \bibfield  {author} {\bibinfo {author} {\bibfnamefont {Y.}~\bibnamefont
  {Chen}}, \bibinfo {author} {\bibfnamefont {J.}~\bibnamefont {Li}}, \bibinfo
  {author} {\bibfnamefont {Z.}~\bibnamefont {Gong}}, \bibinfo {author}
  {\bibfnamefont {X.}~\bibnamefont {Chen}}, \ and\ \bibinfo {author}
  {\bibfnamefont {C.}~\bibnamefont {Lu}},\ }\bibfield  {title} {\enquote
  {\bibinfo {title} {Large eddy simulation and investigation on the
  laminar-turbulent transition and turbulence-cavitation interaction in the
  cavitating flow around hydrofoil},}\ }\href {\doibase
  10.1016/j.ijmultiphaseflow.2018.10.012} {\bibfield  {journal} {\bibinfo
  {journal} {International Journal of Multiphase Flow}\ }\textbf {\bibinfo
  {volume} {112}},\ \bibinfo {pages} {300--322} (\bibinfo {year}
  {2019})}\BibitemShut {NoStop}%
\bibitem [{\citenamefont {Sun}\ \emph {et~al.}(2020)\citenamefont {Sun},
  \citenamefont {Wang}, \citenamefont {Zou},\ and\ \citenamefont
  {Wang}}]{Sun2020}%
  \BibitemOpen
  \bibfield  {author} {\bibinfo {author} {\bibfnamefont {T.}~\bibnamefont
  {Sun}}, \bibinfo {author} {\bibfnamefont {Z.}~\bibnamefont {Wang}}, \bibinfo
  {author} {\bibfnamefont {L.}~\bibnamefont {Zou}}, \ and\ \bibinfo {author}
  {\bibfnamefont {H.}~\bibnamefont {Wang}},\ }\bibfield  {title} {\enquote
  {\bibinfo {title} {Numerical investigation of positive effects of ventilated
  cavitation around a naca66 hydrofoil},}\ }\href {\doibase
  10.1016/j.oceaneng.2019.106831} {\bibfield  {journal} {\bibinfo  {journal}
  {Ocean Engineering}\ }\textbf {\bibinfo {volume} {197}},\ \bibinfo {pages}
  {106831} (\bibinfo {year} {2020})}\BibitemShut {NoStop}%
\bibitem [{\citenamefont {Egerer}\ \emph {et~al.}(2014)\citenamefont {Egerer},
  \citenamefont {Hickel}, \citenamefont {Schmidt},\ and\ \citenamefont
  {Adams}}]{Egerer2014}%
  \BibitemOpen
  \bibfield  {author} {\bibinfo {author} {\bibfnamefont {C.~P.}\ \bibnamefont
  {Egerer}}, \bibinfo {author} {\bibfnamefont {S.}~\bibnamefont {Hickel}},
  \bibinfo {author} {\bibfnamefont {S.~J.}\ \bibnamefont {Schmidt}}, \ and\
  \bibinfo {author} {\bibfnamefont {N.~A.}\ \bibnamefont {Adams}},\ }\bibfield
  {title} {\enquote {\bibinfo {title} {Large-eddy simulation of turbulent
  cavitating flow in a micro channel},}\ }\href {\doibase 10.1063/1.4891325}
  {\bibfield  {journal} {\bibinfo  {journal} {Physics of Fluids}\ }\textbf
  {\bibinfo {volume} {26}},\ \bibinfo {pages} {085102} (\bibinfo {year}
  {2014})}\BibitemShut {NoStop}%
\bibitem [{\citenamefont {Barre}\ \emph {et~al.}(2009)\citenamefont {Barre},
  \citenamefont {Rolland}, \citenamefont {Boitel}, \citenamefont {Goncalves},\
  and\ \citenamefont {Patella}}]{Barre2009}%
  \BibitemOpen
  \bibfield  {author} {\bibinfo {author} {\bibfnamefont {S.}~\bibnamefont
  {Barre}}, \bibinfo {author} {\bibfnamefont {J.}~\bibnamefont {Rolland}},
  \bibinfo {author} {\bibfnamefont {G.}~\bibnamefont {Boitel}}, \bibinfo
  {author} {\bibfnamefont {E.}~\bibnamefont {Goncalves}}, \ and\ \bibinfo
  {author} {\bibfnamefont {R.~F.}\ \bibnamefont {Patella}},\ }\bibfield
  {title} {\enquote {\bibinfo {title} {Experiments and modeling of cavitating
  flows in {V}enturi: Attached sheet cavitation},}\ }\href {\doibase
  10.1016/j.euromechflu.2008.09.001} {\bibfield  {journal} {\bibinfo  {journal}
  {European Journal of Mechanics B/Fluids}\ }\textbf {\bibinfo {volume} {28}},\
  \bibinfo {pages} {444--464} (\bibinfo {year} {2009})}\BibitemShut {NoStop}%
\bibitem [{\citenamefont {Merkle}, \citenamefont {Feng},\ and\ \citenamefont
  {Buelow}(1998)}]{Merkle98}%
  \BibitemOpen
  \bibfield  {author} {\bibinfo {author} {\bibfnamefont {C.}~\bibnamefont
  {Merkle}}, \bibinfo {author} {\bibfnamefont {J.}~\bibnamefont {Feng}}, \ and\
  \bibinfo {author} {\bibfnamefont {P.}~\bibnamefont {Buelow}},\ }\bibfield
  {title} {\enquote {\bibinfo {title} {Computation modeling of the dynamics of
  sheet cavitation},}\ }in\ \href@noop {} {\emph {\bibinfo {booktitle}
  {3$^{rd}$ International Symposium on Cavitation CAV1998}}}\ (\bibinfo
  {address} {Grenoble, France},\ \bibinfo {year} {1998})\BibitemShut {NoStop}%
\bibitem [{\citenamefont {Smith}(1990)}]{Smith90}%
  \BibitemOpen
  \bibfield  {author} {\bibinfo {author} {\bibfnamefont {B.}~\bibnamefont
  {Smith}},\ }\bibfield  {title} {\enquote {\bibinfo {title} {The $k - kl$
  turbulence model and wall layer model for compressible flows},}\ }in\ \href
  {\doibase 10.2514/6.1990-1483} {\emph {\bibinfo {booktitle} {AIAA 90--1483,
  $21^{st}$ Fluid and Plasma Dynamics Conference}}}\ (\bibinfo {address}
  {Seattle, Washington, USA},\ \bibinfo {year} {1990})\BibitemShut {NoStop}%
\bibitem [{\citenamefont {Smith}(1994)}]{Smith94}%
  \BibitemOpen
  \bibfield  {author} {\bibinfo {author} {\bibfnamefont {B.}~\bibnamefont
  {Smith}},\ }\bibfield  {title} {\enquote {\bibinfo {title} {A near wall model
  for the $k-l$ two equation turbulence model},}\ }in\ \href {\doibase
  10.2514/6.1994-2386} {\emph {\bibinfo {booktitle} {AIAA 94--2386, $25^{sh}$
  Fluid Dynamics Conference}}}\ (\bibinfo {address} {Colorado Springs,
  Colorado, USA},\ \bibinfo {year} {1994})\BibitemShut {NoStop}%
\bibitem [{\citenamefont {Goncalvès}\ and\ \citenamefont
  {Decaix}(2012)}]{Goncalves2012}%
  \BibitemOpen
  \bibfield  {author} {\bibinfo {author} {\bibfnamefont {E.}~\bibnamefont
  {Goncalvès}}\ and\ \bibinfo {author} {\bibfnamefont {J.}~\bibnamefont
  {Decaix}},\ }\bibfield  {title} {\enquote {\bibinfo {title} {Wall model and
  mesh influence study for partial cavities},}\ }\href {\doibase
  10.1016/j.euromechflu.2011.08.002} {\bibfield  {journal} {\bibinfo  {journal}
  {European Journal of Mechanics B/Fluids}\ }\textbf {\bibinfo {volume} {31}},\
  \bibinfo {pages} {12--29} (\bibinfo {year} {2012})}\BibitemShut {NoStop}%
\bibitem [{\citenamefont {Charrière}, \citenamefont {Decaix},\ and\
  \citenamefont {Goncalvès}(2015)}]{Charriere2015}%
  \BibitemOpen
  \bibfield  {author} {\bibinfo {author} {\bibfnamefont {B.}~\bibnamefont
  {Charrière}}, \bibinfo {author} {\bibfnamefont {J.}~\bibnamefont {Decaix}},
  \ and\ \bibinfo {author} {\bibfnamefont {E.}~\bibnamefont {Goncalvès}},\
  }\bibfield  {title} {\enquote {\bibinfo {title} {A comparative study of
  cavitation model in a {V}enturi flow},}\ }\href {\doibase
  10.1016/j.euromechflu.2014.10.003} {\bibfield  {journal} {\bibinfo  {journal}
  {European Journal of Mechanics B/Fluids}\ }\textbf {\bibinfo {volume} {49}},\
  \bibinfo {pages} {287--297} (\bibinfo {year} {2015})}\BibitemShut {NoStop}%
\bibitem [{\citenamefont {Reboud}, \citenamefont {Stutz},\ and\ \citenamefont
  {Coutier}(1998)}]{Reboud98}%
  \BibitemOpen
  \bibfield  {author} {\bibinfo {author} {\bibfnamefont {J.-L.}\ \bibnamefont
  {Reboud}}, \bibinfo {author} {\bibfnamefont {B.}~\bibnamefont {Stutz}}, \
  and\ \bibinfo {author} {\bibfnamefont {O.}~\bibnamefont {Coutier}},\
  }\bibfield  {title} {\enquote {\bibinfo {title} {Two-phase flow structure of
  cavitation: experiment and modelling of unsteady effects},}\ }in\ \href@noop
  {} {\emph {\bibinfo {booktitle} {3$^{rd}$ International Symposium on
  Cavitation CAV1998}}}\ (\bibinfo {address} {Grenoble, France},\ \bibinfo
  {year} {1998})\BibitemShut {NoStop}%
\bibitem [{\citenamefont {Decaix}\ and\ \citenamefont
  {Goncalves}(2012)}]{Decaix12}%
  \BibitemOpen
  \bibfield  {author} {\bibinfo {author} {\bibfnamefont {J.}~\bibnamefont
  {Decaix}}\ and\ \bibinfo {author} {\bibfnamefont {E.}~\bibnamefont
  {Goncalves}},\ }\bibfield  {title} {\enquote {\bibinfo {title}
  {Time-dependent simulation of cavitating flow with $k-\ell$ turbulence
  models},}\ }\href {\doibase 10.1002/fld.2601} {\bibfield  {journal} {\bibinfo
   {journal} {Int. Journal for Numerical Methods in Fluids}\ }\textbf {\bibinfo
  {volume} {68}},\ \bibinfo {pages} {1053--1072} (\bibinfo {year}
  {2012})}\BibitemShut {NoStop}%
\bibitem [{\citenamefont {Dandois}(2014)}]{Dandois2014}%
  \BibitemOpen
  \bibfield  {author} {\bibinfo {author} {\bibfnamefont {J.}~\bibnamefont
  {Dandois}},\ }\bibfield  {title} {\enquote {\bibinfo {title} {Improvement of
  corner flow prediction using the quadratic constitutive relation},}\ }\href
  {\doibase 10.2514/1.J052976} {\bibfield  {journal} {\bibinfo  {journal} {AIAA
  Journal}\ }\textbf {\bibinfo {volume} {52}} (\bibinfo {year} {2014}),\
  10.2514/1.J052976}\BibitemShut {NoStop}%
\bibitem [{\citenamefont {Spalart}(2000)}]{Spalart2000}%
  \BibitemOpen
  \bibfield  {author} {\bibinfo {author} {\bibfnamefont {P.~R.}\ \bibnamefont
  {Spalart}},\ }\bibfield  {title} {\enquote {\bibinfo {title} {Strategies for
  turbulence modelling and simulations},}\ }\href {\doibase
  10.1016/S0142-727X(00)00007-2} {\bibfield  {journal} {\bibinfo  {journal}
  {International Journal of Heat and Fluid Flow}\ }\textbf {\bibinfo {volume}
  {21}},\ \bibinfo {pages} {252--263} (\bibinfo {year} {2000})}\BibitemShut
  {NoStop}%
\bibitem [{\citenamefont {Saurel}, \citenamefont {Petitpas},\ and\
  \citenamefont {Abgrall}(2008)}]{Saurel2008}%
  \BibitemOpen
  \bibfield  {author} {\bibinfo {author} {\bibfnamefont {R.}~\bibnamefont
  {Saurel}}, \bibinfo {author} {\bibfnamefont {F.}~\bibnamefont {Petitpas}}, \
  and\ \bibinfo {author} {\bibfnamefont {R.}~\bibnamefont {Abgrall}},\
  }\bibfield  {title} {\enquote {\bibinfo {title} {Modelling phase transition
  in metastable liquids: Application to cavitating and flashing flows},}\
  }\href {\doibase 10.1017/S0022112008002061} {\bibfield  {journal} {\bibinfo
  {journal} {Journal of Fluid Mechanics}\ }\textbf {\bibinfo {volume} {607}},\
  \bibinfo {pages} {313--350} (\bibinfo {year} {2008})}\BibitemShut {NoStop}%
\bibitem [{\citenamefont {Goncalvès}(2013)}]{Goncalves2013}%
  \BibitemOpen
  \bibfield  {author} {\bibinfo {author} {\bibfnamefont {E.}~\bibnamefont
  {Goncalvès}},\ }\bibfield  {title} {\enquote {\bibinfo {title} {Numerical
  study of expansion tube problems: Towards the simulation of cavitation},}\
  }\href {\doibase 10.1016/j.compfluid.2012.11.019} {\bibfield  {journal}
  {\bibinfo  {journal} {Computers \& Fluids}\ }\textbf {\bibinfo {volume}
  {72}},\ \bibinfo {pages} {1--19} (\bibinfo {year} {2013})}\BibitemShut
  {NoStop}%
\bibitem [{\citenamefont {Wallis}(1967)}]{Wallis67}%
  \BibitemOpen
  \bibfield  {author} {\bibinfo {author} {\bibfnamefont {G.}~\bibnamefont
  {Wallis}},\ }\href@noop {} {\enquote {\bibinfo {title} {One-dimensional
  two-phase flow},}\ }\bibinfo {howpublished} {New York: McGraw-Hill} (\bibinfo
  {year} {1967})\BibitemShut {NoStop}%
\bibitem [{\citenamefont {Charri\`{e}re}(2015)}]{these_Charriere2015}%
  \BibitemOpen
  \bibfield  {author} {\bibinfo {author} {\bibfnamefont {B.}~\bibnamefont
  {Charri\`{e}re}},\ }\emph {\bibinfo {title} {Mod\'{e}lisation et simulation
  d'\'{e}coulements turbulents cavitants avec un mod\`{e}le de transport de
  taux de vide; in English: Modelling and simulation of cavitating turbulent
  flows using a transport equation of void ratio.}},\ \href@noop {} {Ph.D.
  thesis},\ \bibinfo  {school} {{Universit{\'e} Grenoble Alpes}, Laboratoire
  des \'{E}coulements G\'{e}ophysiques et Industrielles}, \bibinfo {address}
  {Grenoble, France} (\bibinfo {year} {2015})\BibitemShut {NoStop}%
\bibitem [{\citenamefont {Goncalves}\ and\ \citenamefont
  {Patella}(2009)}]{Goncalves2009}%
  \BibitemOpen
  \bibfield  {author} {\bibinfo {author} {\bibfnamefont {E.}~\bibnamefont
  {Goncalves}}\ and\ \bibinfo {author} {\bibfnamefont {R.~F.}\ \bibnamefont
  {Patella}},\ }\bibfield  {title} {\enquote {\bibinfo {title} {Numerical
  simulation of cavitating flows with homogeneous models},}\ }\href {\doibase
  10.1016/j.compfluid.2009.03.001} {\bibfield  {journal} {\bibinfo  {journal}
  {Computers \& Fluids}\ }\textbf {\bibinfo {volume} {38}},\ \bibinfo {pages}
  {1682--1696} (\bibinfo {year} {2009})}\BibitemShut {NoStop}%
\bibitem [{\citenamefont {Turkel}(1987)}]{Turkel87}%
  \BibitemOpen
  \bibfield  {author} {\bibinfo {author} {\bibfnamefont {E.}~\bibnamefont
  {Turkel}},\ }\bibfield  {title} {\enquote {\bibinfo {title} {Preconditioned
  methods for solving the incompressible and low speed compressible
  equations},}\ }\href {\doibase 10.1016/0021-9991(87)90084-2} {\bibfield
  {journal} {\bibinfo  {journal} {Journal of Computational Physics}\ }\textbf
  {\bibinfo {volume} {72}},\ \bibinfo {pages} {277--298} (\bibinfo {year}
  {1987})}\BibitemShut {NoStop}%
\bibitem [{\citenamefont {Choi}\ and\ \citenamefont {Merkle}(1993)}]{Choi1993}%
  \BibitemOpen
  \bibfield  {author} {\bibinfo {author} {\bibfnamefont {Y.}~\bibnamefont
  {Choi}}\ and\ \bibinfo {author} {\bibfnamefont {C.}~\bibnamefont {Merkle}},\
  }\bibfield  {title} {\enquote {\bibinfo {title} {The application of
  preconditioning to viscous flows},}\ }\href {\doibase 10.1006/jcph.1993.1069}
  {\bibfield  {journal} {\bibinfo  {journal} {Journal of Computational
  Physics}\ }\textbf {\bibinfo {volume} {105}},\ \bibinfo {pages} {207--223}
  (\bibinfo {year} {1993})}\BibitemShut {NoStop}%
\bibitem [{\citenamefont {Spiteri}\ and\ \citenamefont
  {Ruuth}(2002)}]{Spiteri2002}%
  \BibitemOpen
  \bibfield  {author} {\bibinfo {author} {\bibfnamefont {R.~J.}\ \bibnamefont
  {Spiteri}}\ and\ \bibinfo {author} {\bibfnamefont {S.~J.}\ \bibnamefont
  {Ruuth}},\ }\bibfield  {title} {\enquote {\bibinfo {title} {A new class of
  optimal high-order strong-stability-preserving time discretization
  methods},}\ }\href {\doibase 10.1137/s0036142901389025} {\bibfield  {journal}
  {\bibinfo  {journal} {SIAM Journal on Numerical Analysis}\ }\textbf {\bibinfo
  {volume} {40}},\ \bibinfo {pages} {469--491} (\bibinfo {year}
  {2002})}\BibitemShut {NoStop}%
\bibitem [{\citenamefont {Gottlieb}(2005)}]{Gottlieb2005}%
  \BibitemOpen
  \bibfield  {author} {\bibinfo {author} {\bibfnamefont {S.}~\bibnamefont
  {Gottlieb}},\ }\bibfield  {title} {\enquote {\bibinfo {title} {On high order
  strong stability preserving {Runge-Kutta} and multi step time
  discretizations},}\ }\href {\doibase 10.1007/s10915-004-4635-5} {\bibfield
  {journal} {\bibinfo  {journal} {Journal of Scientific Computing}\ }\textbf
  {\bibinfo {volume} {25}},\ \bibinfo {pages} {105--128} (\bibinfo {year}
  {2005})}\BibitemShut {NoStop}%
\bibitem [{\citenamefont {Jameson}, \citenamefont {Schmidt},\ and\
  \citenamefont {Turkel}(1981)}]{Jameson1981}%
  \BibitemOpen
  \bibfield  {author} {\bibinfo {author} {\bibfnamefont {A.}~\bibnamefont
  {Jameson}}, \bibinfo {author} {\bibfnamefont {W.}~\bibnamefont {Schmidt}}, \
  and\ \bibinfo {author} {\bibfnamefont {E.}~\bibnamefont {Turkel}},\
  }\bibfield  {title} {\enquote {\bibinfo {title} {Numerical solutions of the
  {E}uler equations by finite volume methods using {Runge-Kutta} time-stepping
  schemes},}\ }in\ \href {\doibase 10.2514/6.1981-1259} {\emph {\bibinfo
  {booktitle} {AIAA Paper 81--1259, Proceedings of the {AIAA} 14th Fluid and
  Plasma Dynamic Conference}}}\ (\bibinfo {address} {Palo Alto, Californa,
  USA},\ \bibinfo {year} {1981})\BibitemShut {NoStop}%
\bibitem [{\citenamefont {Goncalvès}\ and\ \citenamefont
  {Charrière}(2014)}]{Goncalves2014}%
  \BibitemOpen
  \bibfield  {author} {\bibinfo {author} {\bibfnamefont {E.}~\bibnamefont
  {Goncalvès}}\ and\ \bibinfo {author} {\bibfnamefont {B.}~\bibnamefont
  {Charrière}},\ }\bibfield  {title} {\enquote {\bibinfo {title} {Modelling
  for isothermal cavitation with a four-equation model},}\ }\href {\doibase
  10.1016/j.ijmultiphaseflow.2013.10.015} {\bibfield  {journal} {\bibinfo
  {journal} {International Journal of Multiphase Flow}\ }\textbf {\bibinfo
  {volume} {59}},\ \bibinfo {pages} {54--72} (\bibinfo {year}
  {2014})}\BibitemShut {NoStop}%
\bibitem [{\citenamefont {Goncalves}\ and\ \citenamefont
  {Zeidan}(2018)}]{Goncalves2018}%
  \BibitemOpen
  \bibfield  {author} {\bibinfo {author} {\bibfnamefont {E.}~\bibnamefont
  {Goncalves}}\ and\ \bibinfo {author} {\bibfnamefont {D.}~\bibnamefont
  {Zeidan}},\ }\bibfield  {title} {\enquote {\bibinfo {title} {Simulation of
  compressible two-phase flows using a void ratio transport equation},}\ }\href
  {\doibase 10.4208/cicp.OA-2017-0024} {\bibfield  {journal} {\bibinfo
  {journal} {Computational Physics}\ }\textbf {\bibinfo {volume} {24}},\
  \bibinfo {pages} {167--203} (\bibinfo {year} {2018})}\BibitemShut {NoStop}%
\bibitem [{\citenamefont {Zhou}\ and\ \citenamefont {Wang}(2008)}]{Zhou2008}%
  \BibitemOpen
  \bibfield  {author} {\bibinfo {author} {\bibfnamefont {L.}~\bibnamefont
  {Zhou}}\ and\ \bibinfo {author} {\bibfnamefont {Z.}~\bibnamefont {Wang}},\
  }\bibfield  {title} {\enquote {\bibinfo {title} {Numerical simulation of
  cavitation around a hydrofoil and evaluation of a rng k-epsilon model},}\
  }\href {\doibase 10.1115/1.2816009} {\bibfield  {journal} {\bibinfo
  {journal} {Journal of Fluids Engineering}\ }\textbf {\bibinfo {volume}
  {130}},\ \bibinfo {pages} {011302} (\bibinfo {year} {2008})}\BibitemShut
  {NoStop}%
\bibitem [{\citenamefont {Ducoin}, \citenamefont {Huang},\ and\ \citenamefont
  {Young}(2012)}]{Ducoin2012}%
  \BibitemOpen
  \bibfield  {author} {\bibinfo {author} {\bibfnamefont {A.}~\bibnamefont
  {Ducoin}}, \bibinfo {author} {\bibfnamefont {B.}~\bibnamefont {Huang}}, \
  and\ \bibinfo {author} {\bibfnamefont {Y.}~\bibnamefont {Young}},\ }\bibfield
   {title} {\enquote {\bibinfo {title} {Numerical modeling of unsteady
  cavitating flows around a stationary hydrofoil},}\ }\href {\doibase
  10.1155/2012/215678} {\bibfield  {journal} {\bibinfo  {journal}
  {International Journal of Rotating Machinery}\ ,\ \bibinfo {pages} {215678}}
  (\bibinfo {year} {2012})}\BibitemShut {NoStop}%
\bibitem [{\citenamefont {Ji}\ \emph {et~al.}(2014)\citenamefont {Ji},
  \citenamefont {Luo}, \citenamefont {Arndt},\ and\ \citenamefont
  {Wu}}]{Ji2014}%
  \BibitemOpen
  \bibfield  {author} {\bibinfo {author} {\bibfnamefont {B.}~\bibnamefont
  {Ji}}, \bibinfo {author} {\bibfnamefont {X.}~\bibnamefont {Luo}}, \bibinfo
  {author} {\bibfnamefont {R.}~\bibnamefont {Arndt}}, \ and\ \bibinfo {author}
  {\bibfnamefont {Y.}~\bibnamefont {Wu}},\ }\bibfield  {title} {\enquote
  {\bibinfo {title} {Numerical simulation of three dimensional cavitation
  shedding dynamics with special emphasis on cavitation-vortex interaction},}\
  }\href {\doibase 10.1016/j.oceaneng.2014.05.005} {\bibfield  {journal}
  {\bibinfo  {journal} {Ocean Engineering}\ }\textbf {\bibinfo {volume} {87}},\
  \bibinfo {pages} {65--77} (\bibinfo {year} {2014})}\BibitemShut {NoStop}%
\bibitem [{\citenamefont {Dular}\ and\ \citenamefont
  {Bachert}(2009)}]{Dular2009}%
  \BibitemOpen
  \bibfield  {author} {\bibinfo {author} {\bibfnamefont {M.}~\bibnamefont
  {Dular}}\ and\ \bibinfo {author} {\bibfnamefont {R.}~\bibnamefont
  {Bachert}},\ }\bibfield  {title} {\enquote {\bibinfo {title} {The issue of
  {S}trouhal number definition in cavitating flow},}\ }\href@noop {} {\bibfield
   {journal} {\bibinfo  {journal} {Journal of Mechanical Engineering}\ }\textbf
  {\bibinfo {volume} {55}},\ \bibinfo {pages} {666--674} (\bibinfo {year}
  {2009})}\BibitemShut {NoStop}%
\bibitem [{\citenamefont {Hammond}\ and\ \citenamefont
  {Redekopp}(1998)}]{Hammond1998}%
  \BibitemOpen
  \bibfield  {author} {\bibinfo {author} {\bibfnamefont {D.}~\bibnamefont
  {Hammond}}\ and\ \bibinfo {author} {\bibfnamefont {L.}~\bibnamefont
  {Redekopp}},\ }\bibfield  {title} {\enquote {\bibinfo {title} {Local and
  global instability properties of separation bubbles},}\ }\href {\doibase
  10.1016/S0997-7546(98)80056-3} {\bibfield  {journal} {\bibinfo  {journal}
  {European Journal of Mechanics B/Fluids}\ }\textbf {\bibinfo {volume} {17}},\
  \bibinfo {pages} {145--164} (\bibinfo {year} {1998})}\BibitemShut {NoStop}%
\bibitem [{\citenamefont {Rist}\ and\ \citenamefont
  {Maucher}(2002)}]{Rist2002}%
  \BibitemOpen
  \bibfield  {author} {\bibinfo {author} {\bibfnamefont {U.}~\bibnamefont
  {Rist}}\ and\ \bibinfo {author} {\bibfnamefont {U.}~\bibnamefont {Maucher}},\
  }\bibfield  {title} {\enquote {\bibinfo {title} {Investigations of
  time-growing instabilities in laminar separation bubbles},}\ }\href {\doibase
  10.1016/S0997-7546(02)01205-0} {\bibfield  {journal} {\bibinfo  {journal}
  {European Journal of Mechanics B/Fluids}\ }\textbf {\bibinfo {volume} {21}},\
  \bibinfo {pages} {495--509} (\bibinfo {year} {2002})}\BibitemShut {NoStop}%
\bibitem [{\citenamefont {Gnanaskandan}\ and\ \citenamefont
  {Mahesh}(2016{\natexlab{b}})}]{Gnanaskandan2016}%
  \BibitemOpen
  \bibfield  {author} {\bibinfo {author} {\bibfnamefont {A.}~\bibnamefont
  {Gnanaskandan}}\ and\ \bibinfo {author} {\bibfnamefont {K.}~\bibnamefont
  {Mahesh}},\ }\bibfield  {title} {\enquote {\bibinfo {title} {Numerical
  investigation of near-wake characteristics of cavitating flow over a circular
  cylinder},}\ }\href {\doibase 10.1017/jfm.2016.19} {\bibfield  {journal}
  {\bibinfo  {journal} {Journal of Fluid Mechanics}\ }\textbf {\bibinfo
  {volume} {790}},\ \bibinfo {pages} {453--491} (\bibinfo {year}
  {2016}{\natexlab{b}})}\BibitemShut {NoStop}%
\bibitem [{\citenamefont {Budich}, \citenamefont {Schmidt},\ and\ \citenamefont
  {Adams}(2018)}]{Budich2018}%
  \BibitemOpen
  \bibfield  {author} {\bibinfo {author} {\bibfnamefont {B.}~\bibnamefont
  {Budich}}, \bibinfo {author} {\bibfnamefont {S.~J.}\ \bibnamefont {Schmidt}},
  \ and\ \bibinfo {author} {\bibfnamefont {N.~A.}\ \bibnamefont {Adams}},\
  }\bibfield  {title} {\enquote {\bibinfo {title} {Numerical simulation and
  analysis of condensation shock in cavitating flow},}\ }\href {\doibase
  10.1017/jfm.2017.882} {\bibfield  {journal} {\bibinfo  {journal} {Journal of
  Fluid Mechanics}\ }\textbf {\bibinfo {volume} {838}},\ \bibinfo {pages}
  {759--813} (\bibinfo {year} {2018})}\BibitemShut {NoStop}%
\bibitem [{\citenamefont {Towne}, \citenamefont {Schmidt},\ and\ \citenamefont
  {Colonius}(2018)}]{Towne2018}%
  \BibitemOpen
  \bibfield  {author} {\bibinfo {author} {\bibfnamefont {A.}~\bibnamefont
  {Towne}}, \bibinfo {author} {\bibfnamefont {O.~T.}\ \bibnamefont {Schmidt}},
  \ and\ \bibinfo {author} {\bibfnamefont {T.}~\bibnamefont {Colonius}},\
  }\bibfield  {title} {\enquote {\bibinfo {title} {Spectral proper orthogonal
  decomposition and its relationship to dynamic mode decomposition and
  resolvent analysis},}\ }\href {\doibase 10.1017/jfm.2018.283} {\bibfield
  {journal} {\bibinfo  {journal} {Journal of Fluid Mechanics}\ }\textbf
  {\bibinfo {volume} {847}},\ \bibinfo {pages} {821--867} (\bibinfo {year}
  {2018})}\BibitemShut {NoStop}%
\bibitem [{\citenamefont {Schmidt}\ and\ \citenamefont
  {Colonius}(2020)}]{Schmidt2020}%
  \BibitemOpen
  \bibfield  {author} {\bibinfo {author} {\bibfnamefont {O.~T.}\ \bibnamefont
  {Schmidt}}\ and\ \bibinfo {author} {\bibfnamefont {T.}~\bibnamefont
  {Colonius}},\ }\bibfield  {title} {\enquote {\bibinfo {title} {Guide to
  spectral proper orthogonal decomposition},}\ }\href {\doibase
  10.2514/1.J058809} {\bibfield  {journal} {\bibinfo  {journal} {AIAA Journal}\
  }\textbf {\bibinfo {volume} {58}} (\bibinfo {year} {2020}),\
  10.2514/1.J058809}\BibitemShut {NoStop}%
\bibitem [{\citenamefont {Chu}(1965)}]{Chu1965}%
  \BibitemOpen
  \bibfield  {author} {\bibinfo {author} {\bibfnamefont {B.}~\bibnamefont
  {Chu}},\ }\bibfield  {title} {\enquote {\bibinfo {title} {On the energy
  transfer to small disturbances in fluid flow {(Part I)}},}\ }\href {\doibase
  10.1007/BF01387235} {\bibfield  {journal} {\bibinfo  {journal} {Acta
  Mechanica}\ }\textbf {\bibinfo {volume} {1}},\ \bibinfo {pages} {215--234}
  (\bibinfo {year} {1965})}\BibitemShut {NoStop}%
\end{thebibliography}%

\end{document}